\documentclass[10pt]{iopart}
\usepackage{lineno,bm}
\usepackage{cite}
\usepackage{iopams}
\expandafter\let\csname equation*\endcsname\relax
\expandafter\let\csname endequation*\endcsname\relax
\usepackage{amsmath}
\usepackage{dsfont}
\usepackage{snotez}
\usepackage[dvipsnames]{xcolor}
\DeclareMathOperator{\divi}{div}
\newcommand{\ib}{\textrm{i}}
\newcommand{\gs}{\textrm{g}}

\newcommand{\vn}{\mathbf{n}}
\newcommand{\vx}{\mathbf{x}}
\newcommand{\vE}{\mathbf{E}}
\newcommand{\vr}{\mathbf{r}}
\newcommand{\p}{{\hspace{-.3ex} p}}
\newcommand{\0}{{\hspace{-.5ex}\rm o}}
\newcommand{\1}{{\hspace{-.1ex}\rm o}}
\newcommand{\diag}{\mathrm{diag}}
\newcommand{\sss}{\scriptscriptstyle}

\usepackage{graphicx}
\usepackage{xcolor} \definecolor{darkgreen}{rgb}{0,.5,0}
\usepackage[colorlinks,filecolor=blue,citecolor=darkgreen,unicode]{hyperref}
\begin{document}
\title[Self-action in Gravity]{Self-action in Gravity}
\author{Nail Khusnutdinov$^{1,2}$}
\ead{nail.khusnutdinov@gmail.com}
\address{$^1$ Centro de Matem\'atica, Computa\c{c}\~ao e Cogni\c{c}\~ao, Universidade Federal do ABC, 09210-170 Santo Andr\'e, SP, Brazil}
\address{$^2$ Regional Scientific and Educational Mathematical Center of Kazan Federal University, Kremlevskaya 18, Kazan, Russia 420008}
	
\begin{abstract}
On a particle moving with variable acceleration in the flat space-time affects the self-force due to outgoing radiation. The gravitational fields bring an additional contribution to self-force due to scattering waves on the curved backgrounds. This force is not zero even for a particle at rest. A review of the self-interaction in the gravitational field is presented. We consider the self-force for the particle connected with the vector and scalar fields. Different backgrounds are considered -- black holes, stars, topological defects, and wormholes. 
\end{abstract}

\submitto{CQG}	
\maketitle\tableofcontents
\section{Introduction}

The phenomenon\footnote{The signature $(-,+,+,+)$ is used. The geometrical objects are that as in Hawking and Ellis book \cite{Hawking:1973:TLSSoS} and we use calligraphic notations for them $\mathcal{R}, \mathcal{R}_{\mu\nu}$. The Greek letters include time $\alpha,\beta, \ldots = 0,1,2,3$ and Latin -- the space $i,j,\ldots = 1,2,3$, only.   The $\gs^{(3)} = \det \gs_{ij}$ is determinant of $3D$ metric  and $\cos\gamma = \cos\theta\cos\theta' + \sin\theta\sin\theta' \cos(\varphi-\varphi')$ is cosine of spherical angle. The line element of unite $2$--sphere: $d\Omega_2 = d\theta^2 + \sin^2\theta d\varphi^2$.  $\square = \gs^{\mu\nu}\nabla_\mu\nabla_\nu$ and $\triangle = \gs^{ik}\nabla_i\nabla_k$ are corresponding $4D$ and $3D$ Laplace--Beltrami operator. The semicolon  denotes the covariant derivative, e.g. $u^\mu_{;\nu} = \nabla_\nu u^\mu$.} of self-interaction appears together with the creation of classical electrodynamics and up to now it is a subject of investigation from different points of view and attracts attention in connection with new applications and new ideas in fundamental physics. Considering that the electromagnetic field is an objective reality with own energy and momentum we have to take into account its interaction with electromagnetic particles. As long ago as 1881, Thompson showed in Ref. \cite{Thomson:1881:xoteamepbtmoeb} that the electromagnetic field gives a contribution to the mass of the uniform charged and moving sphere. Then, Fermi \cite{Fermi:1921:seducguespdme} proved that this electromagnetic mass-energy gravitates as usual mass and relativistic relation $m = E/c^2$ holds. To provide this relation, the non-electromagnetic force has to be taken into account \cite{Poincare:1906:slddl}. 

The phenomenon of self-force in Minkowski space-time was considered in detail in monographs \cite{Landau:1975:CTF,Sokolov:1986:RRE,Jackson:1999:CETE,Milonni:1994:QVItQE} and reviews \cite{Erber:1961:tctorr,Plass:1961:ceeomwrr,Klepikov:1985:rdfrcp,Krivitskii:1991:arrfiqe} (in curved space-time see Refs. \cite{Poisson:2004:tmoppics,Khusnutdinov:2005:Psegf,Poisson:2011:tmoppics,Barack:2018:sfarrigr}). The origin of the self-force in flat space-time is connected with the inertial properties of the electromagnetic field. Because the latter possesses energy and momentum then the radiation emitted by an electromagnetic particle carries out the momentum and due to conservation of this quantity, the velocity of the particle decreases. Therefore, the self-interaction force in flat space-time is the reaction to radiation, and as a consequence, the self-force is zero for uniformly moving particle and for a particle at rest.

The covariant self-interaction force acting on a particle with an electric charge, $q$, and velocity, $u^\mu$, in the flat Minkowski space-time is given by the famous Dirac--Lorentz (DL) expression
\begin{equation}
\mathcal{F}^\mu_{\textrm{DL}}=\frac{2q^2}{3}\ddot u^\mu P^\mu_\nu, \label{DL}
\end{equation}
where $P^\mu_\nu = \delta^\mu_\nu + u^\mu u_\nu$ is the projector on the particle's velocity and overdot is derivative with respect to the proper time. The derivation of this formula was made by Dirac \cite{Dirac:1938:ctore}. This force was considered in detail in non-relativistic and relativistic cases and was applied in different branches of physics. The discussion of these applications is out of the scope of our subject. We would like to call attention, for example, for the formula obtained by Pomeranchuk \cite{Landau:1975:CTF} which is applied for charged particles that move through a region filled by electric and magnetic fields. The self-force influences the movement of the particle. For the great velocity of the particle, the self-force prevails over the electromagnetic force, and the motion of the particle is determined by the self-force only. The energy of the particle after crossing the region with the electromagnetic field can not be greater than the threshold value $\mathcal{E}_{c}$ which is defined by the electromagnetic field and the self-force.

Note that the main peculiarities of the DL self-force in Minkowski space-time are given by Eq. (\ref{DL}). First of all, it involves the third derivative of the position of the particle, that is, the derivative of acceleration of the particle and therefore it is zero for a particle at rest, or uniformly moving particle and, additionally, for a uniformly accelerated particle (see the detail discussion in Ref. \cite{Plass:1961:ceeomwrr}). In the latter case, the self-force is zero but the particle loses the energy. The explanation of this contradiction may be found, for example, in Ref. \cite{Sokolov:1986:RRE}. Second, the expression for the self-force (\ref{DL}) is valid for an arbitrary trajectory of the particle and does not depend on the electromagnetic field. It is not necessary to obtain the expression for the self-force for each configuration of the electromagnetic field. Third, the self-force is perpendicular to the velocity which means perpendicularity of the acceleration and velocity of the particle.

To derive the expression (\ref{DL}) we have to make the "classical" renormalization of the particle's mass, which has no dependence on Planck constant. There is no correct explanation for this procedure in the framework of the classical field theory; the corrective renormalization procedure maybe only in the framework of quantum field theory \cite{Vilenkin:1975:tcpitpotseote,Efimov:1977:eseinft,%
Vilenkin:1979:cpitsep,Moniz:1977:rrinqe,Krivitskii:1991:arrfiqe,Higuchi:2006:rrocpitdmicaqe,Higuchi:2006:qrratgfd}. In the Refs. \cite{Vilenkin:1975:tcpitpotseote,Efimov:1977:eseinft,Vilenkin:1979:cpitsep}, the correspondence principle has proved namely, that correction to mass due to the massive operator in the limit $\hbar \to 0$ coincides exactly to that obtained in the classical field theory. It was proved for quantum electrodynamics \cite{Vilenkin:1975:tcpitpotseote} as well as scalar electrodynamics, scalar meson theory \cite{Vilenkin:1979:cpitsep}, and non-local field theory \cite{Efimov:1977:eseinft}. The particular case of a charged particle in an external constant electric and magnetic field was considered correct in the framework of quantum field theory by Ritus in Refs. \cite{Ritus:1978:moeamoiqeoacf,Ritus:1981:tmsoac}. He found the mass shift of the electron in the classical limit. To avoid this problem in the classical level, Dirac in Ref. \cite{Dirac:1938:ctore} suggested using a radiative Green function instead of the retarded Green function. The radiative Green function is defined as the half-difference of the retarded and advanced Green functions. Because both Green functions contain the same divergence, the half-difference of them has no divergence, because it cancels by definition of the radiative Green function. Another approach to self-force was developed by Wheeler and Feynman \cite{Wheeler:1945:iwtaatmor}. It was based on the idea that the self-force appearing due to the media in the Universe. 

In general relativity, the situation with the self-interaction force becomes more complicated. The point is that in the framework of this theory all kind of energies gravitates including fields. Einstein's equations say that the energy-momentum tensor of any matter or fields creates the gravitational field. Due to this, it becomes impossible to obtain the Green function of the electromagnetic field in an arbitrary gravitational field. Thus, we have to calculate the Green function for each configuration of the gravitational field. The local expansion of the Green function in a gravitational field \cite{DeWitt:1965:DTGF} shows us that there exist an infinite number of additional terms, except the standard ones which depend on the external gravitational field. The structure of the singularities of the Green function also changes in space-time of even dimensions -- the logarithmic divergence appears except the standard pole part \cite{Christensen:1978:rracgps}. Ultimately, this leads to violation of the Huygens principle in that sense that the flat or spherical wave being propagated in the curved space-time loses its form, acquiring "tails" \cite{DeWitt:1960:rdiagf}. Free particle, as a local object, tends to move by geodesic line in accordance with the principle of equivalence. The electromagnetic field connected with the particle is a non-local and extensive object. The gravitational field plays the role of the medium for an electromagnetic field and leads to the scattering of the latter. For this reason, the additional gravitationally-induced self-interaction force is appeared alongside with standard DL force (\ref{DL}). Due to the non-linearity of Einstein's equations, the same conclusion may be applied to the gravitation field of the particle -- the gravitational radiation of the particle causes the reaction of radiation. Nevertheless, as we will see later, there is a difference which concerns the origin of the self-forces in flat space-time and in curved space-times. In the latter case the self-force, in general, is not zero, even for a particle at rest in a static space-time when the radiation is absent. For this reason, the gravitationally induced self-force can not be recognized as a radiative reaction. 

In curved space-time the self-force has been considered in the background of black holes (see Sec. \ref{Sec: BH}), the topological defects (see Sec. \ref{Sec:TD}), and recently in wormholes space-times (Sec. \ref{Sec:WHV}) with non-trivial space-time's topology.   



\section{Self-Force. General Considerations}
\subsection{Flat Minkowski space-time}\label{Sec:Flat}

There are different approaches to find the expression for the self-force in flat space-time. We consider the simplest one due to Dirac \cite{Dirac:1931:qsitef}. The Maxwell equations
\begin{equation}
F^{\mu\nu}_{;\nu} = 4\pi J^\mu = \frac{4\pi q}{\sqrt{-\gs}} \int u^\mu(\tau) \delta^{(4)}(x - x(\tau)) d\tau,\ F_{(\mu\nu;\alpha)} =0, 
\end{equation}
in the Lorentz gauge have the following form 
\begin{equation}
 \square A^\mu = -4\pi J^\mu.
\end{equation}
Here, parenthesis means cyclic summation. The solution of these equations may be found with the help of the Green function obtained from the equation
\begin{equation}
 \square G_{\mu\nu'}(x,x') =- \gs_{\mu\nu'}\frac{\delta^{(4)}(x-x')}{\sqrt{-\gs}},
\end{equation}
where $\gs_{\mu\nu'}(x;x')$ is a bivector of parallel transport along geodesic from point $x'$ to $x$. Multiplying this equation by $4\pi J^{\nu'}(x')$ and integrating with respect to the measure $\sqrt{-\gs(x')} dx'$ we obtain 
\begin{equation}
 A_\mu = 4\pi q\int G_{\mu\nu'}(x;x(\tau'))u^{\nu'}(\tau')d\tau'.
\end{equation}
The vector Green function is expressed in terms of the massless scalar Green function $G_{\mu\nu'} = \gs_{\mu\nu'} G$ which obeys the equation 
\begin{equation}
 \square G(x,x') = -\frac{\delta^{(4)}(x-x')}{\sqrt{-\gs}}.
\end{equation}
There are two well-known solutions of this equation, namely, retarded and advanced Green functions
\begin{equation}
 G^{\textrm{ret}}(x;x') = \frac{\theta(t-t')}{4\pi} \delta (\sigma (x,x')),\  G^{\textrm{adv}}(x;x') = \frac{\theta(t'-t)}{4\pi} \delta (\sigma (x,x')),
\end{equation}
where $\sigma = -\tau^2/2$ is the half-square of the geodesic distance\footnote{Synge \cite{Synge:1960:RGT} called this quantity a "world function"} between points $x$ and $x'$. Taking into account the retarded Green function, we obtain the retarded electromagnetic potential in the form below
\begin{equation}
 A_\mu^{\textrm{ret}} = q\int \gs_{\mu\nu'}(x;x(\tau'))u^{\nu'}(\tau')\theta (t-t(\tau')) \delta (\sigma (x,x(\tau')))d\tau'. 
\end{equation}
Integrating over $\tau'$ we obtain the Li\'enard--Wiechert  potentials
\begin{equation}
 A_\mu^{\textrm{ret}} = \left.\frac{qu_{\mu}}{\dot \sigma}\right|_{\tau'=\tau^*}, 
\end{equation}
where the retarded time $\tau^*$ is a solution of the equation
\begin{equation}
 \sigma (x,x(\tau^*)) = 0,\ t-t(\tau^*) >0,
\end{equation}
and the overdot means derivative with respect of $\tau'$. 

\begin{figure}
\centerline{\hspace{-.1cm}\includegraphics[width=4.5truecm]{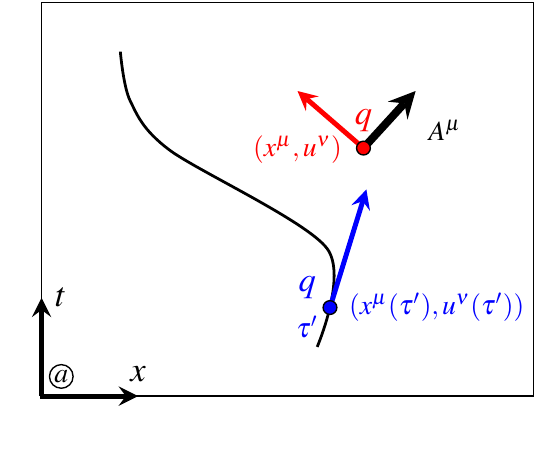}%
\hspace{-.25cm}\includegraphics[width=4.5truecm]{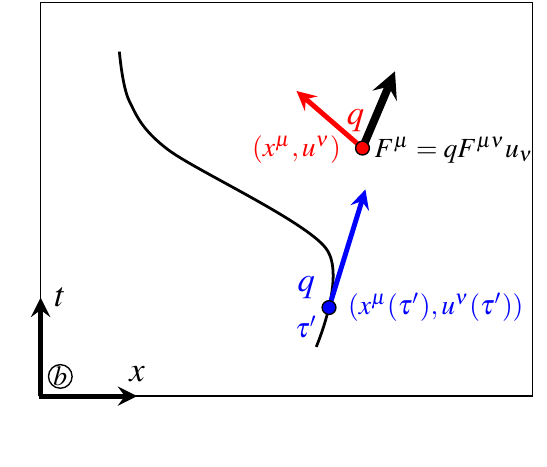}%
\hspace{-.25cm}\includegraphics[width=4.5truecm]{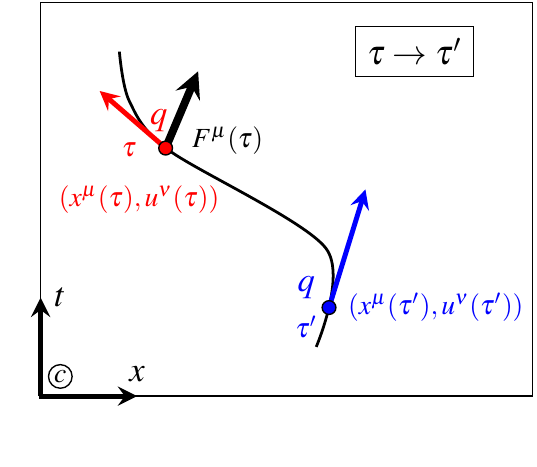}}
\caption{\footnotesize This picture shows us the steps of the procedure for calculating the self-force. (i) In the beginning, the electromagnetic potential is obtained at the place of the probe charge. (ii) Then the calculation of the Lorentz force on this probe particle is done. (iii) Then the probe particle is placed on the trajectory. In other words, the probe particle is considered as a real particle at the later moment and turns both points to each other.}\label{Fig_Method_Flat}
\end{figure}

Therefore, we write out the expression for the electromagnetic $4$-potential $A^\mu$ in the point of observation, $x^\mu$, due to the particle \label{theMethod}(see Fig. \ref{Fig_Method_Flat}a). In what follows let us consider a fictitious particle with charge $q$, velocity $u^\mu$ and calculate the Lorentz force acting on it using the standard procedure, by taking the expression: $F^\mu = q F^{\mu\nu}u_\nu$ (see Fig. \ref{Fig_Method_Flat}b). Then, we put the fictitious particle along the trajectory: $x^\mu \to x^\mu (\tau),\ u^\mu \to u^\mu (\tau)$. Therefore, the fictitious particle represents the real particle at a later moment $\tau$. Thus, to find the self-force we turn the particles to each other, which is $\tau \to \tau'$ (see Fig. \ref{Fig_Method_Flat}c). 

To calculate the self-force let us represent the retarded Green function as a sum of the half-sum (singular) and half-difference (radiative) of the retarded and advanced Green functions, $G^{\textrm{ret}} = G^{\textrm{rad}} + G^{\textrm{sin}}$, where 
\begin{equation}\label{eq:radsin}
G^{\textrm{sin}}(x;x') = \frac{1}{8\pi} \delta (\sigma (x,x')), \  G^{\textrm{rad}}(x;x') = \frac{\varepsilon(t-t')}{8\pi} \delta (\sigma (x,x')),
\end{equation}
with $\varepsilon (x) = \theta(x) - \theta (-x) = x/|x|$ being the sign function. 

Using the above formulas we obtain the force acting on the same particle in the following form
\begin{equation}
{\cal F}_\mu = q F_{\mu\beta}u^\beta = {\cal F}_\mu^{\textrm{rad}} + {\cal F}_\mu^{\textrm{sin}},  
\end{equation}
where
\begin{eqnarray*}
{\cal F}_\mu^{\textrm{rad}}  &=& \frac{q^2}2\int \varepsilon \delta \left\{ \gs_{[\beta |\alpha''|,\mu]}u^{\alpha''}-   \left( \frac{\gs_{[\beta |\alpha''|}\sigma_{,\mu]}u^{\alpha''}}{\dot \sigma}\right)^{\hspace{-1ex}\boldsymbol{\cdot}} \right\}u^\beta d\tau'', \\
{\cal F}_\mu^{\textrm{sin}}  &=& \frac{q^2}2\int \delta \left\{ \gs_{[\beta |\alpha''|,\mu]}u^{\alpha''}-   \left( \frac{\gs_{[\beta |\alpha''|}\sigma_{,\mu]}u^{\alpha''}}{\dot \sigma}\right)^{\hspace{-1ex}\boldsymbol{\cdot}} \right\}u^\beta d\tau''.
\end{eqnarray*}
Here, $\tau'' = \tau - \tau'$ is the time necessary for the particle to move from point $x(\tau')$ to $x(\tau)$ along its trajectory and the overdot is a derivative with respect to time $\tau''$. Now, we expand the integrand over $\tau''$. First of all, we expand the world function:
\begin{equation}
 \sigma (x(\tau),x(\tau -\tau'')) = \sum_{n=0}^\infty \left[\frac{d^n \sigma}{d\tau''^n}\right] \frac{\tau''^n}{n!}.
\end{equation}
Here $[f(x,x')]$ stands for the coincidence limit, $x'\to x$. By using the well-known formulas for coincidence limit \cite{Synge:1960:RGT,DeWitt:1965:DTGF} for derivatives of $\sigma$ and $\gs_{\mu\nu'}$, we obtain 
\begin{eqnarray}
 \sigma &=& -\frac{\tau''^2}{2} + u_\nu \ddot u^\nu \frac{\tau''^4}{24} + \dots  , \nonumber \\
\dot \sigma &=& -\tau''\left\{1 - u_\nu \ddot u^\nu \frac{\tau''^2}{6} + \dots \right\}, \nonumber\\
\sigma_{,\mu} &=& \tau'' \left\{u_\mu - \dot u_\mu \frac{\tau''}{2} + \ddot u_\mu \frac{\tau''^2}{6} + \ldots \right\}, \nonumber\\ 
\gs_{\beta\alpha''} u^{\alpha''} &=& u_\beta - \dot u_\beta \tau'' + \ddot u_\beta \frac{\tau''^2}2 + \ldots , \nonumber\\
 \gs_{\beta\alpha'',\mu} u^{\alpha''} &=& O(\tau''^2) .
\end{eqnarray}
Using the above expansions, we find the following expression for the self-force
\begin{eqnarray*}
{\cal F}^\mu_\textrm{{sin}} &=& \dot u^\mu\left\{-\frac{q^2}2 \int  \frac{\delta(\tau'')}{|\tau''|} d\tau'' \right\}, \\
{\cal F}^\mu_{\textrm{rad}} &=& \frac{2}{3} q^2P^\mu_\nu \ddot u^\nu.
\end{eqnarray*}
All other terms of the expansion with $n\geq 1$ give no contribution since
\begin{displaymath}
 \int \delta (\tau'') \tau''^n d\tau'' =0.
\end{displaymath}

We observe that the singular Green function gives a divergent contribution while the radiative part gives finite contribution. The equation of motion of the particle has the form 
\begin{equation}
 m \dot u^\mu = q F^{\mu\nu}_{\textrm{ext}} u_\nu + {\cal F}^\mu_{\textrm{sin}} + {\cal F}^\mu_{\textrm{rad}}.	
\end{equation}
The divergent term in the right-hand side of this equation is proportional to the acceleration $\dot u^\mu$ as the term on the left-hand side. For this reason, we may include the divergence into the bare mass, that is, we make the "classical renormalization", 
\begin{equation}\label{eq:classrenorm}
 m + \left\{\frac{q^2}2 \int  \frac{\delta(\tau'')}{|\tau''|} d\tau'' \right\} \to m,
\end{equation}
and obtain the well-known DL force
\begin{equation}
 {\cal F}^\mu_{\textrm{DL}}=\frac{2}{3}q^2\ddot u^\nu P^\mu_\nu.
\end{equation}
We have to understand the divergent contribution as the infinite electromagnetic energy of the electron. To avoid the problem with divergence, Dirac \cite{Dirac:1938:ctore} suggested considering the radiative Green function instead of the retarded one. It is easy to see that in this case there is no singular contribution, ${\cal F}^\mu_{\textrm{sin}} = 0$. 

Another approach where have no divergences is based on the Bopp--Podolsky extension of the electrodynamics \cite{Bopp:1940:eltde,Podolsky:1942:agepiq,Podolsky:1948:roage}.  The Lagrangian of this theory has an additional term with higher derivatives 
\begin{equation}\label{eq:BP}
\mathcal{L}_{\rm BP} = \frac{1}{16 \pi} F_{\alpha\beta}F^{\alpha\beta} + \frac{1}{8\pi m_{\rm P}^2} F^{\alpha\nu}_{\ \ ;\alpha} F^\beta_{\ \nu ;\beta},
\end{equation}
where the new parameter $m_{\rm P}$ has the dimension of mass.   The theory is equivalent to two connected vector fields $A_\mu^{\rm(M)}$ and $A_\mu^{\rm (P)}$. The first field is the Maxwell field and the last one is the massive Proca field with mass $m_{\rm P}$:
\begin{equation}
\Box A_\mu^{\rm(M)}= - 4\pi J_\mu,\ \left(\Box - m_{\rm P}^2 \right) A_\mu^{\rm (P)} = -4\pi J_\mu. 
\end{equation}
The tensor $F_{\alpha\beta} = \nabla_\alpha A_\beta - \nabla_\beta A_\alpha$, where $A_\mu = A_\mu^{\rm (M)} - A_\mu^{\rm (P)}$. In the limit $m_{\rm P} \to \infty$, the usual Maxwell electrodynamics is restored. 

The potential of a particle with charge $q$ has the following form:
\begin{equation}\label{eq:Coulomb} 
A_0 = \frac{q}{r} (1-e^{-m_{\rm P} r}),
\end{equation}  
and there is no singularity at particle $A_0|_{r\to 0} = qm_{\rm P}$. Therefore, the mass $m_{\rm P}$ plays the role of a regularization parameter and $m_{\rm P}^{-1} = r_0$ is an effective radius of the particle.  

The vector potential of moving particle is different from the Li\'enard--Wiechert potentials in the Maxwell electrodynamics and expansion over $m_{\rm P} \to \infty$ gives the following expression for self-force \cite{Zayats:2014:siitbectomoacpdoia,Zayats:2016:siibpeswad,McManus:1948:cews}
\begin{equation}\label{eq:Min}
\mathcal{F}^\mu_{\rm em} = -\frac{q^2 m_{\rm P}}{2}\dot{u}^\mu + \frac{2}{3} q^2P^\mu_\nu \ddot u^\nu + O(m_{\rm P}^{-1}).
\end{equation}
The first term is absorbed by the bare mass 
\begin{equation}\label{eq:U_0}
m + \frac{q^2 m_{\rm P}}{2}  = m + U_0 \to m, 
\end{equation}
which looks as regularization of the infinite classical renormalization \eqref{eq:classrenorm} and $U_0 = \frac{q^2}{2 r_0}$ is electromagnetic energy of the sphere with radius $r_0$. This approach has a close relationship with the Pauli--Villars regularization procedure in quantum field theory \cite{Pauli:1949:otirirqt}.

\subsection{Curved space-time}

The first analysis of the electromagnetic self-force for particles moving in the arbitrary gravitational field was made by DeWitt and Brehme in Ref. \cite{DeWitt:1960:rdiagf}. The gravitational field is assumed to be external. The results of this paper were corrected later by Hobbs \cite{Hobbs:1968:avford,Hobbs:1968:rdicfu}. The following equation of motion with self-interaction terms was obtained
\begin{equation}
m\frac{Du^\mu}{d s}=qF_{\textrm{ext}}^{\mu\nu} u_\nu + \frac{2q^2}{3}\frac{D^2u^\nu}{ds^2}P^\mu_\nu +\frac{q^2}{3}\mathcal{R}^\nu_\beta u^\beta P^\mu_\nu + q^2  \int_{-\infty}^s f^\mu_{\,\cdot\alpha\beta'}u^\alpha (s)u^{\beta'} (s')d s',\label{SFGeneral}
\end{equation}
where $P^\mu_\nu = \delta^\mu_\nu + u^\mu u_\nu$ is the projector on the velocity of the particle. The corresponding equation for minimal coupling scalar particle was found by Quinn in Ref. \cite{Quinn:2000:aatrrosppics}. It has the following form 
\begin{equation}
	m\frac{Du^\mu}{d s}=q_s \nabla^\mu \phi_{\textrm{ext}} + \frac{q_s^2}{3}\frac{D^2u^\nu}{ds^2}P^\mu_\nu +\frac{q_s^2}{6}\mathcal{R}^\nu_\beta u^\beta P^\mu_\nu - \frac{q_s}{12} \mathcal{R} u^\mu + q_s^2 \int_{-\infty}^{s} \hspace{-1ex}G^{\mathrm{ret}} (x(s),x(s'))^{;\mu}d s',\label{SFGeneralScalar}
\end{equation}
The first term in the right-hand side of Eq. \eqref{SFGeneral} describes the electromagnetic force acting on the particle due to the external electromagnetic force $F_{\textrm{ext}}$. The second term is the covariant form of the DL self-force (\ref{DL}). The third term, obtained by Hobbs \cite{Hobbs:1968:avford,Hobbs:1968:rdicfu} arises due to the local distribution of the matter-energy. By using Einstein's equations it may be expressed in terms of the energy-momentum tensor, and, finally, the last non-local term depends on the whole prehistory of the particle's motion. The origin of this term is connected with a scattering of the electromagnetic field on the curvature and, as already have been noted, ultimately leads to the violation of the Huygens principle. The last term equals to zero for conformally flat space-time \cite{Hobbs:1968:rdicfu} and plane gravitational wave space-time \cite{Gibbons:1975:qfpipws,Khusnutdinov:2020:sfHp}. The integrand function $f^\mu_{\cdot \alpha\beta'}$ can not be obtained in closed form for arbitrary space-time and it depends on the curvature of the space-time. As will be noted below the function, $f^\mu_{\cdot \alpha\beta'}$, obeys the identity $u_\mu f^\mu_{\cdot \alpha\beta'} = 0$, then the expression obtained for the self-force preserves the perpendicularity of velocity and acceleration of the particle. This function may be expressed in terms of the retarded and advanced Green functions (see Eq. \eqref{eq:fGG}).

We would like to call attention to the main peculiarities of the above expression and its main distinctions gravitationally-induced self-force from the DL self-force (\ref{DL}). (i) First of all, the gravitationally induced self-force contains, in general, both local and non-local contributions which depend on the whole prehistory of the particle's motion. In other, the gravitationally induced self-force depends on the local as well as global properties of space-time. (ii) The self-force depends on the velocity of the particle, and it may be non-zero even for a particle at rest in contrast to the DL self-force which is proportional to the derivative of the acceleration.

The derivation of the expression (\ref{SFGeneral}) is based on the covariant generalization of the Dirac approach \cite{Dirac:1938:ctore} discussed in  Sec. \ref{Sec:Flat}. The basic equation is the equation for the energy balance between particle and field, which has, in the case of absence of forces of non-gravitational origin, the form of the covariant equation for conservation of the summary tensor of the energy-momentum of particle and field
\numparts
\begin{equation}
T^{\mu\nu}_{\ ;\nu}=0,\label{Conservation}
\end{equation}
where the energy-momentum tensor $T^{\mu\nu} = T^{\mu\nu}_{\textrm{mech}} + T^{\mu\nu}_{\textrm{el}}$ consists of the two parts namely, the mechanical,
\begin{equation}
T^{\mu\nu}_{\textrm{mech}}(x) = m_{0} \int \gs^\mu_{\cdot \alpha} (x,x(s)) \gs^\nu_{\cdot \beta}(x,x(s))u^\alpha (s) u^\beta (s)\delta^{(4)} (x - x(s)) \frac{ds}{\sqrt{-\gs}},\label{Tmech}
\end{equation}
and the electromagnetic,
\begin{equation}
T^{\mu\nu}_{\textrm{el}}(x) = \frac 1{4\pi} \left(F^{\mu\alpha} F^\nu_{\ \alpha} - \frac 14 \gs^{\mu\nu} F^{\alpha\beta}F_{\alpha\beta}\right).\label{Tem}
\end{equation}
\endnumparts
Here $\gs^{\mu}_{\cdot \alpha'}(x,x')$ is the bivector of the parallel transport along the geodesic line from the point $x'$ to $x$.

Then we integrate the Eq. (\ref{Conservation}) over the world tube with radius $\varepsilon$ and length $ds$, surrounding the world line of the particle, with the radius of the tube going to zero at the end of calculations. Because of the left-hand side of Eq. (\ref{Conservation}) is a vector at point $x$, at the beginning we have to transport it to a point on the world line of the particle inside the interval $ds$ and then integrate the biscalar obtained. This procedure leads to the following equation of motion  
\begin{equation}
\label{EquMot}
m_0\frac{Du^\mu}{d s} d s = -\lim_{\varepsilon\to 0}\int \gs^{\mu}_{\,\cdot \alpha'}(x,x') T^{\alpha'\beta'}d\Sigma_{\beta'}.
\end{equation}
The left-hand side of this equation appears due to the mechanical part of the energy-momentum tensor (\ref{Tmech}), whereas the right-hand side is the contribution from the electromagnetic part (\ref{Tem}).

The first term of the expansion over $\varepsilon$ in the right-hand side of (\ref{EquMot}) diverges with $1/\varepsilon$. From the physical point of view, it means the contribution from the own infinite electromagnetic energy of the particle. Because the structure of this term coincides with the structure of the left-hand side, then this divergent contribution may be left out by the definition of the "observed"\/ mass $m$ of the particle by the relation
\begin{equation}
\label{RenMass}
m=m_0 + \lim_{\varepsilon\to 0} \frac{q^2}{2\varepsilon}.
\end{equation}
We would like to note that there are no reasonable grounds for this "classical"\/ renormalization procedure in the framework of classical physics. It may be correctly considered only in the framework of quantum field theory where the renormalization procedure is well defined. In Dirac's approach, this problem does not appear because the radiative Green function which was used for calculations of the self-force contains no divergent Coulomb contribution. There is also an axiomatic approach that was developed by Quinn and Wald in Ref. \cite{Quinn:1997:aaateagrropics} in the framework of which the singular contribution did not appear, too (see Sec. \ref{sec:renormalization}). 

Next terms are finite in the limit of zero radii of the tube, namely, as $\varepsilon\to 0$ and they give Eq. (\ref{SFGeneral}), in which the function
\begin{equation}
\label{fmunualpha}
f_{\mu\nu\alpha'} = v_{\mu\alpha' ; \nu} - v_{\nu\alpha' ; \mu}
\end{equation}
is expressed in terms of the bivector $v_{\mu\nu'}$ defined by the Feynman vector Green function in the Hadamard form
\begin{equation}\label{Green}
G^{\textrm{F}}_{\mu\nu'}(x,x')=\frac1{(2\pi)^2}\left\{\frac{\triangle^{1/2}}{\sigma + i0} \gs_{\mu\nu'} + v_{\mu\nu'}\ln (\sigma + i0)  + w_{\mu\nu'} \right\}.
\end{equation}
Here $\sigma (x,x') = \frac 12 s^2$ is the half of the square of the geodesic interval between the points $x$ and $x'$ for shortest geodesics connecting these points; $\gs_{\mu\nu'}$ -- bivector of parallel transport along this geodesics; $\triangle (x,x')= \det(\sigma_{;\mu\nu'}(x,x'))/ \det(\gs_{\mu\nu'}(x,x'))$ -- Van Fleck--Morrett determinant. In this expression, all singularities are shown in manifest form and the quantities $v_{\mu\nu'}$ and $w_{\mu\nu'}$ have no peculiar properties  in the coincidence limit as $\sigma\to 0$. It is possible to express the $f_{\mu\nu\alpha'}$ directly in terms of the retarded Green function by as
\begin{equation}\label{eq:fGG}
f_{\mu\nu\alpha'} = 4\pi\left(G^{\textrm{ret}}_{\nu\alpha';\mu} - G^{\textrm{ret}}_{\mu\alpha';\nu}\right).
\end{equation}
This kind of representation found in Ref. \cite{Quinn:1997:aaateagrropics} is more preferable, because the form of the Green function in the Hadamard form (\ref{Green}) is correctly defined locally and usually used for the analysis of the singularities of Green functions in coincidence limit \cite{Christensen:1978:rracgps,Birrell:1982:QFCS,Fulling:1989:AQFTCS}. The origin of the non-local term in Eq. (\ref{SFGeneral}) is because the retarded Green function \cite{DeWitt:1965:DTGF} which we need for calculation of the right-hand side of (\ref{EquMot})
\begin{equation*}
G^{\textrm{ret}}_{\mu\nu'}(x,x')=\frac{\theta(x,x')}{4\pi }\left\{\triangle^{1/2}\gs_{\mu\nu'} \delta (\sigma)  - v_{\mu\nu'}\theta (-\sigma) \right\}
\end{equation*}
is obtained from (\ref{Green}) and it has both local and non-local terms. Here the function $\theta(x,x')$ is defined in a way that it equals to the unit if the event $x$ is in chronological future of event $x'$ and it equals to zero in the opposite case; $\theta(x)$ is the usual step function.

The general expression (\ref{SFGeneral}) for the self-force is difficult to use for specific situations, but it is important from the fundamental point of view. Nevertheless, it is possible to obtain general results in some specific cases, namely for conformally flat space-times \cite{Hobbs:1968:rdicfu} and in the framework of the weak gravitational field and non-relativistic motion of the particles \cite{DeWitt:1964:fc}. For conformally flat space-times the bivector $v_{\mu\nu'}$ is the gradient of a scalar function: $v_{\mu\nu'} = \Phi_{,\mu\nu'}$. For this reason, $f_{\mu\nu\alpha'} = 0$, and the non-local contribution equals to zero. The only local part of the self-force survives, in this case, \cite{Hobbs:1968:rdicfu}. In the case of a static weak gravitational field approximation and non-relativistic motion of particles, the calculations were made by DeWitt and DeWitt in Ref. \cite{DeWitt:1964:fc}. In this case, the non-local equations (\ref{SFGeneral}) become local. The point is that the retarded effects leading to non-locality become insufficiently for the case of non-relativistic motion \cite{DeWitt:1964:fc}. In this case, the particle is subjected to the additional gravitationally induced self-force
\begin{equation*}
\mathbf{f}_{\textrm{G}} = q^2 \int \frac{\mathbf{x} - \mathbf{x}'}{|\mathbf{x} - \mathbf{x}'|^4} \varrho (\mathbf{x}') d^3 x,
\end{equation*}
alongside with the usual DL self-force (a non-relativistic form of Eq. (\ref{DL}))
\begin{equation*}
\mathbf{f}_{\textrm{DL}} = \frac 23 q^2 \ddot{\mathbf{v}}.
\end{equation*}
Here, the dot denotes the derivative with respect to time, $\dot{\mathbf{v}} = d\mathbf{v}/dt$, and $\varrho (\mathbf{x}')$ is the density of matter which is the source of the weak gravitational field. The complete equation of motion, in this case, has the following form
\begin{equation*}
m (\dot{\mathbf{v}} + \nabla U) =\mathbf{f}_{\textrm{DL}} + \mathbf{f}_{\textrm{G}},
\end{equation*}
where
\begin{equation*}
U(\mathbf{x}) = - \int \frac{\varrho (\mathbf{x}') }{|\mathbf{x} -
\mathbf{x}'|}d^3 x
\end{equation*}
is the gravitational potential of the matter.

\subsection{Self-force in the space-time of black holes} \label{Sec: BH}

The majority of papers were devoted to the consideration of the self-force in the space-times Schwarzschild
\begin{equation}\label{Schwarzschild}
ds^2 = - \left(1 - \frac{2M}r\right) dt^2 + \left(1 - \frac{2M}r\right)^{-1} dr^2 + r^2 (d\theta^2 + \sin^2\theta d\varphi^2),
\end{equation}
Reissner--N\"ordstrom,
\begin{equation}
ds^2 = - \left(1 - \frac{2M}r + \frac{Q^2}{r^2}\right) dt^2 + \left(1 - \frac{2M}r + \frac{Q^2}{r^2}\right)^{-1} dr^2 + r^2 (d\theta^2 + \sin^2\theta d\varphi^2),\label{RaisnerNordstrem}
\end{equation}
Kerr,
\begin{eqnarray} \label{Kerr}
ds^2 &=& - \left(1 - \frac{2Mr}\Sigma\right) dt^2 + \frac\Sigma\Delta dr^2 + \Sigma d\theta^2 - \frac{4JMr\sin^2\theta}\Sigma  dt d\varphi \nonumber\\
&+& \sin^2\theta \left(r^2 + J^2 + \frac{2J^2Mr\sin^2\theta}\Sigma \right) d\varphi^2,
\end{eqnarray}
and Kerr--Newman 
\begin{eqnarray} \label{KerrNewman}
\fl ds^2 &=& - \left(1 - \frac{2Mr - Q^2}\Sigma\right) dt^2 + \frac{\Sigma dr^2}{\Delta + Q^2}  + \Sigma d\theta^2 - \frac{2J(2Mr -Q^2)\sin^2\theta}\Sigma  dt d\varphi \nonumber\\
\fl &+& \sin^2\theta \left(r^2 + J^2 + \frac{J^2(2Mr-Q^2)\sin^2\theta}\Sigma \right) d\varphi^2.
\end{eqnarray}
Here $\Sigma = r^2 + J^2 \cos^2\theta,\ \Delta = r^2 -2Mr + J^2$, with $M,\ Q$, and $J$ being the mass, charge, and angular momentum of black the  hole, respectively. For these spaces, the results obtained in \cite{DeWitt:1964:fc} can not be directly applied, because the gravitational field is not everywhere weak.

In the Refs. \cite{Smith:1980:foascoasbh,Zelnnikov:1982:iogotseocp,Lohiya:1982:csfoaenacrbh} the self-force for the particle at rest in the Schwarzschild space-time was considered. Initially, this question was discussed in the paper of Vilenkin \cite{Vilenkin:1979:siocpitgf} for particles situated far from the black hole. For this case all terms in the right hand side of Eq. (\ref{SFGeneral}) vanishes except the last one, and the particle is repelled by the black hole in the radial direction, and the self-force on a particle at the point $r$, has the following form
\begin{equation}\label{SFS}
\mathcal{F}_{\textrm{em}}^r=\frac{r_sq^2}{2r^3}\sqrt{1 - \frac{r_s}r},\ |\vec{\mathcal{F}}_{\rm em}| = \frac{r_sq^2}{2r^3} ,
\end{equation}
where $r_s = 2M$ is the Schwarzschild radius of the black hole. 

The expression \eqref{SFS} was obtained using local as well as global approaches. The local approach is analogous to the method of DeWitt and Brehme \cite{DeWitt:1960:rdiagf}. It consists of the calculation of the density of the external force ${\cal F}^\mu$ in which we need to support in equilibrium the particle in the Schwarzschild space-time. It has the form of the 4-divergence:
\begin{equation}
{\cal F}^\mu =  T^{\mu\nu}_{\ \ ;\nu}, \label{ForceDensity}
\end{equation}
where the energy-momentum tensor consists of two parts -- mechanical (\ref{Tmech}) and electromagnetic (\ref{Tem}). Then, the expression for the force density (\ref{ForceDensity}) is integrated over a ball with radius $\varepsilon$ around the particle in the frame of the freely falling observer. The radius of the sphere tends to zero at the end of the calculations.

To calculate the self-force, we need the electromagnetic 4-potential of the particle in the Schwarzschild background, which depends on the particle's trajectory. For a particle at rest at the point $x_p$ the electromagnetic potential was obtained by Copson in Ref. \cite{Copson:1928:oeiagf}. The zero component of the vector potential has the following form
\begin{gather}
A_t^{\textrm{part}}  (\mathbf{x},\mathbf{x_p}) = -\frac{q}{r_\p r} \frac{(r - M)(r_\p - M) - M^2 \cos \chi}{R},\nonumber \\
R^2=(r - M)^2 + (r_\p - M)^2 - 2 (r - M)(r_\p - M)\cos\chi - M^2 \sin^2\chi, \label{Copson}
\end{gather}
where $\cos\chi = \cos\theta\cos\theta_p + \sin\theta\sin\theta_p \cos (\varphi - \varphi_p)$ and $\chi$ is an angle between two points on the sphere. This potential has "wrong"\/ behavior at infinity:
\begin{equation*}
A_t^{\textrm{part}}\strut_{\hspace{-1em }r \to \infty} \approx -\frac q{r} \frac{r_\p - M}{r_\p}.
\end{equation*}
The space-time becomes flat infinitely far from the black hole and the electromagnetic potential of the particle has to coincide with the Coulomb one. The potential (\ref{Copson}) has the Coulomb form if the particle is far from the black hole in the asymptotic flat domain. In other words, we may say that the Gauss theorem for this potentials is no longer true. The surface integral over a great sphere around the charge equals $-4\pi q (r_\p - M)/r_\p$ instead of $-4\pi q$.

To solve this problem, it was suggested in Ref. \cite{Leaute:1976:eiarnst} to add to the Copson solution (\ref{Copson}) the solution of the homogeneous equation, which is given by
\begin{equation}\label{Hom}
A_t^{\textrm{hom}} = - \frac{q M}{r r_\p}.
\end{equation}
In this case the total potential $A_t = A_t^{\textrm{part}} + A_t^{\textrm{hom}}$,
\begin{equation}
A_t (\mathbf{x},\mathbf{x_p}) = -\frac{q}{r_\p r} \frac{(r - M)(r_\p - M) - M^2 \cos \chi}{R} - \frac{q M}{r r_\p}, \label{CopsonLinet}
\end{equation}
has the Coulomb form at infinity
\begin{equation}
\label{A0}
A_t|_{r\to \infty} \approx -\frac q r,
\end{equation}
and at the horizon of the black hole, $r=2M$, it is constant
\begin{equation}
\label{A0H1}
A_t|_{r=2M} = -\frac q{r_\p}.
\end{equation}
Therefore, the charge at the black hole's horizon has the Coulomb electric field which should be the case. The potential $A_0$ of the charged particle at rest in the Schwarzschild background taking into account the additional term (\ref{Hom}) was obtained in Refs. \cite{Cohen:1972:notamamogg,Hanni:1973:lofoapcnasbh} as series over spherical harmonics.

To complete the picture, we write out the corresponding formulas for another kind of black hole. The electrostatic potential of a charged particle in the Reissner--N\"ordstrom black hole (\ref{RaisnerNordstrem}) was obtained in Ref. \cite{Leaute:1976:eiarnst} (see also \cite{Zelnnikov:1982:iogotseocp})
\begin{gather}
A_t (\mathbf{x},\mathbf{x_p}) = -\frac{q}{r_\p r} \frac{(r - M)( r_\p - M) - (M^2-Q^2) \cos \chi}{R_Q} - \frac{q M}{r r_\p},\label{CopsonRN} \\
R^2_Q = (r - M)^2 + (r_\p - M)^2 - 2 (r - M)(r_\p - M)\cos\chi - (M^2-Q^2) \sin^2\chi\nonumber,
\end{gather}
It is easy to see that this potential fulfills conditions (\ref{A0}), (\ref{A0H1}),  in which we have to consider the quantity $r_Q = M + \sqrt{M^2 - Q^2}$ instead of the Schwarzschild radius.

The electromagnetic potential for a particle at the symmetry axis of the Kerr black hole (\ref{Kerr}) was found in Ref. \cite{Leaute:1977:eikst}
\begin{eqnarray*}
\fl A_t &=& -\frac{q}{(r^2_\p + J^2)\Sigma} \left[\left(M + \frac{(r - M)( r_\p - M) - (M^2-J^2) \cos \theta}{R_J}\right)\right.\nonumber \\
\fl &\times&\left. (r_\p r + J^2\cos\theta) + J^2 (r-r_p\cos\theta) \frac{(r - M) - (r_\p - M)\cos\theta}{R_J}\right],\nonumber \\
\fl A_\varphi  &=& \frac{qJ}{(r^2_\p + J^2)} \left\{\frac{\sin^2\theta}{\Sigma}\left[\left(M + \frac{(r - M)( r_\p - M) - (M^2-J^2) \cos \theta}{R_J}\right)\right.\right. \\
\fl &\times&\left.\left. (r_\p r + J^2\cos\theta)+J^2 (r-r_p\cos\theta) \frac{(r - M) - (r_\p - M)\cos\theta}{R_J}\right]\right.\nonumber \\
\fl &-&\left.R_J + (r-r_\p \cos\theta)\frac{(r - M) - (r_\p - M)\cos\theta}{R_J} - M (1-\cos\theta) \right\},\nonumber \\
\fl R^2_J&=&(r - M)^2 + (r_\p - M)^2 - 2 (r - M)(r_\p - M)\cos\theta -  (M^2-J^2) \sin^2\theta. \nonumber
\end{eqnarray*}
For a particle at the horizon of the Kerr black hole, $r_\p = r_J = M + \sqrt{M^2 - J^2}$, the potential of charge has the following non-zero components
\begin{equation*}
A_t = -\frac{qr}{\Sigma},\ A_\varphi = \frac{qJr\sin\theta}{\Sigma}
\end{equation*}
in agreement with the electromagnetic potential  out of the Kerr--Newman black hole (see, e.g. \cite{Frolov:1998:bhpbcand}). 

Taking into account the potential $A_t$ in the form above (\ref{CopsonLinet}) in the Eq. (\ref{ForceDensity}) gives us the self-interaction force (\ref{SFS}). To obtain the self-force, we should take infinite classical "renormalization"\/ of the particle's mass. Following this renormalization procedure, the observed mass is the following quantity
\begin{equation*}
m = m_{\textrm{bare}} + \lim_{\varepsilon\to 0} \frac{q^2}{2\varepsilon}.
\end{equation*}
The origin of the infinite contribution is the electromagnetic part of the energy-momentum tensor. 

The global approach is based on the use of the energy conservation law: the work for a virtual radial displacement of charge equals the force multiplied by the displacement. This energy transfers to infinity, and infinitely distant observer measures the variation of mass $\delta M$ of the system, which may be calculated by Carter's formula \cite{Carter:1979:tgtotmeatpobh} via the variation of the surface square of the horizon $\delta A$ and the surface gravity of the  black hole $\kappa$:
\begin{equation*}
\delta M = \frac \kappa{8\pi} \delta A  - \frac 1{8\pi} \delta \int G^0_0 \sqrt{-\gs} d^3 x + \frac 1{16\pi} \int G^{\mu\nu} \delta \gs_{\mu\nu} \sqrt{-\gs} d^3 x.
\end{equation*}
Here $G_{\mu\nu}$ is Einstein's tensor. For slowly moving particle the energy carries no through the horizon and $\delta A = 0$. The last term also equals to zero because we consider a test particle in the Schwarzschild background. Therefore, only one term survives and this term may be rewritten by using  Einstein's equations in the following form
\begin{equation*}
\delta M = - \delta \int T^0_0 \sqrt{-\gs} d^3 x.
\end{equation*}

Both approaches give the same result  concerning the total force to support a particle at rest in the Schwarzschild background. It is given by
\begin{equation*}
\mathcal{F}_{\textrm{em}}^r = \frac {r_s m}{2 r^2} - \frac{r_s q^2}{2 r^3}\sqrt{1 - \frac{r_s}r}.
\end{equation*}
The ratio $\delta$ of the self-force (second term) to gravitational force (first term),
\begin{equation*}
\delta = \frac {q^2}{mr} \sqrt{1 - \frac{r_s}r},
\end{equation*}
riches the maximum at the distance $r_* = \frac 32 r_s$, with the following value at this point
\begin{equation*}
\delta = \frac{2}{3\sqrt{3}}\frac{q^2}{m r_s}.
\end{equation*}
For the critical value of the Schwarzschild radius, $r_s = 2 q^2/3\sqrt{3} m$, the self-force compensates the gravitational attraction of the black hole.

There is another approach for calculation the self-force which is based on simple mechanical consideration. Because the particle holds at rest in the black hole space-time then it has to exist an object which provides the equilibrium -- a string or a strut.  The tension of string or strut equals the total force including the self-force, needed to keep equilibrium. This idea was realized in Refs.  \cite{Krtous:2019:SFoaSPnaBH,LaHaye:2020:sffcswr}. The calculations of the Krtou\v{s} and Zel'nikov \cite{Krtous:2019:SFoaSPnaBH} was based on the exact solution of Einstein equations obtained in Refs.  \cite{Alekseev:2007:ecotcmigr,Alekseev:2008:sofotrns,Alekseev:2012:snoesotcmigr,Manko:2007:drnsatifbtscmigr,Manko:2009:aotdrnsimadgmdabpp} which describes two charged black holes in equilibrium. There is the strut between black holes (see, for example, Ref. \cite{JerryB.Griffiths:2012:estiegr}) with tension $\tau$ which depends on the charges, and masses of the black holes, and the distance between them.  Taking a limit to point-like particle, $\epsilon\to 0$, for one black hole one obtains the system which consists of the black hole with particle and strut between them.  Here $\epsilon$ is a book-keeping parameter to track the series expansion. The expansion of the tension has the following form 
	\begin{equation}\label{eq:WeylExpansion}
	\tau = \mathcal{F}_{\rm gr} \epsilon + \mathcal{F}_{\rm sf, gr} \epsilon^2 + \mathcal{F}_{\rm sf,em} \epsilon^2 + \ldots ,
	\end{equation}    
where the first term is a usual gravitational attractive force and the next two terms are gravitational and electromagnetic self-forces, correspondingly.  All of these terms have no to be renormalized. The gravitational self-force, only, depends on the details of the limit. The result for the electromagnetic part of self-force differs on the results obtained by the renormalization approach \eqref{SFS} by the factor $(1-r_s/2r)/(1-r_s/r)$. 

LaHaye and Poisson \cite{LaHaye:2020:sffcswr} considered perturbation of the Schwarzschild black hole due to mass and charge of the particle. Perturbations lead to the appearance of the string which hangs particle (or strut which supports the particle) in the black hole space-time. The angle deficit $\Delta\varphi$ of the string gives the tension $\tau = \Delta\varphi/8\pi$ of the string which has the same form \eqref{eq:WeylExpansion}. The self-force calculated in this way is in agreement with the above-obtained \eqref{SFS}.    

The self-force for electric ou magnetic dipole in Schwarzschild space-time was calculated by L\'eaut\'e and Linet in Ref. \cite{Leaute:1984:sioaeomditsst} and was corrected in Ref. \cite{LaHaye:2020:sffcswr}. The electric dipole is repulsed by a black hole with radial force \begin{equation}
\mathcal{F}_{\rm em}^{\rm dipole} = \frac{2d^2}{r^5} \left(1 - \frac{3M}{2r}\right).
\end{equation}
For magnetic dipole, the force has the same form with a magnetic dipole instead of the electric one.  L\'eaut\'e and Linet obtained factor $5/2$ instead of  $3/2$. They used the solution of the Maxwell equations with the following $4$-current of electric dipole 
\begin{equation}
J^\mu = q \nabla_\nu \int (d^\mu u^\nu - d^\nu u^\mu) \delta^{(4)} (x - x(s)) \frac{ds}{\sqrt{-\gs}},
\end{equation} 
while in Ref.  \cite{LaHaye:2020:sffcswr} was used as an approach based on the tension of the string.  The self-energy of a dipole in $3D$ static space-time  has considered in Ref. \cite{Frolov:2013:seaoaepditdss}. In the case of ultrastatic space-time without horizons, the self-energy reads
\begin{equation}
\mathcal{E}_{\rm em}^{\rm dipole} = - \frac{d^2}{12} \mathcal{R}.
\end{equation}

Let us consider the massive vector Proca field instead of the electromagnetic field. In this case, the self-force has opposite sign \cite{Vilenkin:1979:siocpitgf,Leaute:1985:SmvfSs} and the particle will be attracted by a black hole. As noted in the Refs. \cite{Vilenkin:1979:siocpitgf,Leaute:1985:SmvfSs} in the limit of mass of Proca field to zero, $m_{\mathrm{P}} \to 0$, the self-force has the same value as for Maxwell filed but opposite sign. In the gravitational field of a spherically massive body, there is
no discontinuity, and self-force has the same value and sign as for the Maxwell field. The discontinuity is connected with horizon existence. The origin of this difference connects with the statement that the finiteness of the energy-momentum tensor at the horizon leads to different boundary conditions for the 4-potential of the massless and massive fields \cite{Teitelboim:1972:notqnoabh,Bekenstein:1972:nobnfsbh}. The electromagnetic energy-momentum tensor (\ref{Tem}) contains only derivatives of the 4-potential, while the energy-momentum tensor of massive Proca field,
\begin{equation}\label{eq:Proca}
T^{\mu\nu}_{\textrm{P}} = \frac 1{4\pi} \left(H^{\mu\alpha} H^\nu_{\ \alpha} - \frac 14 \gs^{\mu\nu} H^{\alpha\beta} H_{\alpha\beta} - \frac{m_{\textrm{P}}^2}{\hbar^2} \left[B^\mu B^\nu - \frac 12 \gs^{\mu\nu} B_\alpha B^\alpha\right] \right),
\end{equation}
contains terms proportional to the square mass of the field, $m_{\mathrm{P}}^2$, and 4-potential in manifest form without derivatives. (Here $H_{\mu\nu}=B_{\mu;\nu} - B_{\nu;\mu}$). Therefore, physical solutions of the Proca equations (i) should obey the boundary the condition $B_0|_{r=2M} = 0$ at the horizon of the black hole \cite{Teitelboim:1972:notqnoabh,Bekenstein:1972:nobnfsbh,Vilenkin:1979:siocpitgf} and (ii) the invariant $B_\mu B^\mu$ should be finite at the horizon of the black hole. To fulfill these conditions, we have to add some solutions to the uniform equation which have no dependence on mass, and which finally changes the sign of the self-force. Even the massless limit in the energy-momentum tensor has no peculiarities it is not the case for boundary conditions. Indeed, for a small mass of the vector field $(m_{\textrm{P}}/\hbar)^{-1} \gg M$ and for distances $M\ll r,r_\p \ll (m_{\textrm{P}}/\hbar)^{-1}$, the time component $B_t$ of the vector potential of the Proca field has the following form \cite{Vilenkin:1979:siocpitgf}, $B_t \approx A_t + \frac{2Mq}{rr_\p}$, where $A_t$ is the zero component of the electromagnetic potential considered above (\ref{CopsonLinet}). Therefore the potential obtained equals zero at the horizon: $B_t|_{r=2M} = 0$ and in agreement with the "no hair"\/ theorem the charge situated on the black hole horizon produces no field outside the hole: $B_t|_{r_\p=2M} = 0$. As it is seen from the above formula, the additional term to the Copson solution has no dependence on the mass $m_{\textrm{P}}$ and has opposite sign compared with the electromagnetic field case. Far from the black hole the Proca field is a superposition of the field of the point charge $q$ and charge $q'=\frac{2Mq}{r_\p}$ situated under black hole horizon: $B_t|_{r\to\infty} \approx -\frac{q^2}{r} + \frac{2Mq}{rr_\p}$.

The origin of this difference between massive and massless vector fields is related to the gauge invariance of the latter. The potential of the electromagnetic field is non-observable due to the gauge invariance of the theory. The measured quantities are the electric and the magnetic fields. The massive vector field of Proca possesses no gauge invariance, and the measured quantity is the potential of the field itself. By using the gauge invariance of the electromagnetic field, we may achieve the "right"\/ behavior of the potential at the infinity and regularity of the energy density at the horizon. The requirement of the finiteness of the energy density of the Proca field at the horizon together with the absence of the gauge invariance of the theory leads to different boundary conditions at infinity which has no dependence on the mass. This distinction closely connects with the "no-hair"\/ theorem of the black hole. The electromagnetic potential of the particle situated at the black hole horizon is not zero but has exactly the Coulomb form, $A_t|_{r_\p=2M} = q/r$. Therefore, the charge crossing the black hole horizon leaves the electric "hair"\/ outside the black hole. In the case of the massive Proca field, the situation is different. The charge crossing horizon produces no field outside the black hole, $B_t|_{r_\p=2M} = 0$.

The self-force equals to zero for a scalar particle at rest with charge $q_s$ and minimally coupled massless field \cite{Zelnnikov:1982:iogotseocp,Mayo:1999:toebatseotoitvoabh,Wiseman:2000:sfoasstcoasbh,Burko:2000:sfosciss,Pfenning:2002:seagsfiwcs} for Schwarzschild background as well as Reissner--N\"ordstrom space-time. This is not the case for a scalar particle moving in the Schwarzschild space-time \cite{Wiseman:2000:sfoasstcoasbh} or particle at rest in the Kerr background \cite{Galtsov:1982:rritkgf,Leaute:1982:sioapcitkst}. Equation of motion of the scalar particle was obtained by Quinn \cite{Quinn:2000:aatrrosppics} and given by Eq. \eqref{SFGeneralScalar}. 

The scalar potential due to a scalar charge $q$ situated at point $x_p$ has the following form \cite{Frolov:1980:tmsfaasbh,Wiseman:2000:sfoasstcoasbh}
\begin{equation}
V^{\textrm{part}}(\mathbf{x},\mathbf{x}_p) = -\frac {q_s}R \sqrt{1 -\frac{2M}{r_p}}. \label{ScalFieldPot}
\end{equation}
Far from the black hole ($r\gg r_p$) one has
\begin{equation*}
V^{\textrm{part}}|_{r\to \infty} \approx -\frac {q_s} r \sqrt{1 -\frac{2M}{r_p}}.
\end{equation*}
The scalar charge at the black hole horizon produces no scalar field outside the black hole: $V^{\textrm{part}}|_{r_\p =2M}=0$, in agreement with the "no-hair"\/ theorem. The potential induced by the charge on the black hole horizon
\begin{equation*}
V^{\textrm{part}}|_{r=2 M} = -\sqrt{1 - \frac{2M}{r_p}} \frac {q_s}{r_p - M - M\cos \gamma},
\end{equation*}
is not constant in the electromagnetic case but depends on the position on the horizon. Finally, the potential (\ref{ScalFieldPot}) gives a zero self-force for a scalar particle at rest in the Schwarzschild space-time. Moreover, the self-force is zero for even-dimensional space-times and is not zero for odd dimensions \cite{Frolov:2012:seoascnhdbh}. For the non-minimal coupled scalar field the self-force is not zero \cite{Zelnnikov:1982:iogotseocp,Pfenning:2002:seagsfiwcs} and proportional to the non-minimal coupling constant $\xi$. But closer inspection made in the Ref. \cite{Cho:2007:tsfoanmcsscoasbh} showed that the self-force for a scalar charge is zero for the arbitrary constant of the non-minimal coupling.

In Parker's papers \cite{Parker:1980:oeaaaposc,Parker:1981:sfaaigf} the self-force was calculated for freely falling atom in the Schwarzschild space-time. It was shown that it equals to zero for the electro-neutral atom. The origin is the following: as was shown above for "right"\/ behavior at infinity of the potential of a charge we need to the Copson solution $A_t^{\textrm{part}}$ \eqref{Copson} to add to the solution of the uniform equation $A_t^{\textrm{hom}}$ \eqref{Hom}. Then the local expansion of the potential close to the charge has an additional term $-qK_i \Delta x^i$ alongside with Coulomb part which may be obtained from Eq. \eqref{Copson}. The origin of this term is due to the addition of the solution of the uniform equation \eqref{Hom}. Here $K_r = 1/r_\p^2$ and the rest terms equal to zero. This term gives not only the self-force but additional interaction with the atomic nucleus. The self-force (repulsive), acting on the electron has the following form: $f_i = q^2 K_i$, while the attractive force to the nucleus with charge $Z$ equals to $f'_i = -Zq^2K_i$. Therefore the total force acting on charges, $\mathcal{F}_i = (1-Z)q^2 K_i$ or for an atom with $Z'$ electrons: $\mathcal{F}_i = (Z'-Z)q^2 K_i$ and it equals to zero for electro-neutral atom. Therefore, in an electro-neutral atom, the self-interaction force (repulsive) acting on the electron's clouds compensates by additional force (attraction to the nucleus).

It was shown also in Ref. \cite{Parker:1981:sfaaigf} that the rest components of the vector potential $A_k$ contains additional term too, which has the following form close to the charge: $-qL_{kj} \Delta x^j$, where $L_{kj}$ is an antisymmetric constant matrix. These terms lead to the appearance of the magnetic field at the position of charge which acts on the charge: $B_i=-q\epsilon_{ijk}L_{kj}$. This magnetic field gives the force which may be called self-torque force with the following components: $f^t_j = -2q\mu_k L_{kj}$, where $\mu_k$ is the magnetic dipole momentum of the electron. Alongside the above consideration, we may reduce that for the electro-neutral atom the self-torque force acting on electron clouds is compensated by a force of the same magnitude due to the interaction with the nucleus. Therefore, we conclude that there is no additional self-force acting on the electro-neutral atom. Note that the qualities $K_i$ and $L_{kj}$ depend only on the global structure of the space-time.

In Refs. \cite{Linet:2000:eseisbhwss,Linet:2000:eboacoaeseibh,Linet:1999:eiasbhpbacs,Linet:1999:ebfacoftknbh} the self-interaction potential was obtained for a particle in spherically symmetric space-times with the following line element
\begin{equation*}
ds^2 = -N^2(r) dt^2 + B^2(r) \left(dr^2 + r^2 d\theta^2 + r^2\sin^2 \theta d\varphi^2\right).
\end{equation*}

The electrostatic self-potential of a particle at rest at the point $r_\p$ in the space-time of black hole with surface gravity $\kappa = N'(r_h)/B(r_h)$ has the following form:
\begin{equation}
\label{LinetSelf}
U^{em}(r_\p) = \frac 12 q^2s[a(r_\p)]^2,\ s = \frac 1{a(r_h)}\left(1 - \frac \kappa{a(r_h)}\right),
\end{equation}
where
\begin{equation*}
a(r) = \int_r^\infty \frac{N(r)}{r^2 B(r)} dr
\end{equation*}
is the potential of a unit charge under the horizon and $r_h$ is the horizon's radius of the black hole. The main point of the approach is the following: the electrostatic potential of a charged particle which are out of horizon is represented as a sum of the solution in the Hadamard form (for Schwarzschild black hole is given by Eq. (\ref{Copson}), and for Reissner--N\"ordstrom is given by Eq. (\ref{CopsonRN}) and solution of Maxwell equation (Eq. (\ref{Hom}) for Schwarzschild black hole). The potential is a symmetric function of the position of the charge and observation point of the field. A solution of the uniform Maxwell equation is represented in the following form $qsa(r)a(r_\p)$, where $a(r) = A_t/q$ is the potential of a unit charge under horizon (out of the black hole it is the spherically symmetric solution of the uniform equations). The coefficient $s$ is found from the Gauss theorem -- flux of field across close surface around charge equals to $4\pi q$. Calculation of the energy of this configuration with infinite renormalization of mass leads to equation (\ref{LinetSelf}).

The calculations of this kind were connected with the estimation of the upper bound of the entropy of the black hole \cite{Bekenstein:1999:bhpaneb,Hod:2000:uebfrs,Linet:1999:ebfacoftknbh,Shimomura:2000:dtgslrebfacs}. Firstly it was calculated by Zaslavskii \cite{Zaslavskii:1992:eaabfcbh}. The point is that if the charged particle crosses the black hole's horizon the self-energy also is absorbed by the black hole and this effect leads to changing the upper boundary of the entropy. It was obtained the following limitation for the black hole's entropy
\begin{equation}\label{Sunusual}
S\leq 2\pi m l + \frac{\pi q^2}{\kappa} (s a(r_h) - 1)a(r_h).
\end{equation}
Taking into account Eq. (\ref{LinetSelf}) one obtains the relation
\begin{equation} \label{Susual}
S\leq 2\pi (m l -\frac 12 q^2).
\end{equation}
Here $m,l$, and $q$ are mass, radius, and charge of the object, respectively. Interestingly, the well-known formula (\ref{Susual}), in which the upper boundary has no dependence on the black hole's parameters is a consequence of more deep formula (\ref{Sunusual}) the upper boundary of which depends on the parameters of the black hole and the self-energy. 

The self-energy and the self-force of the electromagnetic and scalar particle in the space-time of the charged Reissner--N\"ordstrom black hole (\ref{RaisnerNordstrem}) were considered by Zelnikov and Frolov in Ref. \cite{Zelnnikov:1982:iogotseocp}. It was shown that the self-force has the only radial component and it was found the following expression for self-force of the electromagnetic field
\begin{equation*}
\mathcal{F}_{\textrm{em}}^r=\frac{r_sq^2}{2r^3}\sqrt{1 - \frac{r_s}r + \frac{Q^2}{r^2}},\ |\vec{\mathcal{F}}_{\textrm{em}}| = \frac{r_sq^2}{2r^3}.
\end{equation*}

It was a long discussion about self-force for the scalar particle. Because scalar curvature is zero then the Green function has no dependence on the non-minimal coupling, while the energy-momentum tensor of the scalar filed does depend on it even for vacuum case. Firstly was shown in Ref. \cite{Zelnnikov:1982:iogotseocp} that it is proportional to the non-minimal coupling constant. Then more careful calculations \cite{Cho:2007:tsfoanmcsscoasbh} reveals that the self-force is zero $\mathcal{F}_{\textrm{sc}}^r=0$ for arbitrary value of non-minimal coupling  constant. The full energy of a particle at rest in this black hole
\begin{equation*}
E_{\textrm{em}} = m_{\textrm{em}}\sqrt{1 - \frac{r_s}r + \frac{Q^2}{r^2}} +\frac{q^2r_s}{4r^2}
\end{equation*}
is a sum of the rest energy and the self-energy. The infinite own electromagnetic energy is removed by classical renormalization of mass:
\begin{equation*}
m_{\textrm{em}} = m_{\textrm{bare}} + \frac{q^2}{2\varepsilon}.
\end{equation*}

For the scalar particle 
\begin{equation*}
E_{\textrm{sc}} = m_{\textrm{sc}}\sqrt{1 - \frac{r_s}r + \frac{Q^2}{r^2}},
\end{equation*}
where
\begin{equation*}
m_{\textrm{sc}} = m_{\textrm{bare}} - \frac{q_s^2}{2\varepsilon}.
\end{equation*}

The self-force for a particle in the field of the rotated Kerr black hole was calculated by Gal'tsov in Refs. \cite{Galtsov:1982:rritkgf,Galtsov:1986:Pfabh} for massless fields of spin $0,1,2$, and for the particular case of the electromagnetic particle in symmetry axes of the black hole in Ref. \cite{Leaute:1977:eikst,Linet:1977:soecariabhst,Leaute:1982:sioapcitkst}. The method of calculations used in Ref. \cite{Galtsov:1982:rritkgf} is based on using the radiative Green function (half difference of retarded and advanced Green functions) -- the part of the Green function which describes radiation. It was shown that the self-force acting on the particle at rest far from the Kerr black hole has the azimuthal component
\begin{eqnarray}
\mathcal{F}^{\textrm{sc}}_\varphi &=& - \frac 13 Jq_s^2 M^2 \frac{\sin^2\theta}{r^4},\nonumber \\
\mathcal{F}^{\textrm{el}}_\varphi &=& - \frac 23 Jq^2 M^2 \frac{\sin^2\theta}{r^4}, \label{eq:SFKerr}
\end{eqnarray}
corresponds to scalar electromagnetic and massive particles ($r\gg M$). The rotating black hole tends to move the particle in the direction of its rotation. For this reason, the rotation of the black hole is slow down due to the conservation of total angular momentum. This phenomenon was discussed before by Hawking and Hartle \cite{Hawking:1972:eaamfiabh} and it was called tide friction. 

The self-force for a scalar particle at rest in the Kerr--Newman black hole \eqref{KerrNewman} was calculated by Burko and Liu in Ref.  \cite{Burko:2001:sfoascitsoasabh} in the strong-field regime. They found that 
\begin{equation}
\mathcal{F}^{\textrm{sc}}_\mu = - \frac 13 J q_s^2 \frac{\sin^2 \theta \Delta (M^2 -Q^2)}{(\Delta - J^2 \sin^2 \theta )^{5/2} \Sigma^{1/2}} \delta_\mu^\varphi, 
\end{equation}
where $\Delta = r^2 -2 M r + J^2 + Q^2$, and $\Sigma = r^2 + J^2 \cos^2\theta$. For $Q=0$ and far from the black hole we obtain Eq. \eqref{eq:SFKerr}. This result was firstly analytically correctly derived in Ref. \cite{Ottewill:2012:skgficfaaadotsffasscikst} for Kerr black hole. One notes that for Reissner--N\"ordstrom black hole ($J=0$) the self-force equals to zero according to \cite{Zelnnikov:1982:iogotseocp,Cho:2007:tsfoanmcsscoasbh}. 

There is another approach of calculations for Schwarzschild space-time developed in Ref. \cite{Ori:1997:reotccfgoaakbh,Barack:2000:msraftsfibhs,Barack:2000:sfoaspisssvmsrrt,Barack:2000:rrfoappiabh,%
Burko:2001:sfocitsoss,Lousto:2000:patgrribbh,Burko:2000:sfopioaabh} (see also review \cite{Poisson:2011:tmoppics}) which is based on the  renormalization of each term in the potential expansion over orbital momentum. The renormalization consists of the extraction of some terms of asymptotic expansion for large orbital momentum. The force has the form of the divergent sum over angular momentum $\mathcal{F}_\mu = \sum_l \mathcal{F}_\mu^l$. To regularize the force one needs to extract series of $\mathcal{F}_\mu^l = a_\mu l + b_\mu + c_\mu /l$ for large $l$. The next term of expansion is convergent series $\sim \sum_l l^{-2}$. In the framework of this approach, the renormalized self-force is 
\begin{equation}
\mathcal{F}^{\mathrm{ren}}_\mu = \sum_l (\mathcal{F}_\mu^l - a_\mu l - b_\mu - c_\mu l^{-1}) + d_\mu,
\end{equation}
where $d_\mu$ is a finite term, which may be found in an explicit form for simple situations.  

Because the Green function depends on the global structure of the space-time, the self-force has to depend on the internal structure of the spherical solution. Above we considered black holes with horizons which are the final states of collapse.  

The self-force in the space-time of the spherical massive shell was considered in Refs. \cite{Unruh:1976:sfocp,Leaute:1985:SmvfSs,Burko:2001:sfocitsoss}. This space-time has no horizon. Outside the shell of radius $R > r_s = 2M$, the space-time is the Schwarzschild space-time \eqref{Schwarzschild} and inside of shell the space-time is flat and equipped with the metric 
\begin{equation*}
ds^2 = - \left(1 - \frac{2M}R\right) dt^2 + dr^2 + r^2 (d\theta^2 + \sin^2\theta d\varphi^2). 
\end{equation*}

Unruh \cite{Unruh:1976:sfocp} considered self-force for particle inside the spherical shell of radius $R$, where space-time is flat. He found that the self-force is not zero and pushes the charge towards the center of the sphere. At the center of the sphere, it is zero. In particular, if $R\to r_s$ the self-force has the following form
\begin{equation}\label{eq:Unruh}
\mathcal{F}^r_{\mathrm{em}} = - \frac{q^2}{r_s^3} \frac{r}{(1- (r/r_s)^2)^2}.
\end{equation}

Burko, Liu and Soen \cite{Burko:2001:sfocitsoss} considered the same problem for Maxwell and scalar fields in the framework of the mode sum regularization prescription suggested in the Refs. \cite{Ori:1997:reotccfgoaakbh,Barack:2000:msraftsfibhs}. In the limit $R\to r_s$ for scalar particle,  the self-force has the same form as in Eq. \eqref{eq:Unruh} where the scalar charge is understood. 

L\'eaut\'e and Linet \cite{Leaute:1985:SmvfSs} considered the massive Proca field with energy-momentum tensor \eqref{eq:Proca} instead of the Maxwell field in the space-time of spherical massive shell. They interested in behavior the self-force in the limit $m_{\mathrm{P}} \to 0$. They proved that there is no discontinuity in this limit and the self-force tends to that for the  Maxwell field and inside the shell, it is in agreement with Unruh \cite{Unruh:1976:sfocp}. In this limit, in the case of Schwarzschild space-time with the horizon, the opposite sign appears (see discussion above).

The self-force for spherical stars and its dependence on the internal structure of stars were considered in Refs. \cite{Shankar:2007:sfoasecnass,Drivas:2011:dosfoco,Isoyama:2012:sfapois}. It was shown that far from the star, the main term $\sim 1/r^3$ does not depends on the internal structure and coincide with that for the black hole. Dependence on the internal structure appears in the next order $\sim 1/r^5 $ \cite{Isoyama:2012:sfapois}. Close to the surface of the star, the self-force has greater value as for the corresponding black hole case. 

The scalar self-force in the spacetime of the Schwarzschild's star is not zero \cite{Pfenning:2002:seagsfiwcs} as it was in the case of Schwarzschild black hole \cite{Cho:2007:tsfoanmcsscoasbh}.   In the Ref. \cite{Pfenning:2002:seagsfiwcs}, the weak field approximation was considered. For the black hole case, the scalar curvature is zero and that is why there is no dependence on the non-minimal coupling $\xi$. In the weak-filed approximation $\mathcal{R} = 2 \Delta \Phi = 4\pi \rho$, where $\Phi$ is the Newtonian gravitational potential of mass with density $\rho$. The first correction to flat Green function $G_{fl}(x;x')$ contains non-zero term $\sim \xi \int G_{fl}(x;x'')\rho(x'')G_{fl}(x'';x')d^4x''$.  The self-force, in this case, has exactly the same form
\begin{equation}
\mathcal{F}_{\mathrm{sc}}^{(r)} = \xi q_s^2 \frac{r_s}{r^3},
\end{equation} 
as was obtained by Zelnikov and Frolov \cite{Zelnnikov:1982:iogotseocp} and then corrected in Ref. \cite{Frolov:2012:cseaa,Cho:2007:tsfoanmcsscoasbh}.

\subsection{Renormalizations}\label{sec:renormalization}

To obtain the self-force, we have to extract finite terms from the Green functions in the coincidence limit. In the flat space-time, we may obtain a finite result by "classical renormalization"\/ of the bare particle's mass \eqref{eq:classrenorm} in the equation of motion. To avoid this problem, Dirac \cite{Dirac:1938:ctore} suggested represent the retarded Green function as the sum of radiative and singular parts \eqref{eq:radsin} and use the radiative Green function to calculate self-energy and self-force. In this case, the divergent contributions to mass \eqref{eq:classrenorm} come with a different sign from advanced and retarded Green functions and cancel each other. Wheeler and Feynman \cite{Wheeler:1945:iwtaatmor} suggested another approach -- the radiation reaction is the consequence of the interaction with an absorber. 

With the presence of the gravitational field, the problem becomes complicated. The point is that in the curved space-time the Green function has part concentrated on the null cone and the part violated the Huygens principle, which lives inside the cone. Therefore, the radiative Green function has a contribution from the whole interior of the cone. In the Ref. \cite{Detweiler:2003:sfvagfd} Detweiler and Whiting suggested redefining radiative Green function by adding a solution of the homogeneous equation in that way that the final radiative Green function has no support inside the future null cone. The additional term gives no contribution to the self-force \cite{Casals:2012:rossf}. Another approach was suggested by Quinn and Wald \cite{Quinn:1997:aaateagrropics,Quinn:2000:aatrrosppics}. The main "comparison axiom" \cite{Quinn:1997:aaateagrropics} may be formulated in the following way: the difference in the self-force acting on two particles which have the same charge $q$ and the same acceleration is the usual Lorentz force calculated with help of the difference between the electromagnetic fields of the particles. In fact, this axiomatic approach removes divergence which traditionally is removed by the "classical" renormalization of the mass. The results obtained in this approach are in agreement with the results of Refs. \cite{DeWitt:1960:rdiagf} and \cite{Mino:1997:grrtapm}. The corresponding axiom for the scalar case sees in Ref. \cite{Quinn:2000:aatrrosppics}. The methods developed in Refs. \cite{DeWitt:1960:rdiagf,Mino:1997:grrtapm,Quinn:1997:aaateagrropics,Quinn:2000:aatrrosppics} allow obtaining the general structure of the self-force which is important to understand the origin of the self-force. Barack and Ori \cite{Barack:2000:msraftsfibhs} suggested another method based on the averaging of multipole moments that is suitable for the concrete spherical symmetric situation. 

There is another approach in which the finite result is obtained without a renormalization \cite{Krtous:2019:SFoaSPnaBH}.  This approach is based on the exact solution of Einstein equations obtained in Refs. \cite{Alekseev:2007:ecotcmigr,Alekseev:2008:sofotrns,Alekseev:2012:snoesotcmigr,Manko:2007:drnsatifbtscmigr,Manko:2009:aotdrnsimadgmdabpp} which describes two charged black holes in equilibrium. There is the strut between black holes with tension $\tau$ which depends on the charges, and masses of the black holes, and the distance between them.  Taking a limit to point-like particle, $\epsilon\to 0$, for one black hole one obtains the system which consists of the black hole with particle and strut between them.   The tension describes the force to support the particle in equilibrium.  The expansion of the tension has the following form 
\begin{equation}
\tau = |\vec{\mathcal{F}}_{\rm gr}| \epsilon + |\vec{\mathcal{F}}_{\rm sf, gr}| \epsilon^2 + |\vec{\mathcal{F}}_{\rm sf,em}| \epsilon^2 + \ldots ,
\end{equation} 
where the first term is a usual gravitational attractive force and the next two terms are gravitational and electromagnetic self-forces, correspondingly.  All of these terms have no to be renormalized. The gravitational self-force, only, depends on the details of the limit. The result of the electromagnetic part of self-force differs from the results obtained by the renormalization approach \eqref{SFS}. 

There is another approach where the self-energy may be obtained without renormalization. It is based on the Bopp--Podolsky theory \cite{Bopp:1940:eltde,Podolsky:1942:agepiq,Podolsky:1948:roage} (see Eq. \eqref{eq:BP}).  In this case, the electromagnetic potential is the sum of massless Maxwell and massive Proca fields.  The Proca mass plays the role of regularization parameter and the Maxwell electrodynamics is restored in the limit $m_{\rm P} \to \infty$. In this limit, the self-force is the sum of the self-force obtained by mass or Green function  renormalization procedure and the electromagnetic energy of the effective particle with radius $r_0 = m_{\rm P}^{-1}$ (see Eq. \eqref{eq:Coulomb}). This statement is proved for Minkowski space-time  \eqref{eq:Min} \cite{Pauli:1949:otirirqt}, the infinitely thin cosmic string space-time \eqref{eq:BPString}, and point-like global monopole space-time \eqref{eq:BPGM} \cite{Zayats:2016:siibpeswad}. 

Approaches based on consideration of the continuum media have been considered in many papers. We do not touch in this review the self-forces on an extended body and would like to note some new results in the point-like limit case. Gralla, Harte, and Wald \cite{Gralla:2009:rdesf} considered a one-parametric family of the charged bodies that the charge-current vector, matter stress-energy tensor, and the electromagnetic field depend on a parameter $\lambda$.  The limit $\lambda \to 0$ corresponds to the pointwise particle. With the assumptions that the total charge $q$ and mass $m$ of the body tend to zero $\sim \lambda$ with $q/m = const$ the authors obtained DL force without a renormalization scheme with additional contributions due to other macroscopic characteristics of the body -- electromagnetic dipole moment and spin.  The authors of Ref. \cite{Gralla:2009:rdesf} considered the self-force in the flat Minkowski space-time, only.  Harte, Flanagan, and Taylor \cite{Harte:2016:sfsbad,Harte:2018:fotsfpiad} considered self-forces on static extended bodies in arbitrary dimensions. They started from an extended body and used the invariance of the force with respect to the subtracting of electromagnetic self-energy. In this case, the theory does not need additional renormalization.  
 
Ritus in Refs. \cite{Ritus:1978:moeamoiqeoacf,Ritus:1981:tmsoac} found the mass shift of the electron in the classical limit in the framework of quantum field theory with external electric and magnetic fields. The shift appears as a correction due to a massive operator. Therefore, the correct derivation and renormalization of self-force may be obtained in the framework of the quantum field theory with a radiative correction: 
\begin{equation}
(-i\gamma^\mu(x) \widetilde{\nabla}_\mu + m) \Psi(x) + \int M(x;x')\Psi(x') \sqrt{-\gs} d^4 x = 0,
\end{equation}
where $M(x;x')$ is the massive operator. Semiclassical approximation of this equation will give the equation of motion with self-energy correction to the mass of the particle. This program is not realized up to now. 

To obtain the self-energy one has to make UV regularization of the corresponding Green function. Any regularization method may be used for this goal, the point splitting method, the zeta-function approach, dimensional regularization, and so on (see, for example, \cite{Birrell:1982:QFCS,Fulling:1989:AQFTCS}). Traditional for a curved background is the point splitting method.   In the framework of this method,  the regularization reduces to subtraction some first terms of the DeWitt--Schwinger expansion. The general expression for self-energy of scalar and electric particle in static $D$-dimensional space-time with metric
\begin{equation}\label{eq:staticST}
ds^2 =  \gs_{tt} dt^2 + \gs_{ab} dx^a dx^b, 
\end{equation}
was considered in Refs. \cite{Frolov:2012:seoascnhdbh,Frolov:2012:saefossihdmps,Frolov:2012:cseaa,Frolov:2012:aatseoec}. Here, $a,b = \overline{1,D-1}$, and $ \gs_{tt},\gs_{ab}$ do not depend on $t$.  The equations for Maxwell and scalar fields \eqref{eq:MaxSca} in the static space-time have the form of equations for Green function in the $(D-1)$-dimensional space. The energy of the scalar or Maxwell fields formally looks like the Euclidean action of $(D -1)$-dimensional scalar field $\varphi$ ($\psi$ for Maxwell field) interacting with the external dilaton field $\phi = \sqrt{-\gs_{tt}}$ \cite{Frolov:2012:cseaa,Frolov:2012:aatseoec}. In the case of the scalar field, $\varphi = \phi^{1/2} \Phi$ and $\psi = - \phi^{-1/2} A_t$ for the vector field.  

It was shown \cite{Frolov:2012:cseaa,Frolov:2012:aatseoec} that the self-energy has the following form:
\begin{equation}
\mathcal{E} = \phi \Delta m, 
\end{equation} 
where 
\begin{equation}
\Delta m = -  2\pi q^2_s  D^s_{\rm ren}(x,x) = -  2\pi q^2_s \langle \varphi^2 \rangle_{\rm ren},
\end{equation}
for scalar particle, and 
\begin{equation}
\Delta m = 2\pi q^2 D^{\rm em}_{\rm ren}(x,x) = 2\pi q^2 \langle \psi^2 \rangle_{\rm ren} ,
\end{equation}
for electric particle. Here, $D^{s,{\rm em}}_{\rm ren} = \lim_{x' \to x}(D^{s,{\rm em}} - D^{s,{\rm em}}_{\rm DS})$ is the regularized $(D-1)$-dimensional scalar Green function for scalar field $\varphi$ or $\psi$. Regularization consists in subtracting of first terms of DeWitt--Schwinger expansion $D^{s,{\rm em}}_{\rm DS}$. The number of these terms depends on the dimension. For Maxwell field, $D^{\rm em} = - D_{tt'} \phi(x)^{-1/2} \phi(x')^{-1/2}$. The problem of calculation of $\Delta m$ is formally equivalent to the study of the quantum fluctuations of the scalar field in $(D - 1)$-dimensional Euclidean space.      

The energy is invariant under the conformal transformations of the metric $\gs_{ab} =  \overline{\gs}_{ab} \Omega^2$ and the following transformation of dilaton and scalar field
\begin{eqnarray}
\varphi &=& \overline{\varphi} \Omega^{\frac{3-D}{2}} , \phi = \overline{\phi} \Omega^{3-D} , \nonumber\\
\psi &=&  \overline{\psi} \Omega^{\frac{3-D}{2}}, \phi =  \overline{\phi} \Omega^{D-3},
\end{eqnarray}
but the self-energy $\mathcal{E} $ is not, and 
\begin{eqnarray}
\langle \varphi^2 \rangle_{\rm ren} &=& \langle \overline{\varphi}^2 \rangle_{\rm ren} \Omega^{3-D} - B_{\rm s}, \nonumber\\
\langle \psi^2 \rangle_{\rm ren} &=& \langle \overline{\psi}^2 \rangle_{\rm ren} \Omega^{3-D} - B_{\rm em},
\end{eqnarray}
which means that $\mathcal{E}_{s,{\rm em}} = \overline{\mathcal{E}}_{s,{\rm em}} - B_{s,{\rm em}} \Omega^{D-3}$. Frolov and Zelnikov \cite{Frolov:2012:cseaa,Frolov:2012:aatseoec} called $B$ as a classical conformal anomaly. They showed that for even dimensions $B=0$ but for odd dimensions $B\not =0$. This statement is closely connected with that the Hadamard elementary solution has a logarithmic singularity in even dimensions, only \cite{Birrell:1982:QFCS}. If the $(D-1)$-dimensional subspace is conformally flat, then the self-energy $\mathcal{E}_{s,{\rm em}} = - B_{s,{\rm em}} \Omega^{D-3}$. The classical conformal anomaly may be expressed in terms of the DeWitt--Schwinger expansion 
\begin{equation}
B_{s,{\rm em}} = \lim_{x' \to x} \left\{D^{s,{\rm em}}_{\rm DS}  - \frac{\overline{D}^{s,{\rm em}}_{\rm DS}}{(\Omega (x)\Omega (x'))^{\frac{D-3}{2}}}\right\}.
\end{equation} 

The Majumdar--Papapetrou spacetimes which describe extremally charged black holes in equilibrium in a higher dimensional space-times (details see in Refs. \cite{Frolov:2012:cseaa,Frolov:2012:aatseoec}) have conformally flat $(D-1)$ subspace end, therefore, the self-energy is defined by the classical conformal anomaly. In the five-dimensional case 
\begin{equation}
\Delta m = \frac{q^2}{144\pi} \mathcal{R}.
\end{equation} 
About self-force in five-dimensional black holes space-time sees also Refs. \cite{Beach:2014:sfoacoafdbh,Taylor:2015:ssfiafdbhs}.  

Let us consider in detail the renormalization of the time component, $A_t$, of vector potential or scalar field in a static spacetime which is proportional to the Green function. To renormalize Green function, the DeWitt--Schwinger renormalization approach is used (see, for example, \cite{Birrell:1982:QFCS}). Let us consider \cite{Khusnutdinov:2010:Spcsmw,Popov:2011:rftspoascisst} general form of static spacetime with the metric \eqref{eq:staticST}, and charged (scalar $q_s$ or electromagnetic $q$) particle at rest in this spacetime. The equations for Maxwell (in Lorentz gauge $A^\mu_{;\mu} =0$) and scalar fields read 
\begin{eqnarray}\label{eq:MaxSca}
\Box A_\mu - \mathcal{R}^\nu_\mu A_\nu  - m^2_{\mathrm{ef}} A_\mu  &=& - 4\pi J_\mu, \nonumber \\
\Box \Phi - (m_s^2 + \xi \mathcal{R}) \Phi &=& - 4\pi J,
\end{eqnarray}
where $\xi$ is the non-minimal coupling constant. The electromagnetic and scalar current for the particle at rest has the following form
\begin{eqnarray*}
	J_\mu &=&  q \int u_\mu (\tau) \delta^{(4)}(x - x'(\tau))\frac{d\tau}{\sqrt{-\gs}} = - q \sqrt{-\gs_{tt}} \frac{\delta^{(3)}(\vx - \vx')}{ \sqrt{\gs^{(3)}}} \delta_\mu^t, \\
	J &=& q_s \int \delta^{(4)}(x - x'(\tau))\frac{d\tau}{\sqrt{-\gs}} =  q_s\frac{\delta^{(3)}(\vx - \vx')}{ \sqrt{\gs^{(3)}}}.
\end{eqnarray*} 
For electromagnetic field we set $A_\mu = A_t(x^i)\delta^t_\mu$ and for scalar field $\Phi = \Phi (x^i)$. Then, the equations \eqref{eq:MaxSca} has the form of $3D$ equations
\begin{equation}
\widehat{\mathcal{O}}_e A_{(t)} =  -4\pi q\frac{\delta^{(3)}(\vx - \vx')}{ \sqrt{\gs^{(3)}}}, \widehat{\mathcal{O}}_s \Phi =  4\pi q_s\frac{\delta^{(3)}(\vx - \vx')}{ \sqrt{\gs^{(3)}}}.
\end{equation} 
The tetrad component $A_{(t)} = -4\pi q G_e$ and $\Phi = -4\pi q_s G_s$ are proportional to the Green functions of corresponding operators. 

Taking into account approach Bunch and Parker \cite{Bunch:1979:fpicsamsr} in Riemann normal coordinates (see, for example \cite{Petrov:1969:ES}) we obtain (details see in Ref. \cite{Khusnutdinov:2010:Spcsmw,Popov:2011:rftspoascisst}) the following term which has to be subtracted (which are survived in the limits $m_{\textrm{ef}},m_s \rightarrow \infty$ and $\sigma \rightarrow 0$)
\begin{equation} 
 A_{(t)}^{\textrm{DS}}(\vx; \vx') = - q \left(\frac{1}{\sqrt{2 \sigma}} -m_{\textrm{ef}}+\frac{h_{i'}\sigma^{i'}}{2 \sqrt{2 \sigma}}\right), \Phi^{\textrm{DS}}(\vx; \vx') = q_s \left(\frac{1}{\sqrt{2 \sigma}} -m_s+\frac{h_{i'}\sigma^{i'}}{2 \sqrt{2 \sigma}}\right).\label{eq:renorm}
\end{equation}
Here, $h_k = \partial_k \ln \sqrt{-\gs_{tt}}$, $\sigma$ is one half the square of the distance between the points $x^i$ and $x^{i'}$ along the shortest geodesic connecting them, and $A_{(t)}$ is the tetrad component of potential. 

Because $A_{(t)} = - 4\pi q G$ ($G$ is the Green function), then the renormalized potential reads
\begin{equation}
A_{(t)}^{\textrm{ren}} = \lim_{\vx'\to \vx}(A_{(t)} - A_{(t)}^{\textrm{DS}}),\ \Phi^{\textrm{ren}} = \lim_{\vx'\to \vx}(\Phi - \Phi^{\textrm{DS}}), \label{eq:renorm-gen}
\end{equation}
and self-energy has the following form (the sign minus is due to covariant index of the $4$-potential)
\begin{equation}
U_e = - \frac{q}{2} A_{t}^{\textrm{ren}},\ U_s = - \frac{q_s}{2} \Phi^{\textrm{ren}}. \label{eq:self-energy}
\end{equation}

Let us apply the above scheme for the well-known case of the black hole \cite{Vilenkin:1979:siocpitgf,Smith:1980:foascoasbh,Pfenning:2002:seagsfiwcs}.
The electromagnetic potential of a charged particle at rest in the Schwarzschild space-time \eqref{Schwarzschild} was obtained by Copson \cite{Copson:1928:oeiagf} and corrected by Linet in Ref. \cite{Linet:1976:eamitsm}, and given by Eq. \eqref{CopsonLinet}. In the limit of coincided angle variables, we obtain the tetrad component of the potential and singular part (\ref{eq:renorm}) 
\begin{equation*}
A_{(t)} = -\frac{q}{|r-r'|}\sqrt{1 - \frac{2M}{r}},\ A_{(t)}^{\textrm{DS}} =  -\frac{q}{|r-r'|}\sqrt{1 - \frac{2M}{r'}}.
\end{equation*}
To renormalize the potential, we subtract divergent part (\ref{eq:renorm-gen}) and make coincidence limit
\begin{equation*}
A_{(t)}^{\textrm{ren}} = \lim_{r'\to r}(A_{(t)} - A_{(t)}^{\textrm{DS}}) = -\frac{q M}{r^2\sqrt{1 - \frac{2M}{r}}},
\end{equation*}
and then we obtain self-energy
\begin{equation*}
U_e =  -\frac q2 A_{t}^{\textrm{ren}} = \frac{q^2M}{2r^2}.
\end{equation*}
By using this potential in the form $U_e = - q A_t$ we obtain the Lorentz force for the particle at rest 
\begin{equation*}
\mathcal{F}_r = qF_{rt}u^t = - \partial_r U_e \sqrt{\gs_{rr}},
\end{equation*}
and therefore the self-energy has the classical definition of force (tetrad component) as a minus gradient of the potential,
\begin{equation*}
{\cal F}_{(r)} =  - \partial_rU_e,
\end{equation*}
and the sign of the potential give the information about the attractive or repelling character of the force. By using this expression, we recover well-known expression for self-force (tetrad component) in Schwarzschild space-time \cite{Vilenkin:1979:siocpitgf,Smith:1980:foascoasbh,%
	Leaute:1983:poeae,Zelnnikov:1982:iogotseocp,Pfenning:2002:seagsfiwcs}
\begin{equation}\label{eq:bhself}
{\cal F}_{(r)}^{\textrm{em}} =  \frac{Mq^2}{r^3}.
\end{equation}

The potential $\Phi$ of the scalar massless particle was found in Ref. \cite{Linet:1977:soecariabhst,Wiseman:2000:sfoasstcoasbh}. It has the following form
\begin{equation}
\Phi = - \frac{q_s}{R} \sqrt{1 - \frac{2M}{r'}}. 
\end{equation}
From Eq. (\ref{eq:renorm}) we obtain that $\Phi^{\textrm{DS}}$ coincides with the above exact solution, $\Phi^{\textrm{DS}} = \Phi$, and therefore the $\Phi^{\textrm{ren}} =0$ and self-force is also zero which agrees with results obtained before \cite{Zelnnikov:1982:iogotseocp,Pfenning:2002:seagsfiwcs,%
	Wiseman:2000:sfoasstcoasbh,Popov:2011:rftspoascisst,Cho:2007:tsfoanmcsscoasbh}. 

In spacetime (\ref{eq:metricBE}) the singular part of potential reads
\begin{equation}
A_{(t)}^{\textrm{DS}} = -\frac{q}{|\rho -\rho'|}e^{- \frac{1}{2}\alpha(\rho')}\label{eq:singBE}
\end{equation}
and correspondingly in the "drainhole" spacetime we obtain
\begin{equation}
A_{(t)}^{\textrm{DS}} = -\frac{q}{|\rho -\rho'|}.\label{eq:singD}
\end{equation}

\section{Self-force in the space-time of topological defects}\label{Sec:TD}

Topological defects, its creation, evolution, and interaction have been discussed in detail in monographs  \cite{Vilenkin:1994:CSOTD,Anderson:2002:MTCS} and reviews \cite{Vilenkin:1985:CSDW,Hindmarsh:1995:cs}. At the beginning let us briefly discuss the space-time of topological defects and then we consider the phenomenon of the self-force in the space-time of topological defects, namely, cosmic strings and global monopoles. There exist four kinds of topological defects according to the kind of broken symmetry: domain walls, monopoles, cosmic strings, and textures. It is possible to consider also the different kind of hybrid combinations of them. We consider only cosmic strings and global monopoles. 

\subsection{Space-time of cosmic strings}

The infinitely long straight cosmic string represents the thread-like distribution of matter which Lorentz invariant in the direction of the string \cite{Vilenkin:1981:GFVDWS}. The general stationery cylindrically symmetric solution of the vacuum Einstein equations has the following form \cite{Stephani:2003:esoefe}
\begin{gather*}
ds^2 = - e^{2U} (dt + A d\varphi)^2 + e^{-2U}[e^{2k}(dr^2 + dz^2)+r^2 d\varphi^2],\\
e^{2U} = r (a_1 r^n + a_2 r^{-n}),\ n^2 a_1a_2 = -C^2, \\
A = \frac{C}{na_2} \frac{r^n}{a_1 r^n + a_2 r^{-n}} + B,\ e^{2k-2U} = r^{\frac{n^2-1}2}.
\end{gather*}
It was noted \cite{Stephani:2003:esoefe} that the regular solution at origin is only Minkowski space-time. The particular static solution with $A=C=0$ has been found by Levi--Chevita at the end of the 1920s and is given by 
\begin{equation}\label{LeviCivita}
ds^2 = - r^{2m} dt^2 + r^{-2m}[r^{2m^2}(dr^2 + dz^2) + r^2\nu^{-2} d\varphi^2].
\end{equation}
In general, one has two arbitrary constants $m$ and $\nu$. The boost-invariance along the $z$-axis gives the condition for constant $m$. It may be only $m=0$ and $m=2$. The space-time with $m=2$ is nonphysical and has unusual properties, for example, the length of the circle tends to infinity when the radius goes to zero and vice versa, the length tends to zero for a great radius and the space-time becomes effectively three-dimensional. For $m=0$ the space-time is flat and possibly conical if $\nu \not= 1 $ with the following line element:
\begin{equation}\label{ds^2infhin}
ds^2 = -d t^2 + dr^2 + \frac{r^2}{\nu^2} d\varphi^2 + dz^2.
\end{equation}
The main characteristic of this space-time is the flat angle deficit, $\triangle = 2\pi (1 - 1/\nu)$, -- the difference between the circle length of a unit radius in Minkowski space-time and the length of a circle of unit radii around the origin in conical space-time (\ref{ds^2infhin}).

The conical space-time of this kind (\ref{ds^2infhin}) with a flat angle deficit was later investigated in papers by Marder \cite{Marder:1959:FSGF}, Sokolov and Starobinski \cite{Sokolov:1977:sctcs} and Israel \cite{Israel:1977:Lsgr} as an example of space-time with the singular source. The linear energy density of a cylindrical distribution of matter given by Eq. (\ref{LeviCivita}) was considered by Thorn \cite{Thorne:1965:EILCSSGR} from different points of view ($C$-energy, the energy of Levi--Civita). At a later time, it has been shown by Vilenkin in Ref. \cite{Vilenkin:1981:GFVDWS} that if $r\geq 0$ then the space-time (\ref{ds^2infhin}) represents the exact solution of Einstein equations which describes the space-time of an infinitely thin cosmic string.

The space-time of an infinitely thin cosmic string located at the $z$-axis, in cylindrical coordinates, has the form (\ref{ds^2infhin}) and the manifold itself, $\mathbb{M}_4 = \mathbb{M}_2 \otimes \mathbb{C_\nu}$, is a direct product of the two-dimensional Minkowski manifold $\mathbb{M}_2$ with coordinates $t,z \in \mathbb{R}$, and the two-dimensional cone space $\mathbb{C_\nu}$ with a conical singularity at the origin and coordinates $r\in \mathbb{R}^+$ and $\varphi \in [0,2\pi )$.

For the new radial coordinate $r = \nu \rho^{\frac 1\nu}$ the cone part of space-time becomes conformally flat
\begin{equation}
	\label{ds^2infhin2}
ds^2 = -dt^2 +  e^{-4V}(d\rho^2 + \rho^2 d\varphi^2) + dz^2,
\end{equation}
where $V = \frac{\nu -1}{4\nu}\ln \rho^2$. As was shown in Ref. \cite{Sokolov:1977:sctcs} that the scalar curvature of this space-time has to be regarded as a distribution
\begin{equation}
	\label{RSing}
\mathcal{R} =  4\pi \frac{\nu -1}\nu \frac{\delta^{(2)}(x)}{\sqrt {\gs^{(2)}}}
\end{equation}
in order to obey the Gauss--Bonnet theorem. Here $\gs^{(2)} = r^2/\nu^2$ and
\begin{equation*}
	\int_0^\infty\int_{0}^{2\pi} f(r,\varphi)\delta^{(2)}(x) dr
d\varphi = f(0,0).
\end{equation*}
The point is that the Gauss--Bonnet theorem for a general form of the metric
\begin{equation}
ds^2 = -D(r,\varphi)^2dt^2 +  A(r,\varphi)^2 dr^2 + B(r,\varphi)^2 d\varphi^2 + E(r,\varphi)^2 dz^2,
\end{equation}
with asymptotic $ A_{r \to \infty} = 1,\   B_{r \to \infty} \to \frac{r}{\nu},\  A_{r \to 0} = 1,\  B_{r \to 0} \to r$, for two dimensional surface $t=const$ and $z=const$ gives \cite{Ford:1981:gaAe}
\begin{equation}
 \int \mathcal{K} \sqrt{\gs^{(2)}} dS = 2\pi \left(1-\frac{1}{\nu}\right),
\end{equation}
where $\mathcal{K}$ is the Gauss curvature. This result has to be fulfilled for the metric of the cosmic string. The expression (\ref{RSing}) for the scalar curvature may be found by direct calculations by using the metric in the form (\ref{ds^2infhin2}) with $V = f(\rho,\varphi)$. Indeed, in this case,  we obtain the following expression for the scalar curvature
\begin{equation*}
\mathcal{R} = 4e^{4V}\triangle V,
\end{equation*}
where $\triangle = \partial^2_\rho + \frac 1\rho \partial_\rho + \frac 1{\rho^2}\partial^2_\varphi$. Taking into account that
\begin{equation*}
\triangle \ln \rho = 2\pi \frac{\delta (\rho)\delta(\varphi)}{\rho}
\end{equation*}
we obtain the expression (\ref{RSing}).

The energy-momentum tensor corresponding to the metric of a cosmic string (\ref{ds^2infhin}) describes the thread-like distribution of matter along the $z$-axis \cite{Vilenkin:1981:GFVDWS} and it is singular at the string's origin:
\begin{equation}
	\label{EMTString}
T^{\alpha}_{\beta} = -\mu \frac{\delta^{(2)}(x)}{\sqrt {\gs^{(2)}}}\diag (\stackrel{\sss{t}\sss{t}}{1}, \stackrel{\sss{r}\sss{r}}{0}, \stackrel{\sss{\varphi}\sss{\varphi}}{0}, \stackrel{\sss{z}\sss{z}}{1}),
\end{equation}
where $\mu$ is the linear energy density of matter. The matter equation of state of the string is $\mathcal{E} + \mathcal{P} = 0$, and the structure of the energy-momentum tensor (\ref{EMTString}) is defined due to the boost-invariance of the metric along the $z$-axis. The Einstein's equations give the connection of the cone parameter $\nu$ with the linear energy density of energy: $1/\nu = 1 - 4\mu$, or in dimensional units $1/\nu = 1 - 4G\mu/c^2$. Note that the quantity $c^4/G \approx 1.34\cdot 10^{28} gr/cm \cdot c^2$ has a dimension of linear energy density and defines the characteristic scale of the linear mass density of the string $\sim 10^{28} gr/cm$.

There is a specific property of the straight cosmic string space-time (\ref{ds^2infhin}): the Newtonian potential of the string is zero. Indeed, in weak-field approximation, the Newtonian potential $U$ obeys the following equation 
\begin{equation*}
	\triangle U = 4\pi (2T^t_t - T^\alpha_\alpha).
\end{equation*}
The right-hand side of this equation equals zero due to the structure of the energy-momentum tensor (\ref{EMTString}) and therefore there is no source for the Newtonian potential. 

By a simple changing of the angle variable $\varphi\to \nu\phi$, the line element (\ref{ds^2infhin}) transforms into the line element of the Minkowski space-time
\begin{equation}
	\label{ds^2Mink}
ds^2 = -d t^2 + dr^2 + r^2 d\phi^2 + dz^2,
\end{equation}
in which the angle variable belongs to the following domain: $\phi\in [0,2\pi/\nu]$. Therefore, the space-time of infinitely thin cosmic-string may be regarded as the Minkowski space-time without wedge
\begin{equation}
	\label{CutAngle}
\phi\in (0,2\pi(1 - \frac 1\nu))
\end{equation}
with the size equal to the angle deficit and by consequence the identification of the edges of the wedge.

To see the singular structure of the space-time of a cosmic string in manifest form let us turn to the Cartesian coordinates $x^1=r\cos \varphi, x^2 = r\sin\varphi$. The line element now reads ($a,b= 1,2$) 
\begin{equation*}
ds^2 = -dt^2 + \gs_{ab}dx^adx^b + dz^2,
\end{equation*}
where
\begin{equation*}
\gs_{ab} = \delta_{ab} - \frac{\nu^2-1}{\nu^2}\left[\delta^{ab}-\frac{x^ax^b}{r^2}\right].
\end{equation*}
From this expression, we observe the singular behavior of the metric of cosmic string's space-time and the absence of the symmetry for transverse translations. 

\begin{figure}[ht]
\centerline{\includegraphics[width=5truecm]{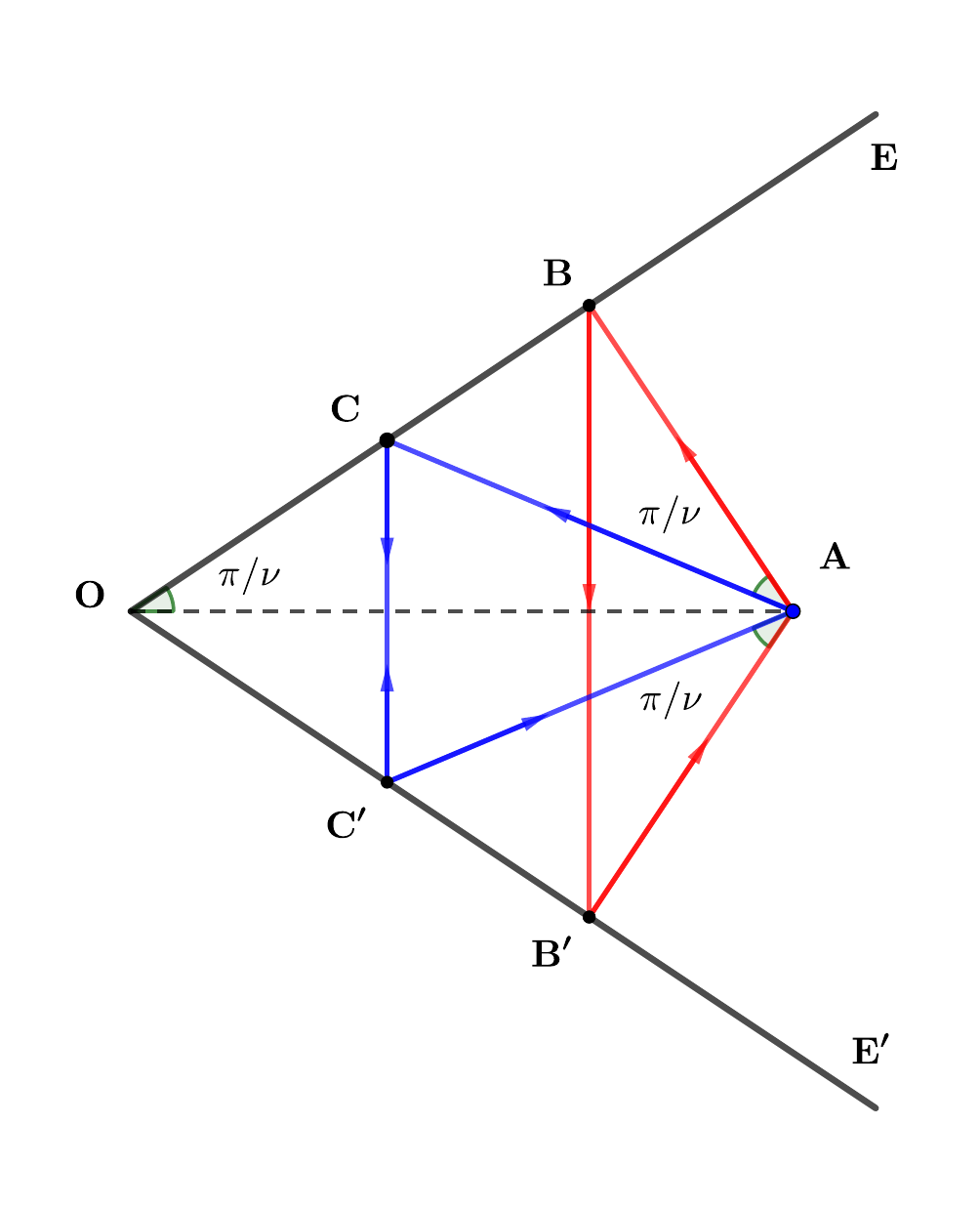}\hspace{4em}\includegraphics[width=4truecm]{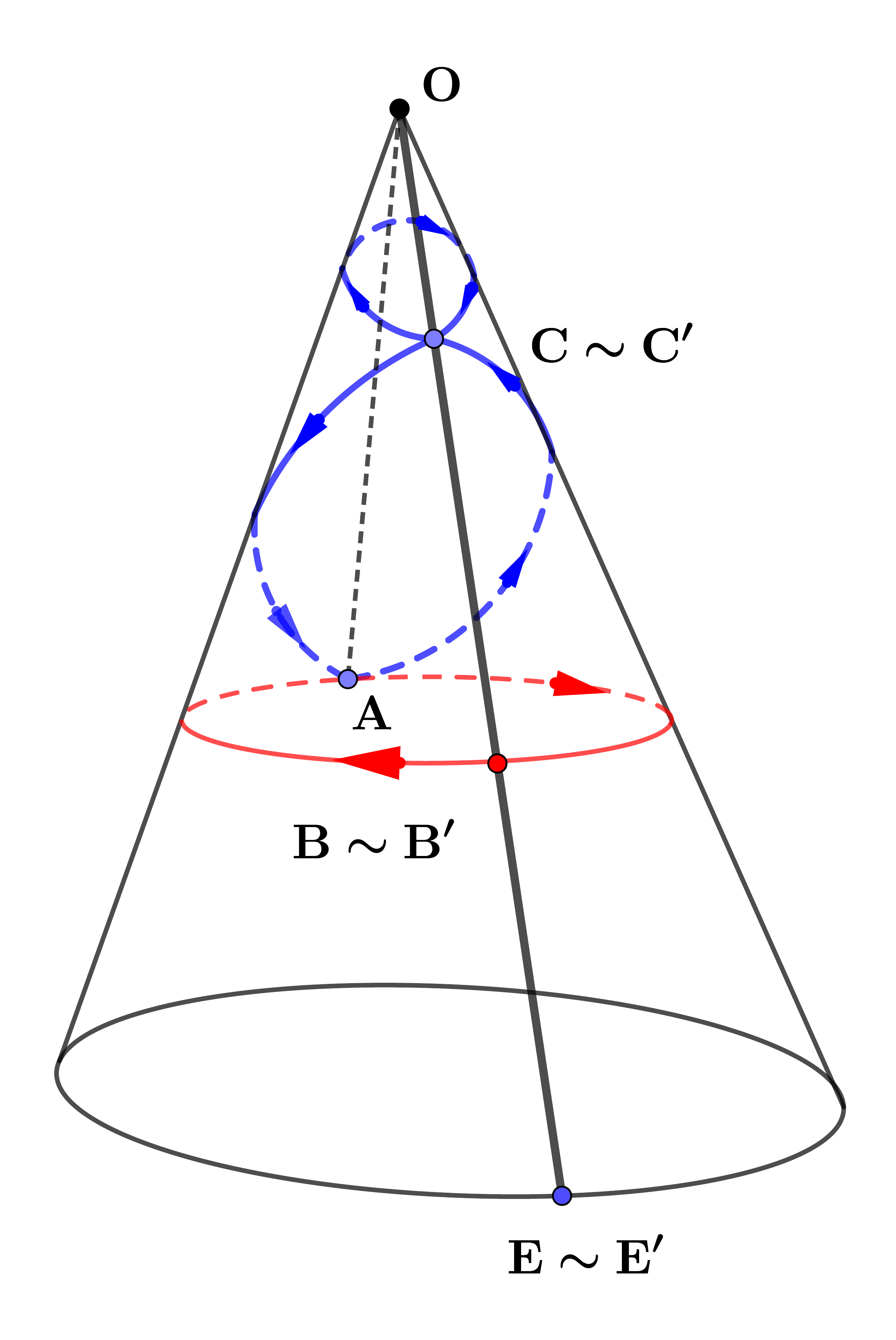}}
\caption{\footnotesize The plots of a cone space with $\nu > 3$. One has to cut the plane along lines $OE$ and $OE'$, and threw away the part of the plane out of sharp angle $EOE'$ and identify the edges of wedge. For $\nu > 3$ there exist two closed geodesics: $A\to B \sim B'\to A$ and $A\to C \sim C' \to C \sim C' \to A$.} \label{Fig:Many}
\end{figure}

It is worth noting that the value $\nu = 2$ is, in some sense, a critical value of $\nu$ \cite{Khusnutdinov:1995:Cpsscs}. In this case, we have to cut and threw away wedge greater than $\pi$. Due to this point in a conical space, $\mathcal{C}_\nu$ with $\nu\geq 2$ the closed geodesics appear. Indeed, let us consider the conical space with $\nu > 3$, that is, with angle $\pi/\nu < \pi/3$. The cone space is represented in Fig. \ref{Fig:Many}. We have to cut the plane along lines $OE$ and $OE'$ and threw away the part out of sharp angle $EOE'$, and identify (glue) edges of cut $OE$ and $OE'$. The string is situated perpendicularly to the plane of the picture at point $O$. It is easy to see that there are two close geodesic lines namely, first one $A \to B \sim B' \to A$, and the second $A\to C \sim C' \to C \sim C' \to A$. The identified points we denoted with sign $\sim$. In a domain $\nu \in (1,2)$ there is no closed geodesics because in this case $\pi/\nu \in (\pi,\pi/2)$ and it is impossible to make perpendicular from point $A$ to ray $OE$ to make closed geodesics. In the domain $\nu\in (2,3)$ there exists single closed geodesics $A\to B \sim B'\to A$. The second geodesics, in this case, is impossible because of the angle $\pi/\nu \in (\pi/2,\pi/3)$ and the angle $\widehat{BAC}$ is greater than the angle $\widehat{BAO}$. In the domain $\nu\in (3,4)$ the second geodesics appears with $\widehat{BAC} = \widehat{BOA}$. The number of closed geodesics $\mathcal{N}$ reads \cite{Khusnutdinov:1995:Cpsscs}
\begin{equation*}
	\mathcal{N} = \left[\nu - 1\right],
\end{equation*}
where $[a]$ means integer part of $a$. The length $L_n$ of the $n$-th closed geodesic is
\begin{equation*}
	L_n = 2r \sin\frac{\pi n}\nu,
\end{equation*}
where $n\leq \nu/2$ and $r$ is the coordinate of the point $A$.

Taking into account the above observation (\ref{CutAngle}) it is possible to construct various kinds of space-times with the conical singularities \cite{Mazur:1986:SCSQE,Deser:1992:Tt,Galtsov:1993:Sscd,Tod:1994:Cst} which are not Lorentz-invariant in the direction of the string by identification of different domains of the Minkowski space-time. The three-dimensional parts of these space-times were considered in a geometrical theory of defects in solid  \cite{Katanaev:1999:Sdcsgtd,Kleinert:1989:GFCMVISVLVISaD,Harris:1977:D}. The space-time of the cosmic dispiration (the cosmic dislocation in terms of the Ref. \cite{Galtsov:1993:Sscd}), where the three-dimensional part is the superposition of the disclination and dislocation, is constructed from the Minkowski space-time by identification of the points by
\begin{equation*}
	(t,r,\phi,Z) = (t,r,\phi + \frac{2\pi}\nu, Z + 2\pi \kappa).
\end{equation*}
The parameters $\nu$ and $\kappa$ describe correspondingly the disclination and dislocation. In new coordinates $\varphi = \phi \nu,\ z = Z - \kappa\nu\phi$, the line element takes the following form:
\begin{equation}	\label{Desperation}
ds^2 = -dt^2 + dr^2 + \frac{r^2}{\nu^2}d\varphi^2 + (dz + \kappa d\varphi)^2.
\end{equation}

The space-time of a cosmic string which is rotated around its symmetry axis with angular momentum $J$ was considered in Refs. \cite{Mazur:1986:SCSQE,Deser:1992:Tt,Galtsov:1993:Sscd,Tod:1994:Cst}. The line element reads
\begin{equation*}
	ds^2 = -(dt + 4 Jd\varphi)^2 + dr^2 + \frac{r^2}{\nu^2}d\varphi^2 + dz^2.
\end{equation*}

The line elements of the space-time of an infinitely thin cosmic string and the string rotated around its symmetry axis may be obtained by adding $dz^2$ to the line elements describing a mass at rest and a rotating mass in $(2+1)$-dimensional case \cite{Staruszkiewicz:1963:GTTS,Deser:1984:TEGDFS,Mazur:1986:SCSQE,Deser:1992:Tt}. As a result of this, it is obvious that the noted above space-times are flat. It may be shown explicitly by simply changing the variables: $(\varphi \to \nu\phi,\ z\to z - \kappa\nu\phi)$ in the space-time of cosmic string and cosmic dispiration and $(\varphi \to \nu\phi,\ t\to t - 4 J\nu\phi)$ in the space-time of a rotated cosmic string. Also, we observe that the space-time of cosmic dispiration is periodic not merely for the angle variable $\phi$, but also the variable $z$ with period $2\pi\kappa\nu$, that is, it has helicoidal structure. The space-time of a rotated string possesses a non-trivial causal structure \cite{Mazur:1986:SCSQE,Mazur:1987:Mr}; it is periodic in time with period $8\pi J\nu$.

The space-time of $N$ parallel cosmic strings \cite{Letelier:1990:MCS} may be found by a simple generalization of Einstein's equations for $N$ point-like masses in the three-dimensional theory of gravity. It is enough to rewrite the space-time of a single cosmic string in the following coordinates:
\begin{equation*}
r = \nu r_c^{1-4\mu},\ x=\alpha + r_c\cos\varphi,\ y=\beta + r_c\sin\varphi ,
\end{equation*}
in which the conical part $\mathbb{C_\nu}$ of the space-time takes the conformally flat form
\begin{equation}\label{MetricaInConfCoord}
ds^2 = -dt^2 + e^{-4V} (dx^2 + dy^2) + dz^2.
\end{equation}

The quantity $V = \mu \ln ([x-\alpha]^2 + [y-\beta]^2)$ represents the Newtonian potential of linear distribution of matter with linear density $\mu$, and $(\alpha, \beta)$ are the coordinates of the string in the hyper-surface $z=const$. The natural generalization of the metric (\ref{MetricaInConfCoord}) for $N$ parallel strings, each of which has linear density $\mu_k$ and situated at the point $(\alpha_k, \beta_k)$, is the space-time with line element (\ref{MetricaInConfCoord}) with the following potential:
\begin{equation*}
	V = \sum_{k=1}^N \mu_k \ln ([x-\alpha_k]^2 + [y-\beta_k]^2).
\end{equation*}
In three-dimensional gravity, this solution describes the gravitational field of $N$ point-like particles. For two particles it was found in Ref. \cite{Staruszkiewicz:1963:GTTS} and for an arbitrary number of particles in Ref. \cite{Deser:1984:TEGDFS}. The derivation of the above metric for $N$ parallel strings is based on the fact that the Einstein equations for the metric with form (\ref{MetricaInConfCoord}) with the source (\ref{EMTString}) reduces to the following single equation
\begin{equation*}
	\triangle_{(2)} V = -4\pi \mu \frac{\delta^{(2)}(x)}r
\end{equation*}
which describes the Newtonian potential of mass $\mu$ situated in the origin in flat Minkowski space-time. The generalization for $N$ mass is obvious.

The space-time of two moving and parallel cosmic strings was obtained by Gott in Ref. \cite{Gott:1991:CtcpbpmcsEs} in connection with appearance in such space-time of closed time-like geodesics. In the Ref. \cite{Deser:1992:Pcsdngctc} was shown that the closed time-like geodesics do not appear if the space-time of a cosmic string obeys the physical condition -- the total angle deficit can not exceed $2\pi$. In the opposite case, the space-time has pathological peculiarities.

The consideration of the non-parallel cosmic strings is more complicated. The general solution of this problem for the case of $N$ non-parallel and arbitrary moving strings was found by Letelier and Gal'tsov in Ref. \cite{Letelier:1993:Mmccs} by using the generalization of the approach suggested in Ref. \cite{Staruszkiewicz:1963:GTTS}.

The metric of this space-time may be obtained from the Minkowski space-time
\begin{equation*}
ds^2 = -dt^2 + dz^2 + dZ dZ^*
\end{equation*}
by the singular transformation of coordinates with help of the Christoffel--Schwartz formula
\begin{equation*}
Z = X + i Y = \int_{\zeta_0}^\zeta \prod_{k=1}^N(\xi - a_k)^{-4\mu_k} d\xi ,
\end{equation*}
where $\zeta = x + i y,\ a_k = \alpha_k + i\beta_k$, and $a_k$ depend on the $t$ and $z$. For the case of $N$ straight and infinitely long strings the functions $a_k$ have the following structure:
\begin{equation*}
a_k = x_{0k} + v_{xk} t + m_{xk}z + i (y_{0k} + v_{yk} t + m_{yk}z),
\end{equation*}
where the coefficients $(x_{0k},y_{0k}),\ (v_{xk},v_{yk}),\ (m_{xk},m_{yk})$ have simple physical sense. They are the coordinates of points where strings cross the surface $z=const$, and the velocity and slope $k$-th string to this surface, respectively.

The space-time metric of the $N$ arbitrary moving strings has the following form:
\begin{equation}
	\label{NonParallel}
ds^2 = -dt^2 + dz^2 + e^{-4V} \left\{[dx + F_1 dt + G_1 dz]^2 + [dy + F_2 dt + G_2 dz]^2\right\},
\end{equation}
where
\begin{eqnarray*}
	V &=& \sum_{k=1}^N \mu_k \ln |\zeta - a_k|^2,\\
F &=& F_1 + i F_2 = \prod_{k=1}^N (\zeta - a_k)^{4\mu_k} \int_{\zeta_0}^\zeta \frac{d\xi}{\prod_{n=1}^N(\xi - a_n)^{4\mu_n}} \sum_{l=1}^N \frac{4\mu_l \dot{a}_l}{\xi - a_l},\\
G &=& G_1 + i G_2 = \prod_{k=1}^N (\zeta - a_k)^{4\mu_k} \int_{\zeta_0}^\zeta \frac{d\xi}{\prod_{n=1}^N(\xi - a_n)^{4\mu_n}} \sum_{l=1}^N \frac{4\mu_l a'_l}{\xi - a_l},
\end{eqnarray*}
and $|\zeta_0|\to \infty$. In the above expression, it was supposed that all $a_k$ are different, otherwise, all strings coincide and equal to one string. The dot and prime denote the derivatives with respect to $t$ and $z$, respectively. 

The source of the space-time (\ref{NonParallel}) is the following energy-momentum tensor: 
\begin{equation*}
	T^{\mu\nu} = e^{4V}\sum_{k=1}^N \mu_k \delta (x - \alpha_k) \delta (y - \beta_k) (e^\mu_t e^\nu_t - e^\mu_z e^\nu_z),
\end{equation*}
where
\begin{equation*}
	e^\mu_t = \delta^\mu_t - F_1 \delta^\mu_x - F_2 \delta^\nu_y,\ e^\mu_z = \delta^\mu_t - G_1 \delta^\mu_x - G_2 \delta^\nu_y.
\end{equation*}
From this expression, it is easy to see that the line element (\ref{NonParallel}) describes $N$ arbitrary moving strings (see also discussion \cite{Clement:2005:TLsr}).

The line element of the space-time with infinitely thin and straight string may be found in the following simple way: we rewrite the space-time in cylindrical or spherical coordinates and insert into the metric the deficit of polar angle. For example, the space-time of a black hole which is intersected by a cosmic string \cite{Aryal:1986:Csbh}, and of Friedman--Robertson--Walker universe with a cosmic string have, respectively, the following forms:
\begin{eqnarray*}
ds^2 &=& -\left(1 - \frac{2M}r\right) dt^2 +  \frac{dr^2}{1 - \frac{2M}r} + r^2 \left(d\theta^2 + \frac 1{\nu^2} \sin^2\theta d\varphi^2\right),\\
ds^2 &=& -dt^2 + a^2(t)\left[\frac{dr^2}{1-kr^2} + r^2 \left(d\theta^2 + \frac 1{\nu^2} \sin^2\theta d\varphi^2\right) \right].
\end{eqnarray*}
The space-time representing the black hole intersected by many cosmic strings was considered in Ref. \cite{Dowker:1992:Pcs,Frolov:2001:Bhpmc,Frolov:2002:Tsbhs}. In Ref. \cite{Frolov:2001:Bhpmc}, it was considered by Frolov and Fursaev a black hole cross on radial cosmic strings. To obtain equilibrium for the system, it is necessary to impose the symmetry conditions with respect to the rotation axis and from this condition, three configurations appear in which the points where the strings cross the black hole are in the tops of tetrahedron, octahedron, and icosahedron. The stable space-time with an arbitrary number of strings that cross the black hole was found in the Ref. \cite{Frolov:2002:Tsbhs}. 

The model of the infinitely thin cosmic string describes the ideal situation -- the string without internal structure. A more realistic model of the cosmic string with a non-trivial internal structure was found by Gott \cite{Gott:1985:GlevsEs} and Hiscock \cite{Hiscock:1985:EGFS}. The model represents the string with non-zero transverse size $r_\0$ with matter of constant energy density $\mathcal{E}=const$ inside it, with the following step-like energy-momentum tensor
\begin{equation}	\label{EMTGotHis}
T^{\alpha}_{\beta} = -\theta (r_\0 - r)\mathcal{E} \diag (\stackrel{\sss{t}\sss{t}}{1}, \stackrel{\sss{r}\sss{r}}{0}, \stackrel{\sss{\varphi}\sss{\varphi}}{0}, \stackrel{\sss{z}\sss{z}}{1})
\end{equation}
and matter equation of state $\mathcal{E} + \mathcal{P} = 0$. The general model with the arbitrary smooth distribution of matter inside the string was considered by Linet in Ref. \cite{Linet:1985:TSMCSDMCS}. The solution of Einstein's equations corresponding to (\ref{EMTGotHis}) has the following simple form: 
\begin{subequations}\label{dsGotHis}
\begin{equation}
	ds_{\textrm{in}}^2 = -dt^2 + d\rho^2 +\frac{\rho_\1^2}{\epsilon^2} \sin^2 \left(\frac{\epsilon\rho}{\rho_\1}\right) d\varphi^2 + dz^2
\label{dsin}
\end{equation}
inside the string, and
\begin{equation}
	ds_{\textrm{out}}^2 = -d t^2 + dr^2 + \frac{r^2}{\nu^2} d\varphi^2 + dz^2
\label{dsout}
\end{equation}
\end{subequations}
outside of it. The manifold is covered by two maps, namely the interior with coordinates $(t,\rho,\varphi,z)$: $(t,z)\in \mathbb{R}$, $\rho\in [0,\rho_\1]$, $\varphi\in [0,2\pi)$ and the exterior with coordinates $(t,r,\varphi,z)$: $(t,z)\in \mathbb{R}$, $r\in [r_\0,+\infty)$, $\varphi\in [0,2\pi)$. The space-time inside the string represents a space-time of constant curvature with the following non-zero components of the Riemann, and Ricci tensors and the scalar curvature
\begin{equation*}
\mathcal{R}^\rho_\cdot{}_{\!\varphi\rho\varphi}= \sin^2\left(\frac{\epsilon\rho}{\rho_\1 }\right) \ ,\ \mathcal{R}^\rho_\cdot{}_{\!\rho} = \mathcal{R}^\varphi_\cdot{}_{\!\varphi} = \frac{\epsilon^2}{\rho_\1 ^2}\ ,\ \mathcal{R}=\frac{2\epsilon^2}{\rho_\1 ^2}.
\end{equation*}
The exterior described by the metric (\ref{dsout}) coincides with the space-time of an infinitely thin cosmic string given by the metric (\ref{ds^2infhin}).

The metric (\ref{dsGotHis}) is $C^1$- and the external curvature is $C^o$- continuous on the surface of the string, at the hypersurface $\rho=\rho_\1$ which implies the absence of the surface energy in accord with Israel's theorem \cite{Israel:1966:Shtsgr}. The radius of string $\rho_\1$ in the internal map (\ref{dsin}) and the radius of string $r_\0$ in the external map (\ref{dsout}), and also the parameters $\epsilon$ and $\nu$ are connected by relations
\begin{equation*}
\frac{r_\0}{\rho_\1} = \frac{\tan \epsilon}\epsilon,\ \cos\epsilon = \frac 1\nu.
\end{equation*}
The parameter $\epsilon$ is defined in term of the "energetic"\/ radius of the string $\rho_* = \frac 1{\sqrt{8\pi \mathcal{E}}}$
\begin{equation*}
	\frac{\rho_\1}{\rho_*} = \epsilon .
\end{equation*}

\begin{figure}
\centerline{\includegraphics[width=5truecm]{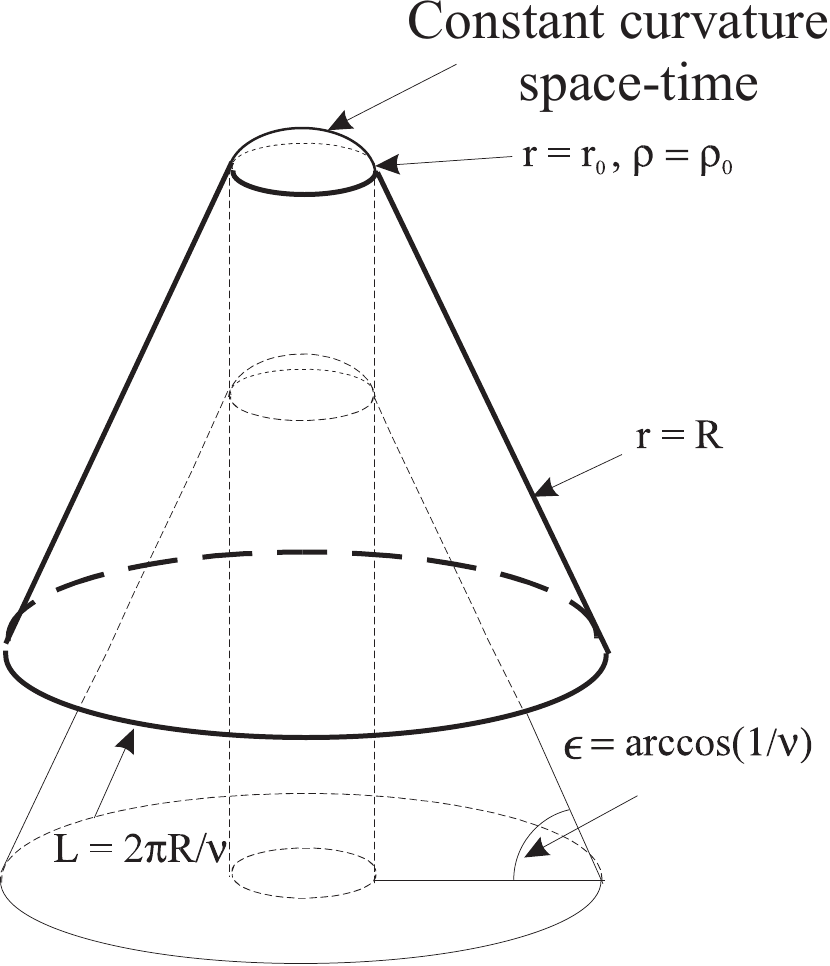}}
\caption{\footnotesize Two dimensional section $(t=const, z=const)$ of the Gott-Hiscock space-time described by the line element (\ref{dsGotHis}).} \label{Fig:Cone}
\end{figure}

The limit of the Minkowski space-time is obtained by taking the energy density inside the string zero $(\rho_* \to \infty )$, provided the radius $\rho_\1$ is fixed. Then, the angle deficit goes to zero because of $\epsilon \to 0$. Therefore, in this limit $\nu = 1/\cos \epsilon = 1,\ r_\0 = \rho_\1$ and the space-time becomes Minkowskian in both maps. From the other side in order to obtain the limit for an infinitely thin cosmic string $(\rho_\1 \to 0)$, provided angle deficit ($\epsilon = const$) is fixed one has to take the energy density ${\cal E}$ at infinity proportional to $\epsilon^2/8\pi\rho_\1^2$. Nevertheless, the energy $\mu$ per unit length of the string, which is the product the energy density $\epsilon^2/8\pi\rho_\1^2$ and the square of the cross-section of the string, is left a constant which does not depend on the radius of the string and is equal to $(1 - 1/\nu)/4$, that is the same quantity as for the infinitely thin cosmic string space-time. The embedding of two dimensional part $(t=const, z= const)$ of the string's space-time (\ref{dsGotHis}) to three-dimensional Euclidean space is plotted in Fig. \ref{Fig:Cone}.

The manifold may be covered by one card. Indeed, extending the external radial coordinate $r$ into the interior by mapping $r=r_0 + ( \rho_0/\epsilon )\tan  (\epsilon\rho /\rho_0 - \epsilon )$. Then the space-time is described by the following line element \cite{Khusnutdinov:1999:Gsemsfbftcs}:
\numparts \label{eq:Thick}
\begin{eqnarray}
	ds_{\textrm{in}}^2 &=& - dt^2\! +\! \frac{dr^2}{[1 + \frac{\epsilon^2(r-r_\0)^2}{\rho_\1^2}]^2}\! +\! \frac{r^2 }{\nu^2}\frac{d\varphi^2}{1 + \frac{\epsilon^2(r-r_\0)^2}{\rho_\1^2}} \! +\! dz^2, \\
ds_{\textrm{out}}^2 &=& -dt^2 + dr^2 + \frac{r^2 }{\nu^2}d\varphi^2 + dz^2,
\end{eqnarray}
\endnumparts
with $r \in [0,r_\0]$ inside the string and $r\in [r_\0, \infty )$ outside of it.

There is another way to cover the manifold by one map by the following mapping
\begin{equation*}
	\sin\left(\frac{\epsilon\rho}{\rho_\1 }\right) =\frac r{r_\0} \sin \epsilon.
\end{equation*}
The line element now reads \cite{Allen:1990:Eccqfcss}:
\begin{equation}
	\label{both}
ds^2=-dt^2 + P^2(r)dr^2 + \frac{r^2}{\nu^2}d\varphi^2 + dz^2\ ,
\end{equation}
where
\begin{equation*}
P(r)=\left\{\begin{array}{ccl}\left[\nu^2 +\frac{r^2}{r_\0^2}(1-\nu^2)\right]^{- 1/2}&,&r\leq r_\0\ ,\\
1&,&r\geq r_\0\ . \end{array}\right.
\end{equation*}

The obvious advantages of the Gott--Hiscock space-time are expressed in the following elements. (i) The metric obtained has simple and manifest form, and (ii) the metric models the strings of the finite cross-section which appear in realistic field theories and it may be used for discussion effects due to the finite string's thickness.

The cosmic strings which have appeared in different kind of field theories have no rigid boundary, and the space-time becomes conical only asymptotically far from the string's origin \cite{Vilenkin:1994:CSOTD}. The simple model in which cosmic strings appear is the Abelian $U(1)$ Higgs model which is described by the Lagrangian 
\begin{equation}
	\mathcal{L} = -\frac 12 D^\mu \phi (D_\mu \phi)^* - \frac\lambda 8 (|\phi|^2 - \eta^2)^2 - \frac 14 F^{\mu\nu}F_{\mu\nu} + \frac 1{16\pi} \mathcal{R},
\end{equation}
where $D_\mu = \nabla_\mu + \ib q A_\mu$, and $\phi$ and $A_\mu$ are the scalar and gauge fields correspondingly. In flat space-time, this model was considered in Ref. \cite{Nielsen:1973:Vmds} to describe the superconductors of the second kind. In the framework of this model, there is a solution representing a thread-like structure that contains a magnetic field with quantized magnetic flux $\Phi = 2\pi n\hbar/q$. This solution describes the well-known phenomenon of penetration of the magnetic field into the superconductor of the second kind \cite{Abrikosov:1957:OMpssg}.

The self-consistent gravitational field of this solution with the following ansatz for the metric:
\begin{equation*}
	ds^2 = - e^A dt^2 + e^B dz^2 + e^C d\varphi^2 + dr^2,
\end{equation*}
where $A,B,C$ are the functions of the radial coordinate only, was considered in Ref. \cite{Garfinkle:1985:GRS,LagunaCastillo:1987:Cfsugcs,Garfinkle:1989:Cgstdpmcs}. Let us note briefly the main results. (i) The energy density of fields exponentially falls dawn outside the core of the string; the scalar field goes to its vacuum value $\eta$, and the gauge field tends to zero. (ii) The space-time asymptotically becomes conical with a flat angle deficit $\sim \eta^2$. In dimensional units, the quantity $\eta^2$ has a linear energy density and the dimensionless combination is $G\eta^2/c^4$. (iii) Transverse scale of the string is defined by the Compton wavelength of the Higgs and gauge bosons with masses $m_\phi = \hbar\sqrt{\lambda/2}\eta,\ m_A = \sqrt{2}e\eta$. The wavelength of the Higgs bosons defines the cross-scale of domain with false vacuum and the wavelength of gauge bosons describes the size of the magnetic tube where mainly the magnetic field concentrates. In the framework of GUT's scales (in dimensional units $l_{\textrm{Pl}} = \sqrt{c^3/G\hbar} = 1.6\cdot 10^{-33} cm$) $\eta^2 c^2 l_{\textrm{Pl}} \sim 10^{16} GeV$ the transverse scale of string has the magnitude $10^{-30} cm$ and linear mass density $\sim 10^{25} gr/cm$. The metrical functions $A,B,C$ were analyzed numerically; there is no expression for them in manifest form.

\subsection{Space-time of global monopole}

The notion of monopole, more precisely magnetic monopole, was introduced in physics by Dirac in 1931 \cite{Dirac:1931:qsitef}. Later,  t'Hooft \cite{tHooft:1974:MMUGT} and Polyakov \cite{Polyakov:1974:Psqft,Polyakov:1975:ISQF} found in manifest form the solution of the Yang--Mills equations with broken \emph{local} $SO(3)$ symmetry which describes magnetic monopole. Magnetic monopole possesses a magnetic charge and radial magnetic field which is expressed in terms of the gauge fields. Global monopoles appear in models with broken \emph{global} $SO(3)$ symmetry. Firstly this space-time was obtained by Barriola and Vilenkin in Ref. \cite{Barriola:1989:GFGM}. The model consists of a Higgs triplet of scalar fields with Lagrangian $(a=1,2,3)$ 
\begin{equation*}
	\mathcal{L} = - \frac 12\partial_\mu\phi^a\partial^\mu\phi^a - \frac 14 \lambda \left(\phi^a\phi^a - \eta^2\right)^2.
\end{equation*}

The gravitational field corresponding to the global monopole is represented by the following spherically symmetric line element
\begin{equation}
	\label{Mon}
ds^2 = -B(r) dt^2 + A(r)dr^2 + r^2\left(d\theta^2 + \sin^2\theta d\varphi^2 \right),
\end{equation}
where the functions $A$ and $B$ have the form below:
\begin{equation*}
	B = A^{-1} = 1 - 8\pi\eta^2 - \frac{2M(r)}r.
\end{equation*}
The parameter $\eta^2$ in dimensional units has the dimension of the linear energy density. In the classical theory of gravity, there is the parameter with the same dimension: $c^4/G \approx 1.34\cdot 10^{28}gr/cm \cdot c^2$. This produces the dimensionless combination $G\eta^2/c^4$.

The scale of the monopole's core is defined by the Compton wavelength of Higgs bosons, $\varrho \sim \lambda^{-1/2}\eta^{-1}$. The parameter $M = \lim_{r\to\infty} M(r)$ which characterize the monopole's mass, has the following order: $M \sim \lambda\eta^4 \varrho^3 \sim \eta/\sqrt{\lambda}$ as was found in the Ref. \cite{Barriola:1989:GFGM}. The more thorough numerical analysis made by Harari and Lousto in Ref. \cite{Harari:1990:Rgegm} showed that this parameter is negative and in wide interval of the $\eta$: $0\leq 8\pi \eta^2 < 1$ approximately is
\begin{equation}\label{Mass}
M\approx - \frac{6\pi \eta}{\sqrt{\lambda}}.
\end{equation}

Therefore asymptotically the space-time constitutes the Schwarzschild space-time with negative mass and additional solid angle deficit, $32\pi^2 \eta^2$. The space-time of point-like monopole is found from the above metric by disregarding the internal structure of the monopole but preserving the solid angle deficit and described by the following line element:
\begin{equation}\label{GlobMon}
	ds^2 = -\alpha^2 dt^2 + \alpha^{-2} dr^2 + r^2 \left(d\theta^2 + \sin^2\theta d\varphi^2\right),
\end{equation}
where the parameter $\alpha^2 = 1 - 8\pi\eta^2$. By changing the radial coordinate $t\to t/\alpha,\ r\to r\alpha$ we obtain metric in the more symmetric form
\begin{equation}\label{GlobMonMetric}
ds^2 = - dt^2 + dr^2 + \alpha^2 r^2 \left(d\theta^2 + \sin^2\theta d\varphi^2\right).
\end{equation}

This space-time is an exact solution of Einstein's equations with the following energy-momentum tensor: 
\begin{equation}\label{EMTMon}
T^{\alpha}_{\beta} = -\frac{1 - \alpha^2}{8\pi\alpha^2 r^2} \diag (\stackrel{\sss{t}\sss{t}}{1}, \stackrel{\sss{r}\sss{r}}{1}, \stackrel{\sss{\theta}\sss{\theta}}{0}, \stackrel{\sss{\varphi}\sss{\varphi}}{0})
\end{equation}
and the non-zero components of the Ricci tensor
\begin{equation*}
	\mathcal{R}^\theta_\theta = \mathcal{R}^\varphi_\varphi = \frac{1 - \alpha^2}{r^2\alpha^2}.
\end{equation*}
Firstly this space-time was considered in Ref. \cite{Sokolov:1977:sctcs} from the geometrical point of view.

The full energy of the monopole's source,
\begin{equation*}
E = - 4\pi \int_0^L T^t_t r^2 dr = \frac{1 - \alpha^2}{2\alpha^2} L
\end{equation*}
is divergent linearly with the radius $L$ of the sphere surrounding the monopole, which is typical for global defects \cite{Durer:1995:Gfdcsf}. Usually, the scale of the cosmological horizon is considered as a natural boundary of the monopole's space-time. The linearly increasing with the distance of energy gives the constant attractive force in a pair monopole-antimonopole which ultimately leads to the annihilation of pair. Though the energy of monopole linearly increases, the ADM mass is finite and reads \cite{Nucamendi:1997:QfstAm}
\begin{equation}
	M_\textrm{{ADM}} = \frac M{(1-8\pi\eta^2)^{3/2}},
\end{equation}
where $M$ is given by Eq. (\ref{Mass}).

The main peculiarity of this space-time is the solid angle deficit $4\pi(1-\alpha^2)$ which is defined as a difference of the solid angle $4\pi$ in flat space-time and the solid angle in the global monopole space-time $4\pi\alpha^2$. For $\alpha < 1$ one has solid angle deficit, and for $\alpha >1$ one has solid angle access. The typical value of the solid angle deficit in the framework of the GUT ($\eta^2 c^2 l_{\textrm{Pl}} \sim 10^{16} GeV$) is the following quantity: $1 - \alpha^2 = 8\pi \eta^2 \sim 10^{-5}$. As in the space-time of the cosmic string, the space-time of global monopole has no Newtonian potential though the curvature of this space-time is not zero. Due to this fact, the energy of the monopole is linearly divergent with distance from the monopole.

At the same time, the global monopole produces strong tidal acceleration $a$ proportional to $1/r^2$ which is important from a cosmological point of view because this property may give the upper boundary of global monopoles in the Universe -- one monopole for a local group of galaxies \cite{Hiscock:1990:Abgm}. The numerical analysis made by Bennet and Rhie \cite{Bennett:1990:Cegmols} shows that the upper boundary is smaller than in Ref. \cite{Hiscock:1990:Abgm}. In fact, we have a few global monopoles per volume of horizon \cite{Bennett:1990:Cegmols,Yamaguchi:2001:Cegm}.

The space-time (\ref{Mon}) with constant mass $M(r) = M = const$ is also the solution of the Einstein's equations with energy-momentum tensor (\ref{EMTMon}) and it describes a black hole with mass $M' = M/\alpha^4$ in the center of which there is global monopole \cite{Dadhich:1998:Sbhgmc}. Indeed, by changing coordinate $t\to t/\alpha,\ r\to r\alpha$ one has the following line element:
\begin{equation*}
	ds^2 = - \left(1 - \frac{2M'}{r}\right) dt^2 + \frac{dr^2}{1 - \frac{2M'}{r}} + r^2\alpha^2 \left(d\theta^2 + \sin^2\theta d\varphi^2\right).
\end{equation*}
Asymptotically for $r\to\infty$ this space-time transforms to the space-time of the point-like global monopole with an angle deficit but locally it represents the space-time of the Schwarzschild. As for the case of infinitely thin cosmic string space-time, the embedding point-like global monopole in different spaces may be done by the following simple procedure. It is necessary to rewrite the background space-time in the spherical coordinates and make the solid angle deficit by introducing parameter in the spherical part of the metric.

The gravitational field of the magnetic monopole was considered in many papers \cite{VanNieuwenhuizen:1976:ORStHMMMCS,Bartnik:1988:PLSEYE,Volkov:1999:GnsbhYf,Kunzle:1990:SssSEf,Straumann:1990:Ibseye,Straumann:1990:Icbhs,Bizon:1990:Cbh,Smoller:1991:SssEYe,Breitenlohner:1992:Gms} (see also review \cite{Volkov:1999:GnsbhYf}). The resulting space-times do not possess a solid angle deficit. The reason for this is very simple. The solutions of Einstein's equations in the case of the spherical symmetry possess the following property
\begin{equation*}
	B = A^{-1} = 1 - \frac{8\pi}r\int_0^r T^t_t r^2 dr.
\end{equation*}
To get asymptotically a solid angle deficit it is necessary that the energy density falls down as $\sim C/r^2$. This condition is valid for the case of the global symmetry only \cite{Barriola:1989:GFGM} whereas local symmetry leads to more fast falling down due to gauge fields.
\subsection{Self-force and self-energy} 

Let us consider a particle in the space-time of an infinitely thin cosmic string with the line element given by Eq. (\ref{ds^2infhin}). To simplify the calculations we consider the Euclidean version of this space-time, $t=\ib\tau$, with the following line element
\begin{equation}
ds^2 = d \tau ^2 + dr^2 + \frac{r^2}{\nu^2} d\varphi^2 + dz^2,\ 0\leq \varphi \leq 2\pi.
\end{equation}
Because space-time is everywhere flat except at the origin the resulting self-force is defined by the non-trivial global structure of the space-time. The self-force in this space-time for a particle with the electromagnetic charge was considered by Linet \cite{Linet:1986:FCSCS,Linet:1986:otweitstoacs} and Smith \cite{Smith:1989:Geiscscqfsv} and for a massive and uncharged particle by Gal'tsov in Ref. \cite{Galtsov:1990:Acsgsmgai}. Because the self-force depends on the trajectory of the particle let us consider the simplest situation -- the particle at rest. In this case, the DL self-force is zero and we obtain in the manifest form a pure gravitationally induced contribution to self-force.

The Maxwell equation for the zero component of vector potential for a particle with charge $q$ at rest situated a distance $r$ from the string has the following form
\begin{equation*}
\triangle A^t = -\frac{4\pi q}{\sqrt{\gs^{(3)}}}\delta(r-r')\delta(\varphi - \varphi')\delta(z-z') ,
\end{equation*}
where $(x',\varphi',z')$ -- the coordinate of the particle; $\gs^{(3)}=r^2/\nu^2$ is the determinant of the three-dimensional spacial part of the metric. Therefore the component $A^t$ proportional to the scalar Green function of the three dimensional Laplace operator in the conical background is given by
\begin{equation*}
A^t(\textbf{r}) = 4\pi q G_\nu^{(3)}(\textbf{r};\textbf{r}').
\end{equation*}
This function describes the vector potential at the point $\textbf{r}$ due to the charged particle at the point $\textbf{r}'$. The potential calculated at the particle position, $\textbf{r}\to \textbf{r}'$, is divergent. This is the standard divergence that originated with the divergence of the electromagnetic self-energy of the particle. To subtract this infinity and obtain the finite result it is enough to subtract the Green function, $G_M(\textbf{r};\textbf{r}')$, of the Minkowski space-time which may be obtained from the Green function in the conical space-time by taking the limit $\nu \to 1$:
\begin{equation*}
G_M^{(3)}(\textbf{r};\textbf{r}') = G_{\nu = 1}^{(3)}(\textbf{r};\textbf{r}') .
\end{equation*}
Therefore, we obtain the renormalized Green function
\begin{equation*}
G^{(3)\textrm{ren}}_\nu(\textbf{r};\textbf{r}') = G_\nu^{(3)}(\textbf{r};\textbf{r}') - G_{\nu = 1}^{(3)}(\textbf{r};\textbf{r}') .
\end{equation*}

The self-potential, $\Phi(\textbf{r})$, the self-energy, $U(\textbf{r})$, and the self-force, $\vec{F}(\textbf{r})$, are defined, in terms of the renormalized Green function in coincidence limit, as respectively,
\begin{subequations}\label{Definitions}
\begin{eqnarray}
\Phi(\textbf{r})&=& 4\pi q G^{(3)\textrm{ren}}_\nu(\textbf{r},\textbf{r}') ,\\
U(\textbf{r})   &=& \frac 12 e \Phi(\textbf{r}) ,\\
\vec{\mathcal{F}}(\textbf{r}) &=& -\vec{\nabla} U(\textbf{r}) .
\end{eqnarray}
\end{subequations}

The three dimensional Green function $G_\nu^{(3)}$ is expressed in terms of the four-dimensional function as an integral over time
\begin{equation*}
G_\nu^{(3)}(\textbf{r};\textbf{r}')=\int G(x;x')d\tau.
\end{equation*}

Taking into account the Green function (\ref{Definitions}) one has the following expression for the self-energy and self-force
\begin{equation}\label{InThStringSF}
U^{\textrm{em}}(r) = \frac{q^2}{2r} L_0(\nu) ,\ \mathcal{F}_r^{\textrm{em}}(r) = \frac{q^2}{2r^2} L_0(\nu) , 
\end{equation}
where
\begin{equation}\label{L0}
L_0(\nu ) = \frac 1\pi \int_0^\infty \frac{\nu\coth   (\nu x) - \coth (x)}{\sinh  (x)} dx  . 
\end{equation}
Numerical evaluation of the function $L_0$ shows the approximately linear dependence of this quantity on $\nu$. For small angle deficit, $|\nu -1|\ll 1$, one has from Eq. (\ref{L0})
\begin{equation}
L_0(\nu ) \approx \frac\pi 8 (\nu -1) . \label{SmallDef}
\end{equation}
Therefore, the charged particle is repelled by the string in the transverse direction for an arbitrary value of $\nu$. The force for the small-angle deficit reads
\begin{equation}
\mathcal{F}^{\textrm{em}}_r \approx \frac{\pi q^2}{16 r^2}(\nu -1). \label{FemSmallAngle}
\end{equation}

In the framework of the Bopp--Podolsky theory \eqref{eq:BP}, the self-energy has obtained in Ref. \cite{Zayats:2016:siibpeswad}. The energy has the following form 
\begin{equation}\label{eq:BPString}
U^{\rm em}_{\rm BP} = \frac{\nu U_0}{\pi r m_{\rm P}} \int_0^\infty dx \frac{\coth \nu x}{\sinh x}(1- \cos (2r m_{\rm P} \sinh x)),
\end{equation}
where $U_0 = q^2/2r_0$ is electromagnetic energy of the effective particle (see Eq. \eqref{eq:U_0}). This expression was obtained without any renormalization procedure.  At the origin, it is finite which corresponds to  the finiteness of the Coulomb potential \eqref{eq:Coulomb} in the Bopp--Podolsky theory.  In the case of absence the angle deficit, $\nu =1$, the self-energy $U^{\rm em}_{\rm BP} = U_0$. In the limit $m_{\rm P} r\to \infty$, the self-energy tends to that given by Eq.  \eqref{InThStringSF} with additional contribution $U_0$:
\begin{equation}
U^{\rm em}_{\rm BP} = U^{\textrm{em}} + U_0.
\end{equation}

The self-force for arbitrary moving particles was considered in Refs. \cite{Khusnutdinov:1994:Sfpcst,Khusnutdinov:1995:Cpsscs,Khusnutdinov:1995:Sfcpstscs}. The following approach has been used which was already described in Sec \ref{Sec:Flat}. Let us consider the particle with charge $q$, moving on the trajectory $x^\mu (\tau')$ with $4$-velocity $u^\mu (\tau')$. First of all, we write out the expression for the electromagnetic $4$-potential $A^\mu$ in the point of observation, $x^\mu$, due to the particle (see Fig. \ref{Fig_Method_Flat}a). Then we put at this point a fictitious particle with charge $e$, velocity $u^\mu$ and calculate the Lorentz force, acting on it by the standard way: $\mathcal{F}^\mu = q F^{\mu\nu}u_\nu$ (see Fig. \ref{Fig_Method_Flat}b). Then we put the fictitious particle on the trajectory: $x^\mu \to x^\mu (\tau),\ u^\mu \to u^\mu (\tau)$. Therefore, the fictitious particle represents the real particle at a more later moment $\tau$. Then to find the self-force we turn the particles to each other, which is $\tau \to \tau'$ (see Fig. \ref{Fig_Method_Flat}c). It is well-known from the general theory of the Green functions in curved space-time (see, for example, \cite{DeWitt:1965:DTGF}) that the retarded Green function consists of two parts -- local and nonlocal (tail):
\begin{equation*}
G^{\textrm{ret}}(x;x') = \theta(x,x') \frac{\triangle^{1/2} \delta(\sigma)}{4\pi} + \frac{\triangle^{1/2} \theta(-\sigma)}{8\pi} \sum_{n=1}^\infty (-1)^{n-1} \frac{a_n(x;x')\sigma^{n-1}}{2^{n-1}(n-1)!}.
\end{equation*}
Here $\triangle$ is the DeWitt--Morette determinant, $a_n$ are the DeWitt--Schwinger coefficients (heat kernel coefficients), and the function $\theta(x,x')$ is zero if the event $x'$ is in the future of the event $x$ and it is unit in the opposite case. The local part is proportional to $\delta (\sigma)$, where $\sigma$ is the half-square the interval between points $x^\mu (\tau)$ and $x^\mu (\tau')$, gives (after appropriate classical mass renormalization) the local part of the self-force -- the DL force and local contribution from the energy (see (\ref{SFGeneral})). The nonlocal part of the Green function, proportional to the $\theta (-\sigma)$ gives a nonlocal part of the self-force.

It was shown in the particular case of a particle at rest that the increase in the parameter $L_0(\nu)$ with increasing the angle deficit $\nu$ is due to increasing the number of closed geodesics on the cone, the number of which $N$ is defined as an integer part of $\nu-1$:
\begin{equation}
N = \left[\nu -1\right]. \label{NGeo}
\end{equation}
For supermassive string ($\nu\to \infty, \mu \to 1/4$) the following estimation was obtained \cite{Khusnutdinov:1995:Cpsscs}
\begin{equation*}
L_0 \approx \frac\nu\pi \ln \frac{ 2\nu}\pi .
\end{equation*}
In this limit case the exact field equations give the metric of the cylindrical space \cite{Linet:1990:OsUgcs}. The supermassive cosmic strings were discussed in the context of topological inflation \cite{Vilenkin:1994:Ti,deLaix:1998:Timw,Ellis:1999:TRi,Izawa:1999:Rti,Sakai:1999:Tiibncmsf,Kawasaki:2000:Tis,Kawasaki:2002:Astim}.

The more general situation was considered by Linet in Ref. \cite{Linet:1986:otweitstoacs}. The self-force was calculated for the electric particle as well as for particles of massive scalar and vector fields. In these cases, the self-energy has the following form
\begin{equation*}
U^{\mathrm{em}}(r) = \varepsilon \frac{q^2}{2r} L(\nu, mr) ,
\end{equation*}
where
\begin{equation*}
L(\nu, mr) = \frac{\nu\sin (\pi\nu)}\pi \int_0^\infty \frac{e^{-2mr\cosh x}}{\cos (\pi\nu) - \cosh  (2\nu x)} \frac{dx}{\cosh  (x)},
\end{equation*}
$q$ is the charge of the scalar or massive (massless) vector particle, and $\varepsilon$ equals to $-1$ for the scalar field and $+1$ for the vector field.

Therefore, a scalar particle is attracted by the string but vector massive/massless particle is repelled by the string. The difference connects with the form of the interactive term in the Lagrangian of the scalar field \cite{Anderson:1967:PRP}. The self-energy, in this case, has the following form: 
\begin{equation*}
U^{\mathrm{em}}(x) = \varepsilon \frac q2 \Phi (x) .
\end{equation*}
For the electromagnetic field $(m=0,\ \varepsilon = +1)$ we obtain the formula (\ref{InThStringSF}).

The self-energy of massless fields is the inverse of the distance from the string (\ref{InThStringSF}), while in massive case \cite{Linet:1986:otweitstoacs} the dependence has the exponential form
\begin{equation*}
U^{\mathrm{em}}(x) \approx \varepsilon \frac{m\nu q^2}{4\sqrt{\pi}}\cot (\frac{\pi\nu}{2}) \frac{e^{-2mr}}{(mr)^{3/2}} ,\ mr \to \infty .
\end{equation*}

The gravitational self-force for uncharged probe particles with mass $m$ was calculated in Refs. \cite{Smith:1989:Geiscscqfsv,Galtsov:1990:Acsgsmgai} in linear over gravitational constant $G$ approximation. It was shown that the particle is attracted to the string with force
\begin{equation}
\mathcal{F}^{\textrm{gr}}_r(r) = -\frac{m^2}{2r^2}L_0(\nu) .\label{FgrSmallAngle}
\end{equation}
The difference with electromagnetic case (\ref{InThStringSF}) connects with the statement that the gravitational masses (charges) are always attractive while the like charges repel each other. 

Electromagnetic and gravitational self-forces for a particle at rest in the space-time of the cosmic desperation \cite{Galtsov:1993:Sscd,Tod:1994:Cst,Moraes:2000:Cmpalgc} the line element of which is given by Eq. (\ref{Desperation}) was investigated in Ref. \cite{DeLorenci:2002:Vpsc}. It was found the following expression for self-energy
\begin{equation*}
U^{\textrm{em}} = \frac{q^2}{2r} L(\nu,\kappa),\ U^{\textrm{gr}} = -\frac{m^2}{2r} L(\nu,\kappa),
\end{equation*}
where function $L(\nu,\kappa)$ is expressed as a coincidence limit of renormalized Green function: 
\begin{eqnarray*}
\fl L(\nu,\kappa) &=& -\frac{\ln 2}{\pi} - 2\sum_{n=1}^\infty \int_0^\infty \frac{x^2 - \pi^2 (4n^2/\nu^2 - 1)}{[x^2 + \pi^2 (2n/\nu + 1)^2][x^2 + \pi^2 (2n/\nu -
1)^2]}\\
\fl &\times& \frac{dx}{\sqrt{\cosh ^2 (x/2) + (\pi n \kappa/r)^2}}. 
\end{eqnarray*}

Remarkably, the quantity $L(\nu,\kappa)$ may be as positive as well as negative \cite{DeLorenci:2002:Vpsc} -- self-force may change its sign for different distances from the string. The natural scale for distance in this space-time is the declination $\kappa$. For infinitely small declination (or a great distance from the string), for $\kappa/r \ll 0$, the self-energy transforms as expected to the expression obtained before for the infinitely thin cosmic string (\ref{L0}): $L(\nu,\kappa) \to L_0(\nu)$. In the opposite situation great declination (or for small distances from the string), for $\kappa/r \to \infty$, we obtain 
\begin{equation*}
L(\nu,\kappa) \approx -\frac{\ln 2}\pi.
\end{equation*}
Therefore, for a great declination or small distances from the string, the self-energy has no dependence on the string's parameters. The situation is opposite for the case of the infinitely thin string namely, electromagnetic self-force attracts the particle but gravitational repels.

The self-force may be obtained by DeWitt and DeWitt \cite{DeWitt:1964:fc} approach developed for weak gravitational fields even the space-time of cosmic string is flat. In this case, the role of the parameter which characterizes the weakness of the gravitational field plays the small-angle deficit. The calculations of this kind we made in Ref. \cite{Boisseau:1996:Esswgfcs}. It was found the following expression (for the first degree of the gravitational constant $G$)
\begin{equation}
U^{\textrm{em}}(x) = \frac{\pi q^2}{4}\int \frac{2T^0_0 - T^1_1 - T^2_2 - T^3_3}{\rho (x,x')}d^2 x',\label{Uel}
\end{equation}
where $\rho (x,x')$ -- Euclidean distance in the plane perpendicular to the string, and $T^\mu_\nu$ -- energy-momentum tensor of the string. For infinitely thin string,
\begin{equation*}
T^0_0 = T^1_1 = \mu \delta^{(2)}(x'),\ T^2_2 = T^3_3 = 0,
\end{equation*}
and from (\ref{Uel}) one obtains the famous expression for the electromagnetic self-energy (see (\ref{FemSmallAngle}))
\begin{equation*}
U^{\textrm{em}} = \frac{\pi q^2\mu}{4 r} = \frac{\pi q^2 (\nu - 1)}{16 r}.
\end{equation*}

The electromagnetic self-force for space-time of the multiple cosmic strings was investigated in Refs. \cite{BezerradeMello:1998:Ssmcs,BezerradeMello:1998:Sciscss,Grats:2000:Elfscs,Grats:1996:Ti+gCf,Galtsov:1995:Vptscms,Galtsov:1993:PfgsCssc,Galtsov:1994:Crscs}. The results obtained valid for three-dimensional spaces with the following metric $(a,b,c = 1,2)$
\begin{equation*}
ds^2 = -dt^2 + \gamma_{ab}(x^c)dx^adx^b ,
\end{equation*}
or for four-dimensional spaces with line element
\begin{equation*}
ds^2 = -dt^2 + dz^2 + \gamma_{ab}(x^c)dx^adx^b .
\end{equation*}
Effective two dimensions of space part admitted by the following form of the currency density
\begin{equation*}
J^\mu(x^c) = (J^t(x^c),0,0,J^z(x^c)) .
\end{equation*}

Because two dimensional Riemann surfaces are conformally flat it is possible to introduce coordinates in which 
\begin{equation*}
\gamma_{ab} = e^{-\Omega (x^c)}\delta_{ab}.
\end{equation*}

In Ref. \cite{BezerradeMello:1998:Ssmcs} was calculated the linear energy density ${\cal U}$ and force density $\vec{{\cal F}}$ for stationer currency along the string direction via the point $(x^1_1,x^2_1)$ 
\begin{equation*}
J^\mu(x^c) = (J^t,0,0,J^z)\frac{\delta(x^1-x^1_1)\delta(x^2-x^2_1)}{\sqrt{\gamma (x)}} .
\end{equation*}
It was shown also that
\begin{eqnarray*}
{\cal U} &=& -\frac 12 \Omega (J^t{}^2 + J^z{}^2) , \\
\vec{{\cal F}} &=& \frac 12 e^{\frac\Omega 2}\vec{\nabla}\Omega (J^t{}^2-J^z{}^2) .
\end{eqnarray*}

Concerning the multiconical spaces of $N$ parallel strings in Refs. \cite{Staruszkiewicz:1963:GTTS,Letelier:1990:MCS,Deser:1984:TEGDFS}, with conformal factor
\begin{equation*}
\Omega = \sum_{k=1}^N2(1-\nu_k^{-1})\ln |\vec{x} - \vec{x}_k| ,
\end{equation*}
were obtained the following expressions for self-energy and self-force
\begin{subequations}
\begin{eqnarray}
{\cal U} &=& - (J^t{}^2 + J^z{}^2)\sum_{k=1}^N(1-\nu_k^{-1})\ln |\vec{x} - \vec{x}_k|  ,\label{Umultistrings} \\
\vec{{\cal F}} &=& e^{\frac\Omega 2}(J^t{}^2 - J^z{}^2)\sum_{k=1}^N(1-\nu_k^{-1}) \frac{\vec{n}_k}{|\vec{x} - \vec{x}_k|} , \label{Fmultistrings}
\end{eqnarray}
\end{subequations}
where
\begin{equation*}
\vec{n}_k = \frac{\vec{x} - \vec{x}_k}{|\vec{x} - \vec{x}_k|} .
\end{equation*}
Therefore, the self-energy (\ref{Umultistrings}) is an additive quantity while the self-force (\ref{Fmultistrings}) is not due to dependence on the conformal factor.

For the particular case of one string $(N=1)$, the self-force of currency parallel to the string was obtained in Ref. \cite{BezerradeMello:1995:Semlsscs}:
\begin{gather*}
\vec{{\cal F}} = (J^t{}^2 - J^z{}^2)(\nu - 1) \frac{\vec{\rho}}{\rho^2},
\end{gather*}
where $\rho = \nu|\vec{x}|^{1/\nu}$. Nominally this force coincides with the interaction force of two currents in Minkowski space-time: $J^\mu$ and the "induced"\/ current $J^\mu_{\textrm{ind}} = \frac 12 (\nu -1)J^\mu$. Sign of force depends on the sign of square of 4-current: space-like current are attracted and time-like currents are repelled.

If the non-trivial internal structure of currents takes place the induced momentum appears \cite{BezerradeMello:1998:Sciscss,Grats:2000:Elfscs} except induced currencies. The self-energy of dipole momentum coincides with the energy interaction of dipole $\vec{d}$ and induced dipole momentum
\begin{equation*}
\vec{d}_{\textrm{ind}} = -\frac 1{24}(\nu^2 - 1)\vec{d} .
\end{equation*}
For magnetic momentum $\vec{\mu}$ and quadrupole $D^{ab}$ momentum we have
\begin{equation*}
\vec{\mu}_{\textrm{ind}} = -\frac 1{24}(\nu^2 - 1)\vec{\mu} ,\ D^{ab}_{\textrm{ind}} = \frac 1{1440}(11 + \nu^2)(\nu^2 - 1) D^{ab} .
\end{equation*}
The internal structure leads to the appearance of the momentum of the self-force \cite{Grats:2000:Elfscs}.  


The self-force above is divergent with moving particles to the string. The origin of this is in the taken model of the string. It is expected that the non-trivial internal structure leads to an essential modification of the picture obtained. These calculations were made in Ref. \cite{Khusnutdinov:2001:Ssstcs} for a cosmic string with constant matter density $\mathcal{E}$ inside it \cite{Gott:1985:GlevsEs,Hiscock:1985:EGFS}. The line element of this string has the form given by Eq. \eqref{eq:Thick}. Because for zero angle deficit, the self-energy, and self-force vanish it is more suitable to extract the factor $\frac{q^2}{2r_\0} \frac{\nu^2-1} \nu$, that is the self-energy $U$ and level of the energetic barrier $U_{\textrm{max}}$ of the charged particle in the following form
\begin{subequations}
\begin{eqnarray}
U&=&\frac{q^2}{2r_\0}\frac{\nu^2-1}\nu{\cal U}(\nu,r/r_\0) ,\label{Ucal}\\
U_{\textrm{max}}&=&\frac{q^2}{2r_\0}\frac{\nu^2 - 1}\nu{\cal U}_{\textrm{max}}(\nu).
\label{Umax1}
\end{eqnarray}
\end{subequations}
The height of the energetic barrier is defined by the value of the self-energy in the center of the string, in the place where it is maximal. The function ${\cal U}(\nu,r/r_\0)$ for $r\to \infty$ tends to that for infinitely thin cosmic string and distance of two diameters of the string both quantities coincide. But it has the finite value ${\cal U}_{\textrm{max}}(\nu)$ in the center of the string. For $\nu =1$ the maximum ${\cal U}_{\textrm{max}}(1) \approx 0.39$ and it grows up with growing $\nu$ \cite{Khusnutdinov:2001:Ssstcs}. 

The self-force repels the particle out the string and attract inside one, and it has the only radial component. Close to the string's surface, for $r\approx r_\0$, the self-force is logarithmically divergent as follows
\begin{equation}
\mathcal{F}_r\approx -\frac{q^2}{2r_\0^2}\frac{\nu^2-1}8\ln \left|\frac{r}{r_\0}-1\right|, \label{Fdiv}
\end{equation}
but the work against the self-force is finite and equals to the barrier's height $U_{\textrm{max}}$. This divergence originates, with the taken model of the string: the metric is $C^1$-smooth, but the curvature has jumped on the surface of the string (inside the string the space-time is the constant curvature space-time, while outside the string the space-time is flat). The height of the barrier for GUT's string is $U_{\textrm{max}} = 2.8\cdot 10^5 \textit{GeV}$.

The self-energy for the space-time of the string with finite cross-section is important in the context of the string catalysis of the baryon's decay \cite{Brandenberger:1988:Ces,Brandenberger:1989:CSB,Alford:1989:AICSM,Alford:1989:Ebnvdtcs,Perkins:1991:Sfcs,Perkins:1991:TtsNv,Fewster:1993:Mdbdebcs}. The point is that the self-force prevents the penetration of the particle to the interior of the string, where the baryons decay may tale place. The energetic barrier appears. This question was considered quantitatively in Ref. \cite{Perkins:1991:TtsNv}. The detailed consideration of the electromagnetic self-force for a charged particle in the space-time of cosmic string with finite thickness is in Ref. \cite{Khusnutdinov:2001:Ssstcs}. It is expected that the situation quantitatively will not change for a string with another internal structure.

The space-time of infinitely thin cosmic string surrounded by a cylindrical shell was considered in the Refs. \cite{Celis:2016:eoactsomotesfoac,Tomasini:2019:acmsfictss}. The metric of the space-time both inside and outside the shell has the form \eqref{ds^2infhin} but with different parameters $\nu_{in}$ and $\nu_{out}$ and with relation $\nu_{in} r_{in} = \nu_{out} r_{out}$ on the shell. The self-energy crucially depends on the sign of Israel shell energy
\begin{equation}
\mathcal{E} = - \frac{\nu_{in} - \nu_{out}}{8\pi \nu_{out} r_{in}}.
\end{equation} 
The shell is made of an ordinary matter if $\nu_{out} > \nu_{in}$ and of an exotic matter if $\nu_{in} > \nu_{out}$. The self-energy is the sum
\begin{equation}
U^{\textrm{em}}(r) = \frac{q^2}{2r} L_0(\nu) + U^{\textrm{em}}_{\mathrm{sh}}(r),
\end{equation}
where the first term is the self-energy for infinitely thin string \eqref{InThStringSF} and the second term is shell contribution. Here $\nu = \nu_{in}$ inside the shell and $\nu = \nu_{out}$ outside the shell. 

The electromagnetic and gravitational self-potential for charged particles in the global monopole background with the line element given by Eq. (\ref{GlobMon}) was considered in Ref. \cite{BezerradeMello:1997:TNspbgm}. We remind that this space-time is curved in contrast to the cosmic string space-time. For calculations of the self-energy, the approach developed in Refs. \cite{Linet:1986:FCSCS,Linet:1986:otweitstoacs,Smith:1989:Geiscscqfsv} was adopted. Renormalization of the Green function was made by subtracting the Green function of the Minkowski space-time. The electromagnetic and gravitational potentials of self-interaction have the following form:
\begin{equation}
U^{\textrm{em}}(r) = \frac{q^2}{2r} S(\alpha) ,\ U^{\textrm{gr}}(r) = -\frac{m^2}{2r} S(\alpha) ,\label{MonopoleSF}
\end{equation}
where
\begin{equation*}
S(\alpha) = \sum_{l=0}^\infty \left\{\frac{2l+1}{\sqrt{\alpha^2 + 4l(l+1)}} - 1 \right\} .
\end{equation*}
For small solid angle deficit, $|1-\alpha|\ll 1$, this expression gives
\begin{equation*}
S(\alpha)\approx \frac\pi 8 (1-\alpha) .
\end{equation*}
Therefore like in the case of the infinitely thin cosmic string the electromagnetic self-force repels the particle and gravitational self-force attracts. 

In the case of the Bopp-Podolsky theory (see Eq. \eqref{eq:BP}) the self-energy may be obtained without any renormalization procedure  \cite{Zayats:2016:siibpeswad}. The self-energy reads
\begin{equation}\label{eq:BPGM}
U^{\textrm{em}}_{\rm BP} = \frac{U_0}{\alpha r m_{\rm P}} \sum_{l=0}^\infty \frac{2l+1}{s} \left\{ 1 -  \frac{s}{\alpha} I_{\frac{s}{2\alpha}}(r m_{\rm P}) K_{\frac{s}{2\alpha}}(r m_{\rm P})\right\},
\end{equation}
where $s=\sqrt{\alpha^2 + 4l(l+1)}$
It is finite at the origin due to finiteness the Coulomb potential \eqref{eq:Coulomb} in this theory. In the limit $rm_{\rm P} \to \infty$ the self-energy 
\begin{equation}
U^{\textrm{em}}_{\rm BP} = U^{\textrm{em}} + U_0,
\end{equation} 
where $U^{\textrm{em}}$ is given by Eq. \eqref{MonopoleSF}. 

The gravitationally induced self-force may give some interesting effects. Because in the space-time of the string it is not zero for a particle at rest and it has the form of the Coulomb interaction of charge $q$ with fictitious charge $q' = q L_0(\nu)/2$ (see (\ref{InThStringSF})), situated on the string, then it leads to scattering of the particle on the string the cross-section of which is proportional to Rutherford section ($\varepsilon$ -- the energy of impact particle):
\begin{equation}
d\sigma_{\textrm{self}} = \frac 14 L_0^2 (\nu) d\sigma_{R} = \frac 14 L_0^2 (\nu) \left(\frac{q^2}{2\varepsilon}\right)^2 \frac{\cos\frac{\theta}{2}}{\sin^3\frac{\theta}{2}} d\theta . \label{CrossSelf}
\end{equation}
The same effect takes place in the space of global monopole. It is enough to only change $L_0(\nu)$ for $S(\alpha)$ (\ref{MonopoleSF}). Alongside this differential cross-section, there is another kind of cross-sections that were considered in Refs.  \cite{Everett:1981:CSUGT,Alford:1989:AICSM,deSousaGerbert:1989:Cqssc}.

In Ref. \cite{Everett:1981:CSUGT} Everett has obtained the section of scattering of scalar particles over the unit length of the string with a finite cross radius $r_\0$, which is given by
\begin{equation}
\frac{d\sigma}{d\theta dl} = \frac{\pi\hbar}{2p_\perp} \frac 1{\ln^2 (p_\perp r_\0/\hbar)}.\label{Cross1}
\end{equation}
Here $p_\perp$ -- perpendicular with respect the string component of the moment of the particle. This section appears for direct interaction of the particles with a string of the non-zero transfer size, and it has no dependence on the connection constant. In the Ref. \cite{Everett:1981:CSUGT} the model interaction of the scalar particles with a scalar field inside the string was considered. The analogous expression was obtained in Ref. \cite{Perkins:1991:Sfcs} for the scattering of the charged fermions connecting with the magnetic field inside the string. Taking into account these results Vilenkin \cite{Vilenkin:1985:CSDW} found the expression for energy loss of the string moving with velocity $v$ through the matter
\begin{equation*}
\frac{d E}{dl dt} \approx \frac{\hbar n v^2}{\ln^2 (p_\perp r_\0/\hbar)},
\end{equation*}
where $n$ -- number particles density.

The scattering section obtained in the Refs. \cite{Alford:1989:AICSM,deSousaGerbert:1989:Cqssc} completely identify to the Aharonov--Bohm cross section \cite{Aharonov:1959:Sepqt}
\begin{equation}
\frac{d\sigma}{d\theta dl} = \frac{\hbar}{2\pi p_\perp} \frac{\sin^2( \pi\alpha)}{\sin^2 \frac\theta 2}\label{Cross2}
\end{equation}
and it gives the following energy loss: 
\begin{equation*}
\frac{d E}{dl dt} = \frac{2\hbar n v^2}{\sqrt{1 - \frac{v^2}{c^2}}},
\end{equation*}
where $\alpha = e\Phi/2\pi$ -- flux of the field in the units of elemental flux.

The origin of this interaction is considered as follows. As shown in Ref. \cite{Alford:1989:AICSM}, the strings appeared in field models due to spontaneous symmetry breaking have inside them the magnetic field. Outside the string interior, the field transforms into a pure gauge without a magnetic field. Therefore, the configuration of fields is identical to the solenoid \cite{Aharonov:1959:Sepqt} which gives us the section considered above.

Both sections (\ref{Cross1}) and (\ref{Cross2}) have no connection with the conical structure of the space-time, but they are consequence either a specific distribution of fields as source the string or interaction of particles with fields inside the string. The noted sections have calculated with the assumption that the space-time is the Minkowski space-time. The origin of the section (\ref{CrossSelf}) is directly connected with the conical structure of the space-time.

The acceleration due to self-force leads to radiation of the electromagnetic waves. Indeed, the self-energy of a particle in the global monopole background has the form of the Coulomb interaction of a particle with charge $q_1=q$ with some imaginary particle with charge $q_2=q S(\alpha)/2$ and situated in the center of monopole (\ref{MonopoleSF}). For infinitely thin string (\ref{InThStringSF}) the fictitious charge has the value $e_2 = e L_0(\nu)/2$. Interaction of Coulomb kind leads to bremsstrahlung \cite{Landau:1975:CTF}. In an ultrarelativistic case \cite{Landau:1975:CTF} the following expression takes place for the energy emitted for all time of moving
\begin{equation*}
E = \frac{\pi q_1^4 q_2^2\gamma^2}{4m^2\rho^3},
\end{equation*}
where $\gamma$ and $\rho$ -- relativistic factor and impact distance, correspondingly. 

At the same time, there is another way for the radiation of the electromagnetic waves in the space-time of the topological defects, and which is not connected with acceleration. It was shown in Refs. \cite{Aliev:1989:Garsgcs,Serebryanyi:1989:Pegfcs,Audretsch:1991:QftpncsTpl,Audretsch:1991:Cbcss,Audretsch:1992:Pcdmpnfacs,Skarzhinsky:1994:Qegfcs,Audretsch:1994:Bgfcs} that the particles in the space-time of a cosmic string radiate even for a particle on the geodesic line. This process is forbidden in the Minkowski space-time, but it is possible in the space-time of the cosmic string. These phenomena were investigated in detail in the above Refs. for infinitely thin cosmic string space-time. The analogous phenomenon for a charged particle in the space-time of another topological defect -- pointlike global monopole with line element (\ref{GlobMon}) was considered in Ref. \cite{Bezerra:2002:Bgfgm}. For ultrarelativistic particles, the energy radiated due to self-force is much smaller than the radiation due to its movement on the geodesic line \cite{Bezerra:2002:Bgfgm}.

The gravitationally induced self-force may give some quantum phenomena. Because the particle has additional energy even for the rest we have to take into account this energy on the state of the particle. Additional energy may influence no merely to the scattering states but maybe the origin for boundary states. This question for non-relativistic spinless particles was considered in Ref. \cite{Gibbons:1990:TNCPC} and for relativistic particles with spin $0$ and $1/2$ in Ref. \cite{Bordag:1996:Rbscs}.

Because the electromagnetic (\ref{FemSmallAngle}) and gravitational (\ref{FgrSmallAngle}) self-forces have the same structure, it is possible to consider their joint influence for quantum particle. The presence or absence of the boundary states is defined by the value of the parameter $Z = q^2 - m^2$. For positive values of $Z$ (attraction is smaller then repulsive force), the boundary states are absent. In opposite case for negative $Z$ (attraction is greater then repulsive force) as, for example, for massive uncharged particles, the boundary states exist. For a particle with spin $1/2$ the spectrum which is defined by localized solutions of the Dirac equations ($U=Z/r$)
\begin{equation*}
\left[-\ib \gamma^\mu (x) \widetilde{\nabla}_\mu + m + U\right]\Psi = 0,
\end{equation*}
have the following form:
\begin{equation}
E_{N,M} =\pm \sqrt{m^2 + p_3^2} \left[1 - \frac{m^2}{m^2 + p_3^2 } \frac{Z^2}{(N + \sqrt{\nu^2 M^2 + Z^2})^2}\right]^{1/2},
\label{SpectrSpinor}
\end{equation}
where $N=0,1,\ldots$ (main quantum number) and $M = n + 1/2$ (full orbital moment, half-integer), $n=0,\pm 1,\ldots$. The quantity $p_3$ is the longitudinal (along with the string) component of the momentum. For spinless relativistic scalar particles, the wave function obeys the Klein--Gordon equation
\begin{equation*}
\left\{\Box - (m + U)^2\right\} \Psi = 0,
\end{equation*}
the spectrum has a similar structure, we have to replace only $N\to N + 1/2$ and $M\to n,\ n=0,\pm 1,\ldots$ (the full orbital momentum, integer): 
\begin{equation}
E_{N,n} =\pm \sqrt{m^2 + p_3^2} \left[1 - \frac{m^2}{m^2 + p_3^2 } \frac{Z^2}{(N + \frac 12 + \sqrt{\nu^2 n^2 + Z^2})^2}\right]^{1/2} \label{SpectrScalar}.
\end{equation}
For the non-relativistic limit, we obtain the following value of the energy of the boundary states (without rest energy):
\begin{equation*}
E_{N,n} = -\frac{m Z^2}{2(N + \frac 12 + \nu |n|)^2},
\end{equation*}
in agreement with \cite{Gibbons:1990:TNCPC}.

With $p_3 = 0,\ \nu =1$ and $Z = - m m_0$ the spectrum obtained coincides with spectrum of the boundary states of electron of mass $m$ in the Newtonian gravitational field of mass $m_0$ \cite{Soff:1973:SDespiiap,Greiner:1985:QESF}. It has to be noted the radical difference in the spectrum's origin (\ref{SpectrSpinor}) and obtained in \cite{Soff:1973:SDespiiap,Greiner:1985:QESF}. For the case the flat space-time we have the only attraction of the electron with mass $m_0$, which leads to boundary states considered in Ref. \cite{Soff:1973:SDespiiap,Greiner:1985:QESF}. For this reason the quantity $Z$ always negative and equals $-m m_0$. In space-time of cosmic string as it is well-known \cite{Vilenkin:1985:CSDW} the Newtonian interaction is absent. The origin of spectra (\ref{SpectrScalar}) and (\ref{SpectrSpinor}) are connected with specific interaction particles with string -- self-interaction. The mass, as well as charge, gives the interaction of this kind. The boundary states are absent if the electromagnetic self-force (repulsion) prevails over gravitational self-force (attraction). In the opposite case, the boundary states appear because the total self-force becomes attractive. 

Some numerical estimations. The energy scale of the spectrum obtained which characterizes the distance between levels is the following quantity:
\begin{equation*}
E_{\textrm{scale}} = Z^2 mc^2.
\end{equation*}
For neutral particles we have
\begin{equation*}
E_{\textrm{scale}}= L_0^2 \left(\frac{Gm^2}{\hbar c}\right)^2 mc^2 = L_0^2 \left(\frac{m}{m_{\textrm{Pl}}}\right)^4 mc^2 = 0.2 \cdot 10^{-96} \left(\frac{L_0}{L_0^{\textrm{GUT}}}\right)^2 \left(\frac{m}{m_{\textrm{el}}}\right)^5 eV.
\end{equation*}
Here $L_0^{\textrm{GUT}}\approx \frac\pi 8 10^{-6}$ -- the characteristic value of parameter $L_0$ for GUT, $m_{\textrm{Pl}} = \sqrt{\hbar c/G}$ the Planckian mass, and $m_{\textrm{el}}$ -- the electron's mass.

Therefore, for an elementary particle the spectrum obtained is practically uniform. We may expect the essential effect for very massive particles. For example, for a particle with mass $m = 60 m_{\textrm{Pl}} \approx 10^{-3} \textit{gr}$ the energy scale $E_{\textrm{scale}}$ equals $1eV$. 
\section{The self-force in the space-time of wormholes} \label{Sec:WHV}

Wormholes are topological handles in space-time linking widely separated regions of a single universe, or "bridges"\/ joining two different space-times. Interest in these configurations dates back at least as far as 1916 \cite{Flamm:1916:BzEG} with revivals of activity following both the classic work of Einstein and Rosen in 1935 \cite{Einstein:1935:PPGTR} and the later series of works initiated by Wheeler in 1955 \cite{Wheeler:1955:G} (see also books \cite{Wheeler:1960:Ngg,Wheeler:1962:G}). More recently, interest in the topic has been rekindled by the works of Morris and Thorne \cite{Morris:1988:wisatufitatftgr} and Morris, Thorne, and Yurtsever \cite{Morris:1988:wtmatwec}. These authors constructed and investigated a class of objects they referred to as "traversable wormholes." Their works led to a flurry of activity in wormhole physics \cite{Visser:1995:LWfEtH,Lobo:2007:esigrtwawds}. From the point of view of self-force, the wormhole space-time is an interesting example of space-time with nontrivial topology with bridges. It is interesting to reveal the role of the nontrivial topology for self-force.

The definition of a static space-time of a traversable wormhole is based on the existence of a $2$-dimensional hypersurface of the minimal area taken in one of the constant-time spatial slices \cite{Hochberg:1997:gsotgstwt}. This minimal hypersurface is called the throat of a wormhole. From this point of view may be considered the wormholes with trivial $\mathds{R}^2$ topology or with $\mathds{S}^2$ topology, or with the form of a handle between two different universes or as a bridge in a single universe. Different kind of metrics of wormhole's space-time may be found in Visser's book \cite{Visser:1995:LWfEtH}. 

\subsection{Bronnikov--Ellis wormhole}

Here, we consider the spherically symmetric wormhole with line element close to Schwarzschild's form of metric
\begin{equation}
ds^2 = -A(\rho)dt^2 + B(\rho) \left(d\rho^2 + r^2(\rho) d\Omega_2\right),
\end{equation}
where $\rho \in (-\infty,+\infty)$ and function $r(\rho)$ describes a profile of the wormhole's throat. The throat, $a$, is defined by relations
\begin{equation}
a = r(0), r'(0) = 0,\ r''(0) > 0.
\end{equation}

The line element of this kind was found in Refs. \cite{Bronnikov:1973:Stasc,Ellis:1973:eftadapmigr} as the solution of Einstein equations with the massless scalar field with opposite sign of kinetic term. The metric reads 
\numparts\label{eq:metricBE}
\begin{equation}
ds^2 = -e^{-\alpha(\rho)} dt^2 + e^{\alpha(\rho)}d\rho^2 + r^2(\rho) d\Omega_2,
\end{equation}
where
\begin{eqnarray}
r^2(\rho) &=& (\rho^2 + n^2 -m^2)e^{\alpha(\rho)},\\
\alpha(\rho) &=& \frac{2m}{\sqrt{n^2-m^2}} \left\{ \frac{\pi}{2} - \arctan \left(\frac{\rho}{\sqrt{n^2-m^2}}\right) \right\}.
\end{eqnarray}
\endnumparts
The square of the sphere of radial coordinate $\rho$, $S=4\pi r^2(\rho)$, is minimized for $\rho = m$. At infinity, the spacetime becomes Minkowskian:
\begin{eqnarray*}
\lim_{\rho\to + \infty}\frac{r^2}{\rho^2} &=& 1,\ \lim_{\rho\to + \infty}\alpha  = 0\\
\lim_{\rho\to - \infty}\frac{r^2}{\rho^2}&=& e^{\alpha_{max}},\ \lim_{\rho\to - \infty}\alpha= \alpha_{max} = \frac{2\pi m}{\sqrt{n^2-m^2}}.
\end{eqnarray*}

The Ricci tensor has a single component
\begin{equation}
\mathcal{R}^{\rho}_{\rho} = \mathcal{R} = - \frac{2n^{2}}{(\rho^2 + n^2 -m^2)^2}e^{-\alpha}.
\end{equation}

There are two parameters, $n$ and $m$, in the metric (\ref{eq:metricBE}) with relation $m^2 < n^2$. We consider the case $m>0$ without the loss of generality. To reveal the role of the parameter $m$ let us consider the gravitational acceleration, $a$, acting for a particle. In the spherically  symmetric case, it has the well-known form \cite{Frolov:1998:bhpbcand,Doroshkevich:2008:tportaw}
\begin{equation}
a=\sqrt{\gs_{\rho\rho}} \, a^{\rho} = -\frac{1}{\sqrt{\gs_{\rho\rho}}}\frac{d\ln{\sqrt{-\gs_{tt}}}}{d\rho}.
\end{equation}
For the metric under consideration (\ref{eq:metricBE}) we obtain 
\begin{equation}
a = -\frac{m}{\rho^2 + n^2 -m^2}e^{-\frac{\alpha}{2}}.
\end{equation}
In the limit $\rho\to +\infty$ we have the Newtonian gravitational attraction, $a\approx-{m}/{\rho^2}$, to the wormhole throat with effective mass $m$. For $\rho\to -\infty$ we obtain Newtonian gravitational repulsion, $a\approx-{m^*}/{\rho^2}$, with negative effective mass $m^* = -me^{-\alpha_{max}/2}$, where $\alpha_{max} = \alpha_{\rho\to -\infty} = 2\pi m/\sqrt{n^2-m^2} $. Therefore the parameter $m$ characterizes the effective mass of the wormhole. 

Let us define new radial coordinate (proper distance is $|L|$)
\begin{equation}
L = \int^{\rho}_m e^{\frac{\alpha(x)}{2}} dx,\ L_{\rho\to +\infty} =\rho,\ L_{\rho\to -\infty} = e^{\frac{\alpha_{max}}{2}} \rho. \label{eq:l}
\end{equation}
With this definition the metric reads
\begin{equation}\label{eq:metricBE1}
ds^2 = -e^{-\alpha(\rho)} dt^2 + dL^2 + r^2(\rho) d\Omega_2,
\end{equation}
and $r^2_{L\to \pm \infty} = L^{2}$. In this coordinate the acceleration has the following behavior at infinity
\begin{equation}
a_{L\to\pm \infty} = - \frac{m}{L^2},
\end{equation}
and therefore $m^* = -m$.

The parameter $n$ of the wormhole may be understood in terms of a charge of some massless scalar field going through the wormhole (for electric charge, see Ref. \cite{Shatskii:2008:admotwatmm}). Indeed, the equation for the scalar massless field in the background with line element (\ref{eq:metricBE}) admits radial non-singular solution for the scalar field. The derivative of this field with respect of the radial coordinate reads $\phi'(\rho) = -q/(\rho^2 + n^2 -m^2)$ with some parameter $q$ which corresponds to the scalar charge from the point of view of a distant observer. The stress-energy tensor of this field has the following form $8\pi T^\nu_\mu = q^2 e^{-\alpha}/(\rho^2 + n^2 -m^2)^2 \textrm{diag}(-1,1,-1,-1)$. On the other hand, the metric (\ref{eq:metricBE}) is a non-singular solution of the Einstein equations with scalar massless field with opposite sign of the stress-energy tensor which reads $8\pi T^\nu_\mu = -n^2 e^{-\alpha}/(\rho^2 + n^2 -m^2)^2 \textrm{diag}(-1,1,-1,-1)$. Therefore, the parameter $n$ plays the role of the scalar charge of the phantom scalar field which passes through the wormhole space-time.

For $m=0$ we obtain from metric (\ref{eq:metricBE}) the wormhole which was called as massless "drainhole" \cite{Ellis:1973:eftadapmigr} with throat radius $n$. The metric has the following form
\begin{equation} 
ds^2 = -dt^2 + d\rho^2 + r(\rho)^2 d\Omega_2,\label{eq:metricD}
\end{equation}
where $t,\rho\in \mathds{R}$, $\theta\in [0,\pi)$ and $\varphi\in [0,2\pi)$. The function $r(\rho)$ describes a profile of the throat. The first universe belongs to domain of positive $\rho>0$ and negative $\rho<0$ is an another universe. The radius of throat $a$ is defined as minimal value of function $r$, that is the following relations are fulfilled
\begin{equation}
r(0) = a, r'(0) =0, r''(0) >0.
\end{equation}
We suppose that far from the wormhole's throat the space-time becomes Minkowskian, which is
\begin{equation}
\lim_{\rho\to\pm\infty} \frac{r^2(\rho)}{\rho^2} = 1.\label{limitrho}
\end{equation}

The non-zero components of the Ricci tensor and scalar curvature have the following form
\begin{equation}\label{eq:R}
	\mathcal{R}^\rho_\rho = -\frac{2r''}{r},\ 	\mathcal{R}^\theta_\theta = \mathcal{R}^\varphi_\varphi = -\frac{-1+r'^2 + r r''}{r^2},\ \mathcal{R} = -\frac{2(-1+r'^2 + 2r r'')}{r^2}.
\end{equation}
The energy-momentum tensor corresponding to this metric has a diagonal form from which we observe that the source of this metric possesses the following energy density and pressure:
\begin{equation*}
	\varepsilon = -\frac{-1 + r'^2 + 2rr''}{8\pi r^2},\ p_\rho =\frac{-1 + r'^2}{8\pi r^2},\ p_\theta = p_\varphi = \frac{r''}{8\pi r}.
\end{equation*}

Three dimensional section, $t=const$, of this spacetime is conformally flat. Indeed, let us in $3D$ flat space in spherical coordinates with line element  
\begin{equation*}
dl_{\textrm{fl}}^2 = dR^2 + R^2 d\Omega_2,
\end{equation*}
change radial coordinate $R\to \rho$: $R = r(\rho) e^{\sigma(\rho)}$, where
\begin{equation*}
\sigma = \int^\rho \frac{d\rho}{r(\rho)} - \ln r(\rho). 
\end{equation*} 
Then we obtain claimed result
\begin{equation}
dl_{\textrm{fl}}^2 = e^{2\sigma} \left(d\rho^2 + r^2(\rho) d\Omega_2\right).\label{eq:conf_flat}
\end{equation}

Let us consider specific kinds of throat's profile. The simplest model of a wormhole is that with an infinitely short throat \cite{Khusnutdinov:2002:Gsews},
\begin{equation}\label{eq:infshort}
r = a + |\rho|. 
\end{equation}
The space-time is flat everywhere except for the throat, $\rho =0$, where the curvature has a delta-like form,
\begin{equation}\label{eq:infshort_curvature}
\mathcal{R }= -8\frac{\delta (\rho)}{a}. 
\end{equation}
We may define new coordinate $\rho + a$ in the first universe and $-\rho + a$ in the second one and the metric will get the Minkowskian form with radial coordinate $r \geq a$. In fact, the cross section $t=const, \theta = \pi/2$ is $2$-dimensional and two side plane with hole of radius $a$.

\begin{figure}[ht]
	\centering 
		\hspace{-2em}\includegraphics[width=4truecm]{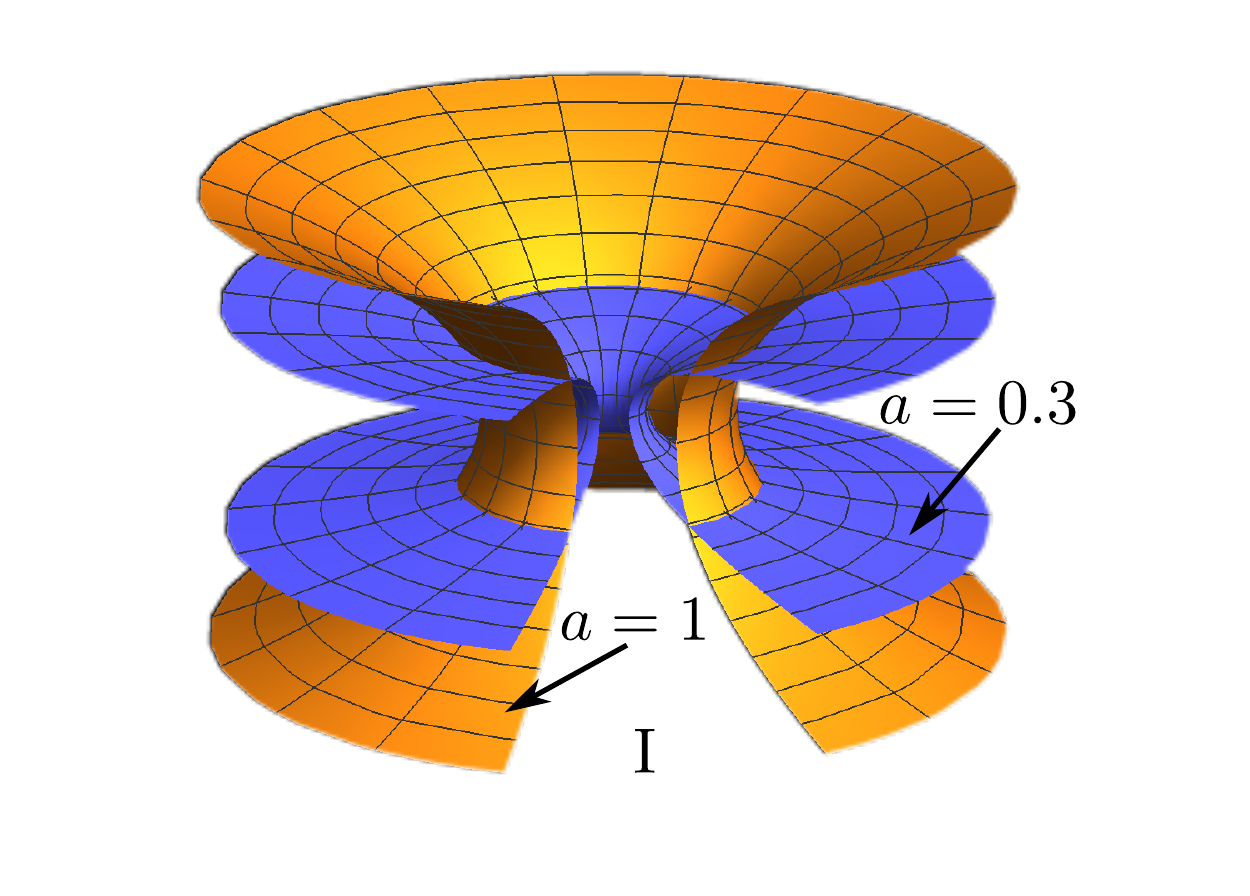}%
		\hspace{-2em}\includegraphics[width=4truecm]{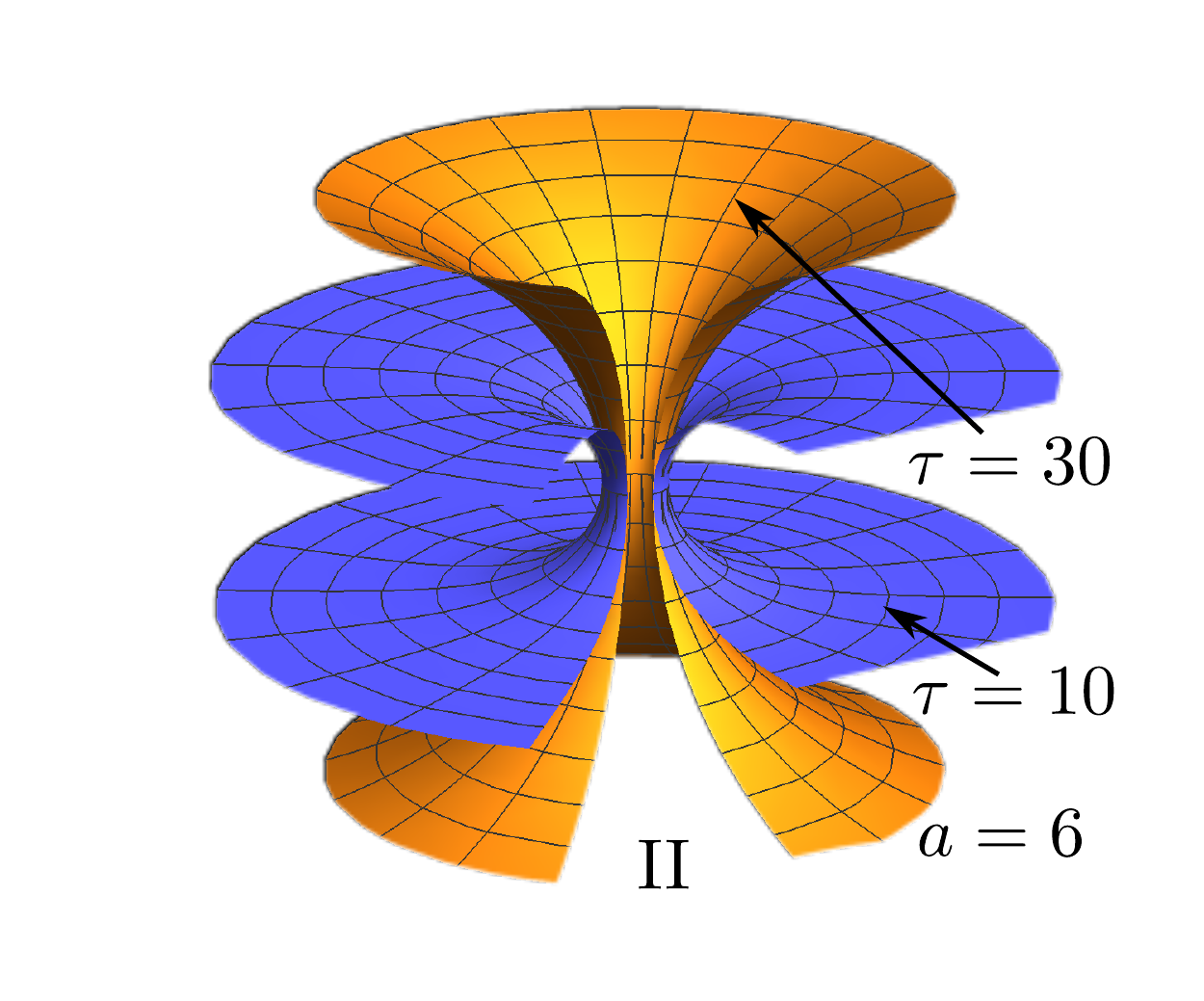}%
		\hspace{-2em}\includegraphics[width=4truecm]{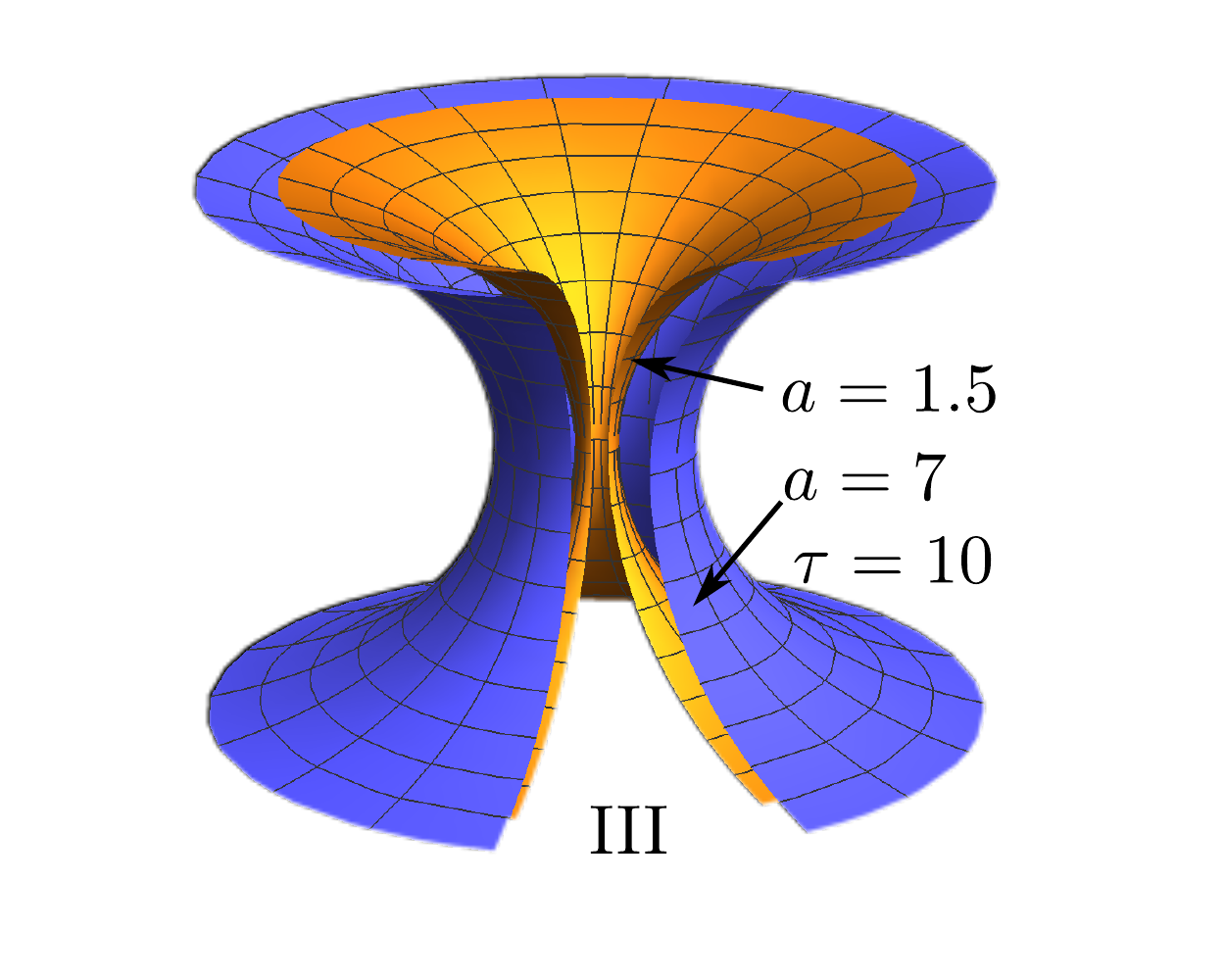}%
		\hspace{-3em}\includegraphics[width=4truecm]{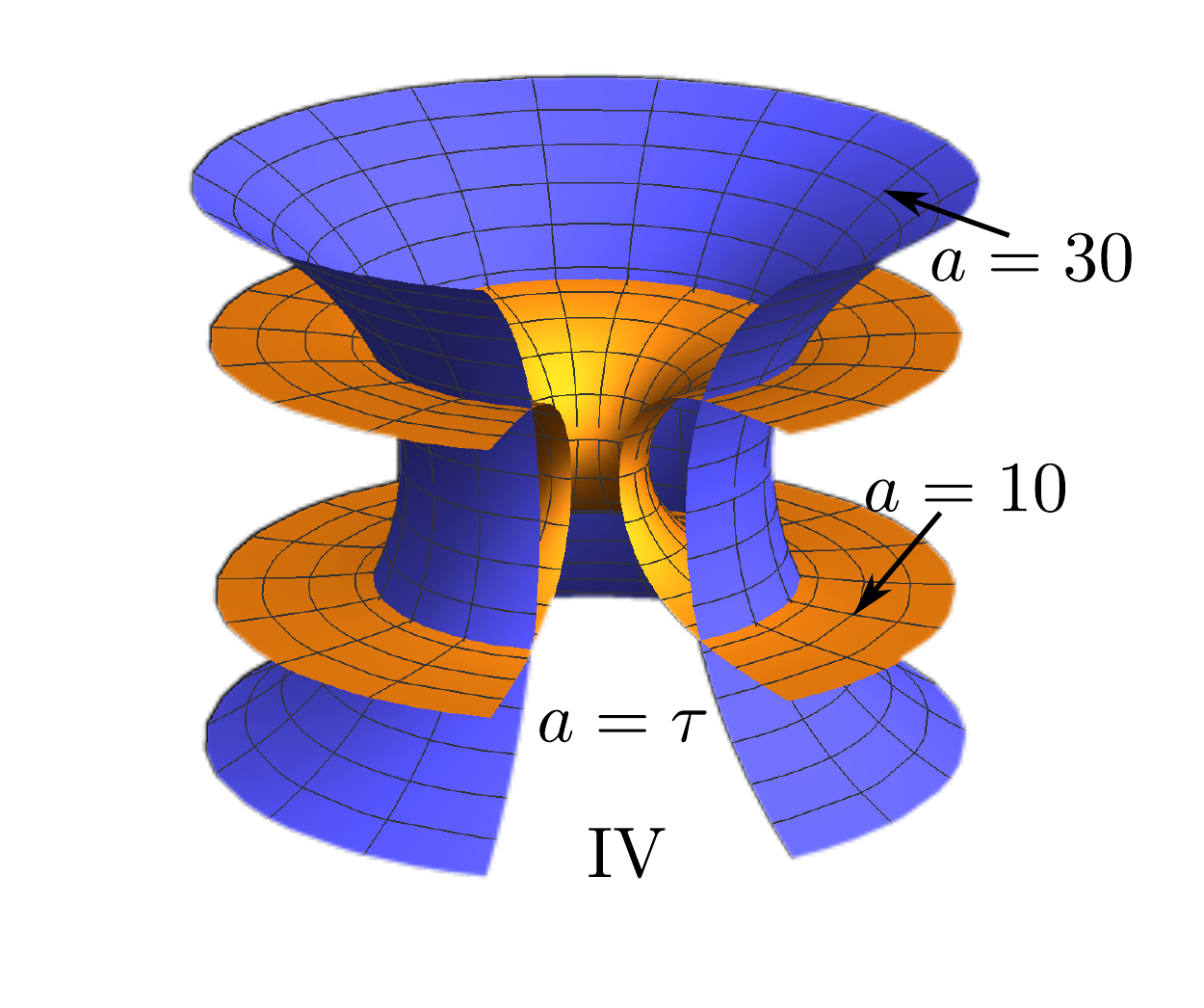}
		\caption{First figure (I) represents the section $t=const,\theta = \pi/2$ of wormhole's space-time with profile function $r(\rho)=\sqrt{\rho^2 + a^2}$ for two different values of the radius of the throat. Three next figures illustrate the wormhole with profile of throat $r(\rho) = \rho \coth(\rho/\tau) - \tau + a$. Figure (II) and (III) illustrate that the $a$ and $\tau$ are radius and length of the throat, accordingly. In last figure two wormholes with different $a$ but with the same ratio radius and length of the throat are depicted. 
	} \label{fig:worms}
\end{figure}

The next and more realistic model is the profile of the throat which has the following form:
\begin{equation}
r(\rho) = \sqrt{\rho^2 + a^2},\label{eq:profileDH}
\end{equation}
where $a$ is the radius of the throat which characterizes wormhole's size. The embedding into the three-dimensional Euclidean space of the section of the space-time by surface $t=const, \theta = \pi/2$ is plotted in Fig.\ref{fig:worms}(I) for two different values of the radius of the throat. In Euclidean space with cylindrical coordinates $(r,\varphi,z)$ this surface may be found in parametric form from relations: $r=r(\rho),z'(\rho) = \sqrt{1-r'^2}$. 

The section $t=const, \theta = \pi/2$ with metric 
\begin{equation}
	ds^2 = d\rho^2 + r^2(\rho) d\varphi^2, \label{eq:cross}
\end{equation}
is conformal to cylinder with a radius $b$. Indeed, by changing coordinate $\rho = f(z)$, where function $f(z)$ obeys to equation
\begin{equation*}
f'^2(z) b^2 = r^2(f(z)),
\end{equation*}
we obtain the metric (\ref{eq:cross}) in the following form 
\begin{equation}
ds^2 = f'^2(z) (dz^2 + b^2 d\varphi^2).
\end{equation}
For particular case of drainhole (\ref{eq:profileDH}) we have $\rho = a \sinh \frac{z}{q}$ and 
\begin{equation}
ds^2 =\frac{a^2}{b^2} \cosh^2 \frac{z}{q}  (dz^2 + b^2 d\varphi^2). \label{eq:crossDHcylinder}
\end{equation}

The spacetime  \eqref{eq:metricD} maybe written  \cite{Linet:2007:Ewg,Boisseau:2013:eiaswr} in isotropic form by introducing new radial coordinate $l>0$: $2l = \rho + r(\rho),\ \rho = l - a^{2}/{4l}$. In these coordinates the metric has the following form
\begin{equation}
ds^2 = - dt^2 + \left(1 + \frac{a^{2}}{4l^2}\right)\left(dl^2 + l^2 d\Omega_2\right).  
\end{equation}

Another model has been considered in Ref. \cite{Sushkov:2001:Dwws} and it is characterized by the following profile of the throat
\begin{equation}\label{Second}
r(\rho) = \rho \coth \left(\frac\rho\tau\right) - \tau + a.
\end{equation}
This model possesses a more reach structure. There are two parameters $\tau$ and $a$. The latter parameter is the radius of the throat. In this model, we may introduce another parameter which may be called the length of the throat. The point is that the function $r(\rho)$ turns into linear function of $\rho$ starting from distance $\rho > \tau/2$ and the space-time becomes approximately Minkowskian. Therefore, the length of throat $l=\tau$. Using new variables $y=\rho /a,\ \alpha = \tau/a$ one rewrites the function $r$ in the form below
\begin{equation*}
r(y) =a\left(y\coth \left(\frac y\alpha \right) - \alpha  + 1\right).
\end{equation*}
The parameter $\alpha $ is the ratio of the length and radius of the throat. It allows us to consider wormholes of different form, that is with different ratio of the radius and length of the throat. The wormhole with profile \eqref{Second} was used in Refs. \cite{Popov:2010:sfoaspcitlt,Popov:2013:sfoascitltoaw,Popov:2015:ssfoscialt} to calculate self-force in the long length of throat approximation, $a/\tau \ll 1$.

In Fig.\ref{fig:worms}(II-IV) the sections $t=const, \theta =\pi/2$ of this wormhole space-time are shown for different values of $a$ and $\tau$. Namely in Fig.\ref{fig:worms}(II) we represent two wormholes with the same radius of the throat but with different lengths and vice verse in Fig.\ref{fig:worms}(II) we depict two wormholes with the same length of the throat but with different radii of the throat. In the last picture, Fig.\ref{fig:worms}(IV) two wormholes with the same ratio of length and radius of the throat, but with different values of throats' radii are depicted. Therefore, the size of the wormhole with the same ratio of length and radius throat is managed by parameter $a$. The parameter $\alpha $ describes wormhole's form. 

In this section we consider massless and massive wormholes with metrics \eqref{eq:metricD} and \eqref{eq:metricBE} for Maxwell field.  

\subsubsection{The Gauss theorem}

\begin{figure}[ht]
\centerline{\includegraphics[width=6truecm]{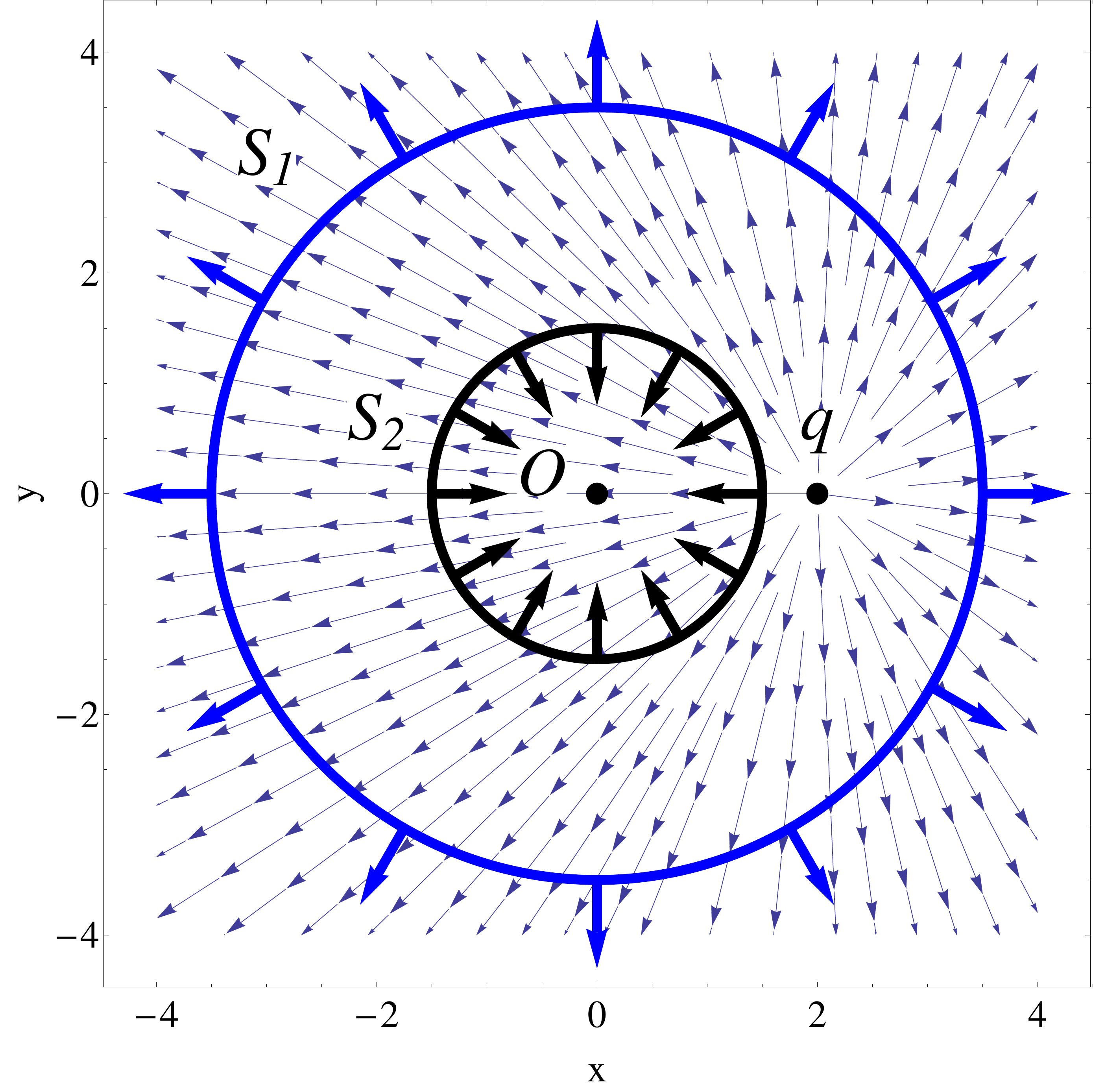}}
\caption{The charge $q=1$ is at rest at point $\vr' = (2,0,0)$. The sphere $S_2$ maybe shrank to point $O$ and flux of electric field $\vE = (\vr - \vr')/|\vr - \vr'|^3$ through this sphere is zero.  }\label{fig:flat}
\end{figure}

Let us discuss the Gauss theorem for spacetime with throat, where the homotopy group $\pi_2$ is not trivial. Let us start with flat space and consider charge $q$ at the point $\vr'$. The electric field $\vE = q(\vr - \vr')/|\vr - \vr'|^3$ obeys to Maxwell equation $\divi \vE = 4\pi q\delta^{(3)}(\vr,\vr')$. We surround origin of coordinate $O$ and charge by sphere $S_1$ and radius $R_1$ (see Fig. \ref{fig:flat}) and integrate this equation over the ball with radius $R_1$.  
Integrating the divergent by part (Stokes theorem) we obtain, in general, two surface integrals
\begin{equation}
\Pi = \Pi_1 + \Pi_2 = \int_{S_1} (\vE\cdot \vn_1) dS + \int_{S_2}(\vE\cdot \vn_2) dS = 4\pi q, 
\end{equation}
where $\vn_1$ is outward normal for $S_1$ and $\vn_2$ is inward normal for $S_2$. The sphere $S_2$ in the second integral may be shrunk to point $O$ and therefore flux through this sphere is zero and we arrive with famous Gauss theorem 
\begin{equation}
\Pi =  \int_{S_1} (\vE\cdot \vn_1) dS  = 4\pi q. 
\end{equation}  
	
\begin{figure}[ht]
	\begin{center}
		\includegraphics[width=5truecm]{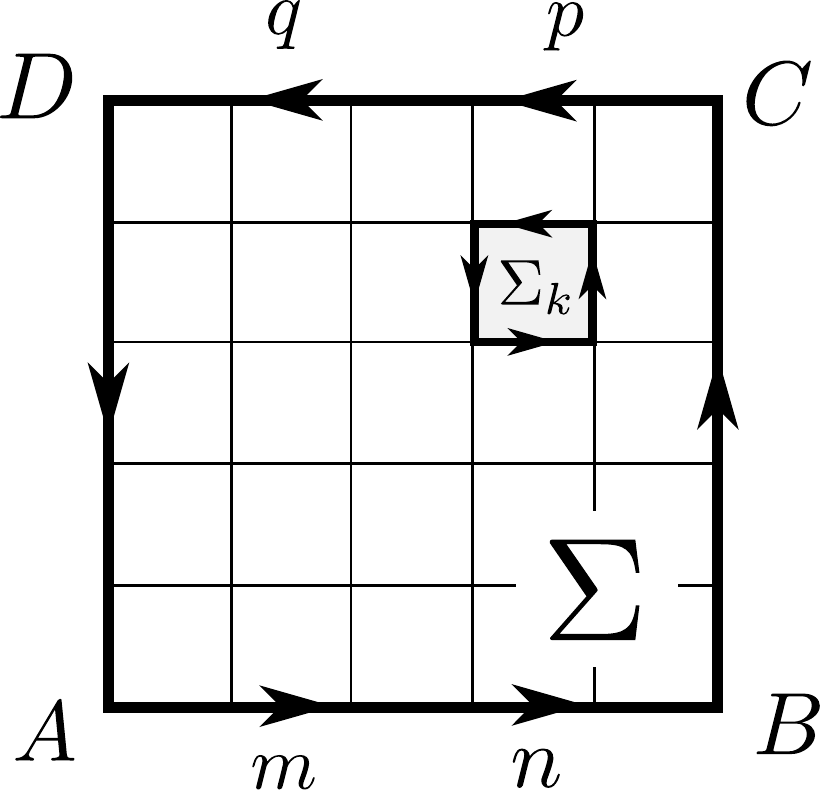}\hspace{4em}\includegraphics[width=5truecm]{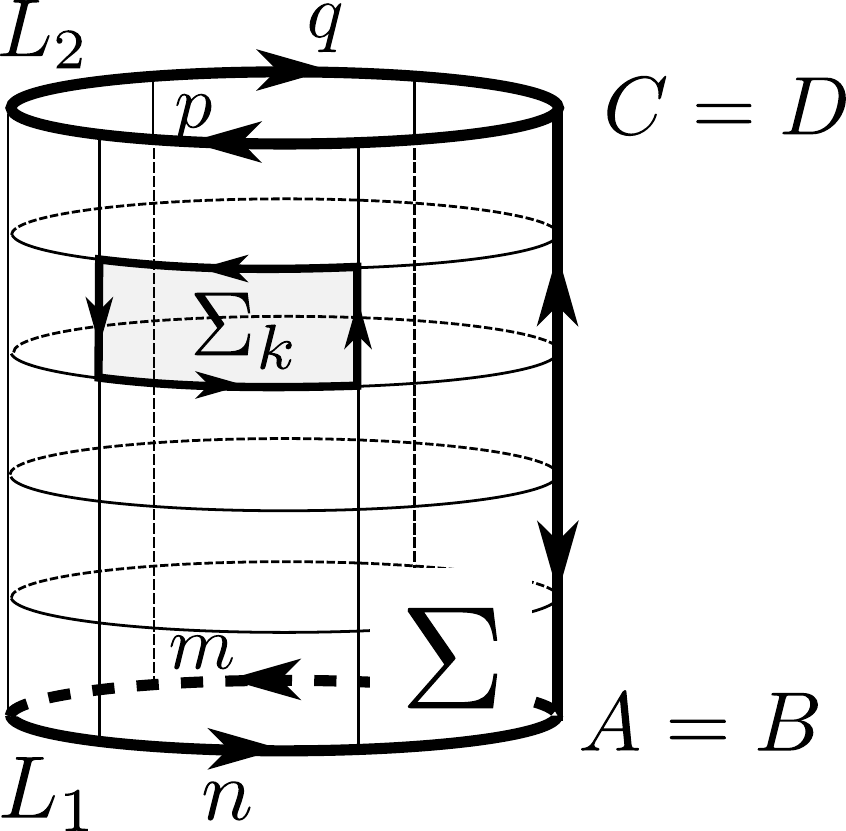}
	\end{center} \caption{Left panel: Sum over all boundaries of sub-areas $\Sigma_k$ gives the integral over boundary $\partial\Sigma =ABCD$ of the total rectangle. Right panel: Integral over cylinder has two disconnected boundaries -- circles $\partial\Sigma_1 = L_1 = AmnA$ and $\partial\Sigma_2 = L_2 = CpqC$. } \label{fig:stokes}
\end{figure}

Let us proceed to wormhole background. The cross section $t=const,\theta = \frac{\pi}{2}$ of wormhole spacetime is conformal to cylindric space (see Eq. (\ref{eq:crossDHcylinder})). Let us consider the Stokes theorem in this case for an differential form $\omega$. For $2D$ flat, non-cylindrical space (see, for example \cite{Arnold:1989:MMoCM}) we divide rectangle domain $\Sigma$ on set of small rectangles $\Sigma_k$ and full integral is represented as sum of integrals over these rectangles (see left panel of Fig. \ref{fig:stokes}).
The integrals over internal edges are canceled and we obtain the Stokes theorem
\begin{equation}
\int_{\partial \Sigma} \omega = \int_\Sigma \mathbf{d}\omega, 
\end{equation}    
where $\mathbf{d}$ is external differential and boundary $\partial \Sigma$ is polyline $ABCD$. In cylindrical case (see right panel of Fig. \ref{fig:stokes}) we use the same approach and obtain the Stokes theorem with boundary $\partial \Sigma$ composed of two disconnected parts: $\partial \Sigma_1$ is circle $L_1 = AmnA$ and $\partial \Sigma_2$ -- circle $L_2 = CpqC$ because lines $DC$ and $CD$ were glued and integrals over these line are canceled. If we change direction of integration in second integral we obtain the Stokes theorem in the following form
\begin{equation}
\int_{\partial \Sigma} \omega = \int_{\partial \Sigma_1} \omega -  \int_{\partial \Sigma_2} \omega= \int_\Sigma \mathbf{d}\omega, \label{eq:stokes}
\end{equation}
where $\partial \Sigma_1 = L_1 = AmnA$ and $\partial \Sigma_2 = L_2 = DqpD$. The particular case the Stokes theorem in the form (\ref{eq:stokes}) is the Newton--Leibniz formula for $0$-form, where we have two boundaries (points). We may realize the line as a cylinder of zero radii and therefore the boundary circles become two points.

Generalization for full $4D$ wormhole spacetime is obvious. Let us consider $2$-form $\omega$ and integrate this form over a ball of finite volume which is bounded by two concentric spheres $S_1$ and $S_2$, then Stokes theorem has the form Eq. (\ref{eq:stokes}), where two boundaries are spheres, $\partial \Sigma_1 = S_1$ and $\partial \Sigma_2 = S_2$. If there is no throat inside of spheres, then we may shrink the second sphere to point and integral over $S_2$ becomes zero. In the opposite case, there is a sphere of minimal radius, and we can not shrink the second sphere to point and have to use the two integral in Eq. \eqref{eq:stokes}.  

Now we apply the Stokes theorem for Maxwell equations which read (see, for example, \cite{Misner:1973:G}) 
\begin{equation}
\mathbf{d} * \mathbf{F} = 4\pi *\mathbf{J},
\end{equation}
where $*$ is the Hodge star, $\mathbf{F}$ is the Maxwell $2$-form and $\mathbf{J}$ is $1$-form of current:
	\begin{equation}
	\mathbf{F} = F_{\mu\nu} dx^\mu\wedge dx^\nu\ (\mu<\nu),\ \mathbf{J} = J_\mu dx^\mu. 
	\end{equation} 
We surround the charge by concentric spheres $S_1$ and $S_2$ with a finite volume between spheres, then integrate $3$-forms and use the Stokes theorem (\ref{eq:stokes}). The result depends on the position of the sphere:
\numparts\label{eq:gauss_gen}
\begin{eqnarray}
\Pi &=& \int_{S_1} * \mathbf{F} - \int_{S_2} * \mathbf{F} = 4\pi q,\ \textrm{throat is inside}\ S_1, \label{eq:gauss_gen_1}\\
\Pi &=& \int_{S_1} * \mathbf{F} = 4\pi q,\ \textrm{throat is outside}\ S_1,\label{eq:gauss_gen_2}
\end{eqnarray}   
\endnumparts
where sphere $S_2$ is situated in another part of wormhole spacetime. In the case (\ref{eq:gauss_gen_1}) we can not shrink sphere $S_2$ to point because we have the throat with minimal area. In the second case (\ref{eq:gauss_gen_2}) we can do it.
	
Therefore, the full flux consists of two parts $\Pi = \Pi_1 + \Pi_2$ where
\begin{equation}
\Pi_1 = \int_{S_1} * \mathbf{F}, \ \Pi_2 = - \int_{S_2} * \mathbf{F}. 
\end{equation}
If wormhole's throat is outside the $S_2$ we may shrink the second sphere to point and obtain $\Pi_2 =0$. If we surround the charge and throat by sphere $S_1$ and calculate the flux through this sphere we obtain $\Pi_1 \not = 4\pi q$ which means that the part of the electric field goes through the throat. The charge observed by flux $\Pi_1$: $q' = \Pi_1/4\pi < q$ and it depends on the position of charge $q$. 
	
If $\gs_{\rho\rho} =1$ as in the spacetime (\ref{eq:metricD}) we may write out the Gauss theorem in a simple form (see Ref. \cite{Dubrovin:1984:MG-MaAPITGoSTGaF})
\begin{equation}
\Pi = \Pi_1 + \Pi_2 = \iint_{S_1}  (\vE\cdot \vn_1) dS_1 + \iint_{S_2}  (\vE\cdot \vn_2) dS_2 = 4\pi q,\label{eq:gauss}
\end{equation}
where $q$ is charge between sphere $S_1$ of radius $R_1$ and normal $\vn_1 = (+1,0,0)$ and sphere $S_2$ of radius $R_2$ and normal $\vn_2 = (-1,0,0)$.

\subsubsection{General static sourceless solution}\label{sec:GSS}

Let us consider the Maxwell equations with current
\begin{equation}
F^{\mu\nu}_{;\nu} = 4\pi J^\mu
\end{equation}
on the background of the massive wormhole (\ref{eq:metricBE}). By using vector potential, $F_{\mu\nu} = \nabla_\mu A_\nu - \nabla_\nu A_\mu$, in Lorentz gauge $A^\nu_{\ ;\nu} = 0$ we obtain the following equations
\begin{equation}
\square A_\mu - \mathcal{R}_\mu^\nu A_\nu = - 4\pi J_\mu.
\end{equation}

For the particle at rest in point $(\rho',\theta',\varphi')$ we can set $A_\rho = A_\theta = A_\varphi = 0$ and $A_t = A_t (\rho,\Omega)$. The Lorentz gauge is satisfied identically, and the Maxwell equations read
\begin{equation}
A_t'' + \frac{2r'}{r} A_t' -\frac{e^{\alpha} }{r^2} \hat{L}^2 A_t = -4\pi e^{\alpha} J_t,
\end{equation}
where 
\begin{equation*}
\hat{L}^2 = - \partial^2_{\theta\theta} - \cot \theta \partial_\theta - \frac{1}{\sin^2\theta} \partial^2_{\varphi\varphi}
\end{equation*}
is the operator of square momentum with eigenvalues $l(l+1)$ and spherical eigenfunctions $Y_{l,m}(\Omega)$. We have used above that $\mathcal{R}_t^\mu =0$. 

The vector potential is defined ambiguously, up to the solution of the homogeneous equation  
\begin{equation}
\square A_\mu - \mathcal{R}^\nu_\mu A_\nu = 0.
\end{equation}
For static solutions with $A_k =0,\ A_t = A_t (\vx)$ we obtain equation   
\begin{equation}
A_t'' + \frac{2r'}{r} A_t' -\frac{e^{\alpha} }{r^2} \hat{L}^2 A_t = 0.
\end{equation}

The general solution of this equation reads
\begin{equation}
A_t = \sqrt{4\pi}\sum_{l=0}^\infty \sum_{m=-l}^{+l} Y_{l,m}(\Omega) z_l(\rho),
\end{equation}
where radial function, $z_l(\rho)$, obeys to the following equation
\begin{equation*}
z_l'' + \frac{2r'}{r} z_l' -\frac{l(l+1)}{r^2} e^{\alpha} z_l = 0.
\end{equation*}

Due to relation  
\begin{equation*}
\iint_S Y_{l,m} d\Omega = \sqrt{4\pi} \delta_{l0}\delta_{m0},
\end{equation*}
the flux of this field through the sphere $S_1, \rho = R_1$ with normal $\vn_1 = (+1,0,0)$ is defined by monopole component of radial function
\begin{equation*}
\Pi_1 = \int_{S_1} *\mathbf{F} = \iint_{S_1} A_t'(R_1,\Omega) r^2(R_1) d\Omega = 4\pi r^2(R_1) z'_0(R_1),
\end{equation*}
which obeys to the following equation 
\begin{equation*}
z_0'' + \frac{2r'}{r} z_0' = 0.
\end{equation*}

The general solution of this equation reads
\begin{equation}
z_0  = Q \int_0^\rho \frac{dx}{r^2(x)} + C,
\end{equation}
where $Q$ and $C$ are the constants of integration. Therefore, the flux of field through sphere $S_1$ equals to 
\begin{equation}
\Pi_1 = 4 \pi Q.
\end{equation}
The constant of integration $Q$ has the dimension of charge. The straightforward calculations by using the Gauss theorem (\ref{eq:gauss}) give $\Pi_1 = - \Pi_2$ which means that local observers on opposite sides of wormhole's throat will see the same charge $Q$ but with opposite signs. The total flux $\Pi =0$ which should be the case because there is no charge between spheres. An observer who is taking into account only one flux $\Pi_1$ will observe charge $Q_1 = q$. Another observer in the opposite part of the wormhole will take into account flux $\Pi_2$ and he will observe charge with opposite sign $Q_2 = - q$. This field is a realization of Wheeler's idea \cite{Wheeler:1960:Ngg} "charge without charge". 

Therefore, the general static solution of homogeneous equations read
\begin{equation}
A_t = Q \int^\rho_0 \frac{dx}{r^2(x)} + C + \sqrt{4\pi}\sum_{l=1}^\infty \sum_{m=-l}^{+l} Y_{l,m}(\Omega) z_l(\rho),
\end{equation}
and the corresponding electric field has the following form
\begin{equation}
E_\rho = A_t' = \frac{Q}{r^2} + \sqrt{4\pi}\sum_{l=1}^\infty \sum_{m=-l}^{+l} Y_{l,m}(\Omega) z_l'(\rho). 
\end{equation}

Let us consider the rest multipole terms. If the profile of the throat at infinity tends to Minkowskian, $r \approx \alpha |\rho|$ then two independent solutions will get the following behavior $\sim A_l\rho^l$ and $\sim B_l \rho^{-l-1}$. From the requirement the finiteness potential at infinity $|\rho | \to \infty$ we obtain that $A_l = 0$. Then the condition that potential must be $C^1$ continuous on the throat when $\rho =0$ we obtain that all $B_l =0$. Therefore, we obtain a general expression for potential which is finite at infinity
\begin{equation}
	A_t = Q \int^\rho_0 \frac{dx}{r^2(x)} + C, E_\rho = A_t' = \frac{Q}{r^2(\rho)}. \label{eq:static_hom}
\end{equation}
In Minkowsky spacetime, without wormhole, the conditions of finiteness at origin and zero at infinity give the only possibility $A_t = const$. Therefore, we may conclude that the general static solution of non-homogeneous Maxwell equations in the wormhole background is defined non-unique up to solution (\ref{eq:static_hom}) of corresponding homogeneous equations.  

Some additional conditions were suggested to avoid this ambiguity \cite{Krasnikov:2008:eioapcwaw,Beskin:2011:eamfitaow,Boisseau:2013:eiaswr}. It was suggested in Ref. \cite{Krasnikov:2008:eioapcwaw} to consider flux through the wormhole's throat as a fixed parameter, "charge" of the wormhole, which characterize the wormhole itself and it differs one wormhole from another one. This suggestion leads to modification of the self-force (see Eq. \eqref{eq:KrasnikovCoerrection}). The authors of Ref. \cite{Beskin:2011:eamfitaow} imposed a condition that the electric flux at the infinity $\rho \to \infty$ is exactly equal to $4\pi q$ as expected in the usual case at the spatial infinity. 

$Z_2$--symmetry was used in Ref. \cite{Boisseau:2013:eiaswr} which means identification $(\rho,\Omega) = (-\rho,\Omega)$. This symmetry has been proposed as black hole foils in Ref. \cite{Damour:2007:wabhf} for massive wormholes. As shown in Ref. \cite{Boisseau:2013:eiaswr} the $Z_2$--symmetry is equivalent to the condition in Ref. \cite{Beskin:2011:eamfitaow} because the electric field at the throat is zero and therefore the flux is zero, too. In the framework of $Z_2$-symmetry we have to consider two identical charges with the same distance from the throat. The symmetric charge gives an additional contribution to flux which exactly compensates flux to the second part of wormhole's spacetime. In the end, we obtain that the flux through the sphere $S_1$ is $4\pi q$.

\subsubsection{Electromagnetic self-force in the space-time of massless wormholes} \label{sec:elecmassless}

In this section we develop the approach to find self-force for a charged particle at rest in the point $\rho',\theta',\varphi'$ in the space-time with metric (\ref{eq:metricD}). 

The Maxwell equation for the zero component of the potential reads
\begin{equation*}
\triangle A^t = - 4\pi q \frac{\delta^{(3)} (\vx - \vx')}{\sqrt{\gs^{(3)}}}. 
\end{equation*}
It is obvious that electrostatic potential $V = A^t$. The radial part, $z_l$, of potential $V$ defined by relation (here $\Omega = (\theta,\varphi)$) 
\begin{equation*}
V = 4\pi q \sum_{l=0}^\infty \sum_{m=-l}^l Y_{lm}(\Omega) Y_{lm}^*(\Omega') z_l(\rho,\rho'),
\end{equation*}
is subject to the equation
\begin{equation}
z''_l + \frac{2 r'}r z'_l - \frac{l(l+1)}{r^2} z_l =- \frac{\delta (\rho - \rho')}{r^2(\rho)}.\label{eq:maineq}
\end{equation}

To make renormalization we have to subtract from potential its singular part which was found in Sec. \ref{sec:renormalization} (see Eq. \eqref{eq:singD}) which is, in fact, the potential in the Minkowski space-time. Now we are in a position to consider several specific cases in detail and after that to proceed to the general profile of the throat. \bigskip

\textbf{I. Profile} $r = |\rho| + a$
\bigskip

The electrostatic potential $V$ reads \cite{Khusnutdinov:2007:Scpwst}
\begin{eqnarray*}
\fl V_{\rho > \rho'} &=& \frac q{\sqrt{r(\rho)^2 -2r(\rho)r(\rho')\cos\gamma + r(\rho')^2}}- \frac{q}{2a}\ln\left|1+\frac{2t}{1-t+\sqrt{t^2 -   2t\cos\gamma+1}}\right|, \\
\fl V_{\rho <0} &=& \frac{q t}{a\sqrt{t^2 - 2t\cos\gamma+1}} -   \frac{q}{2a}\ln\left|1+\frac{2t}{1-t+\sqrt{t^2 -   2t\cos\gamma+1}}\right|, 
\end{eqnarray*}
where $t = \frac{a^2}{r(\rho) r(\rho')}$ and the particle position $\rho' >0$. 

\begin{figure}[ht]
	\centerline{\includegraphics[width=6truecm]{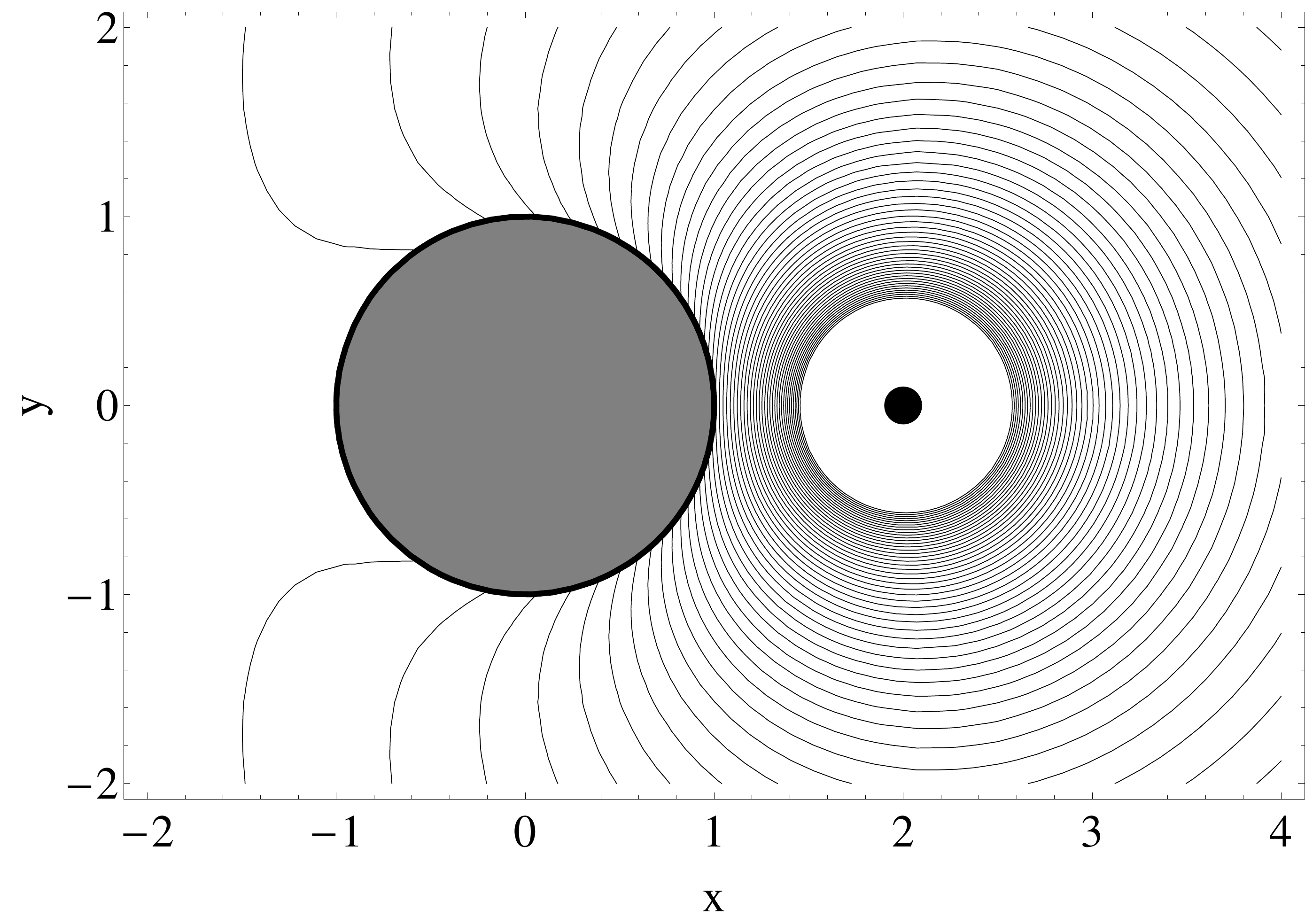}\includegraphics[width=6truecm]{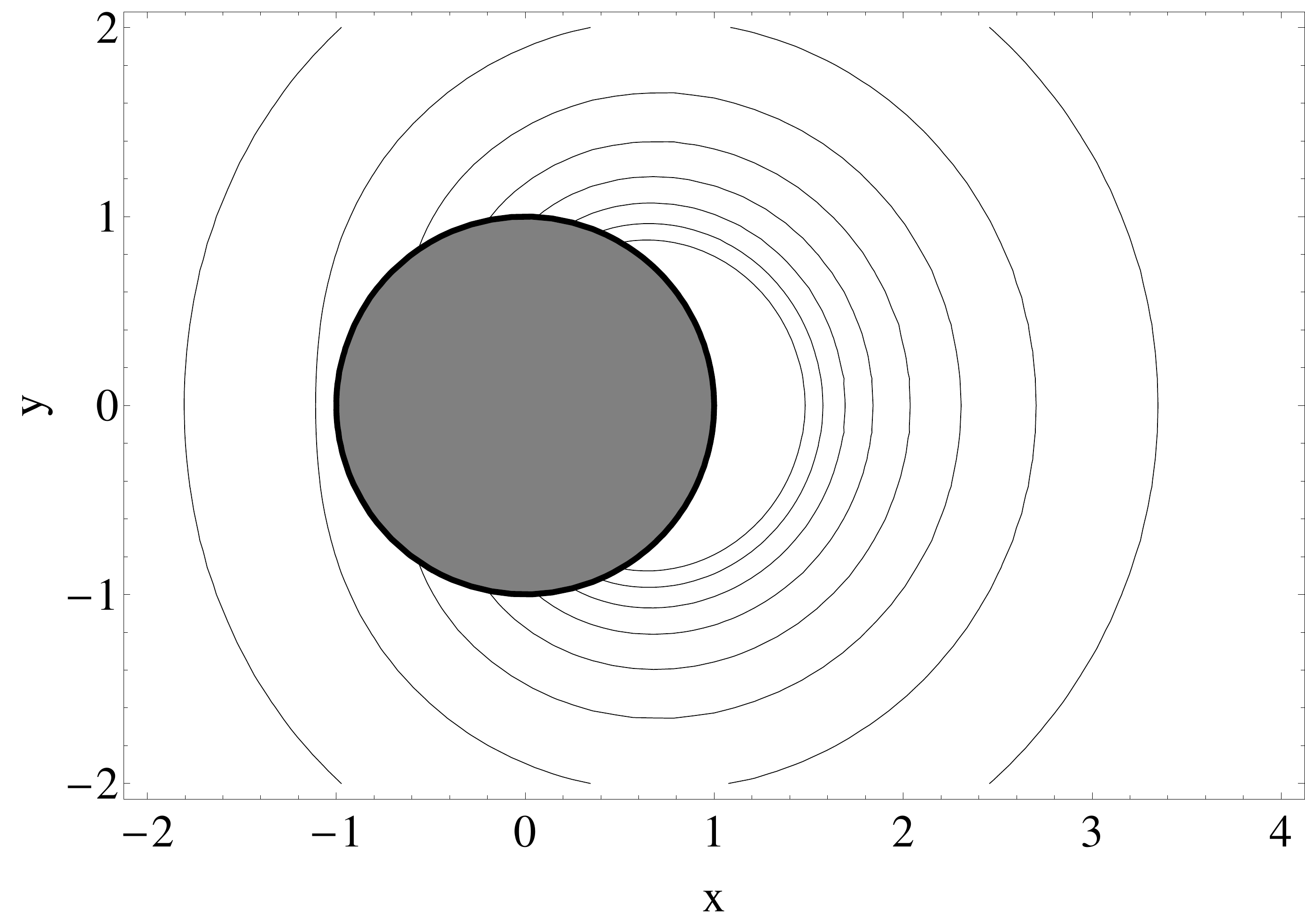}}
	\caption{The level line of electrostatic potential $V = A^t = 4\pi q G(x;x')$. The radius of throat $a=1$ and charge $q=1$. The particle is situated at the point $(x',y',z') = (2,0,0)$. Left panel is for $\rho >0$ and right panel is for $\rho <0$. }
	\label{fig:whcontour}
\end{figure}

\begin{figure}[ht]
	\centerline{\includegraphics[width=6truecm]{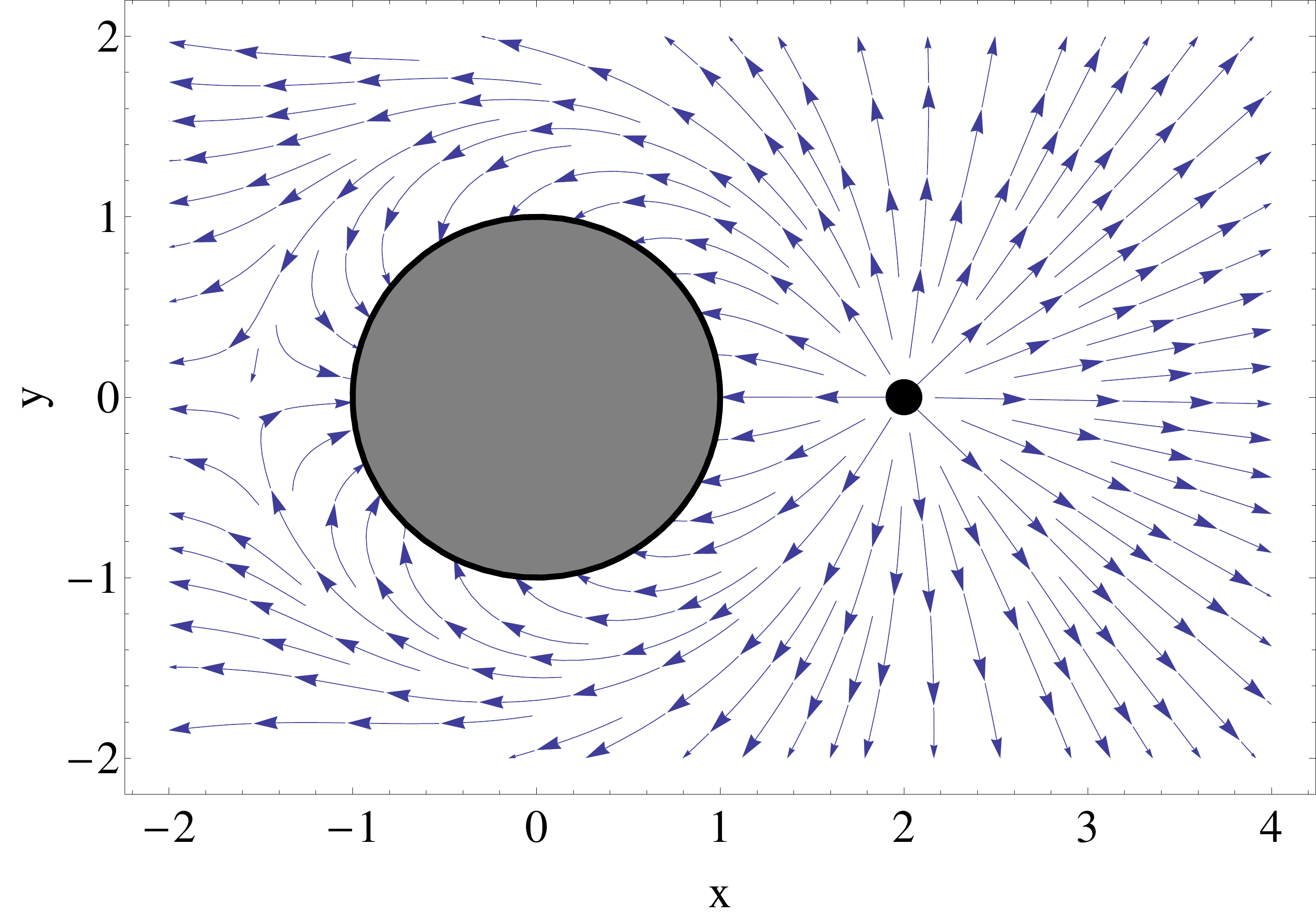}\includegraphics[width=6truecm]{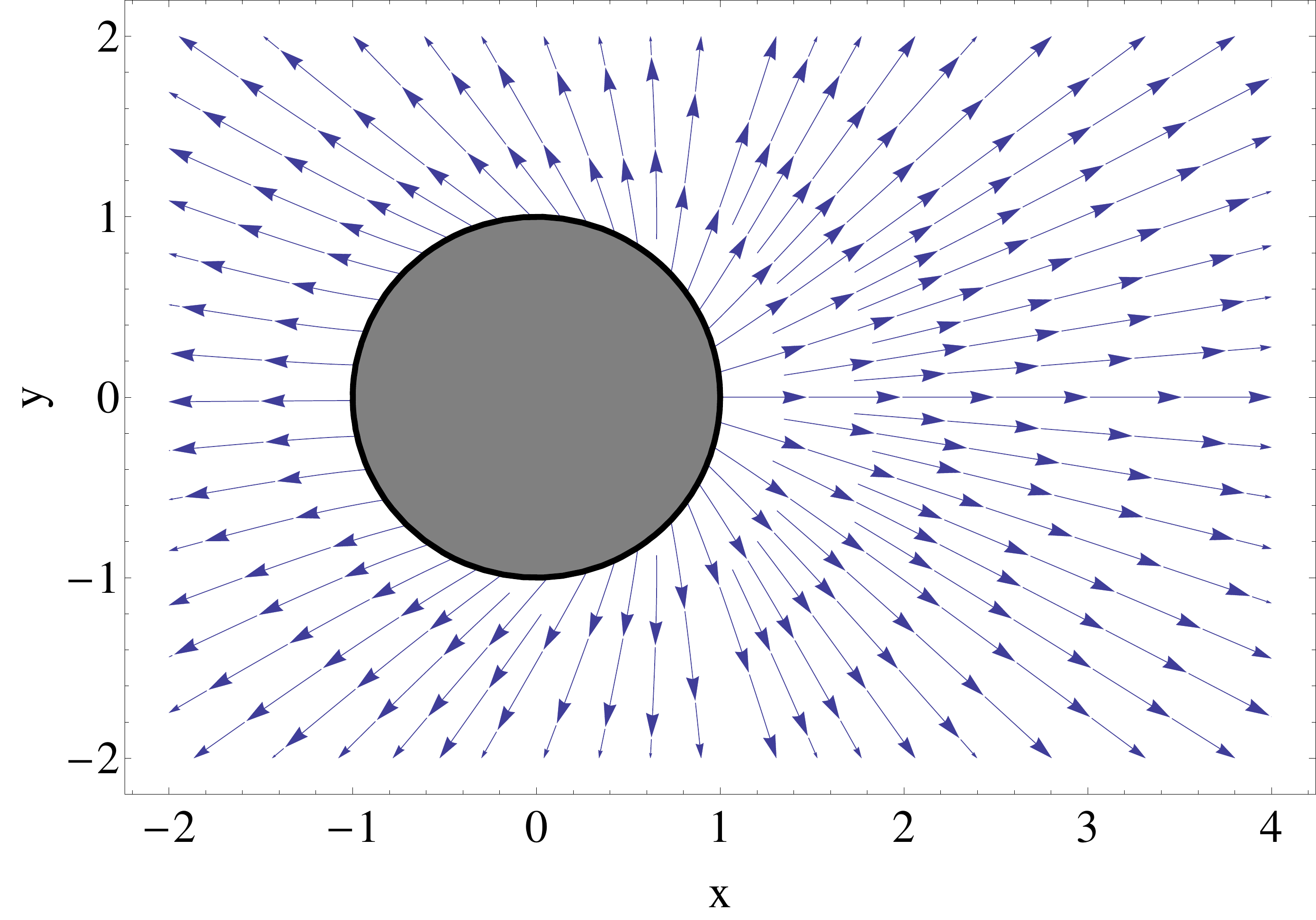}}
	\caption{The lines of electric field $\vE = - \nabla A^t$. We set radius of throat $a=1$ and position of charge $(x',y',z') = (2,0,0)$. Left panel: $\rho>0$, right panel: $\rho <0$. }
	\label{fig:whelectric}
\end{figure}

In Fig. \ref{fig:whcontour} we plot the level lines of electrostatic potential $V$ of charge at the point $x'=(1,\frac{\pi}{2},0)$ in the spacetime of the wormhole of the unit radius the throat, $a=1$.
The lines of the electric field, $\vE = - \nabla A^t$, are plotted in Fig. \ref{fig:whelectric} for both parts of the wormhole spacetime. 
We observe that the electric field goes through the throat to the opposite side of the wormhole spacetime. 

The flux of this field through the spheres of radius $R_1$ (for $\rho >0$) and $R_2$ (for $\rho < 0$) reads 
\begin{equation}
\Pi_1 = \left\{ 
\begin{array}{ll}
4\pi q - \frac{2\pi a }{r(\rho')}q, & 0 < \rho' < R_1, \\
- \frac{2\pi a }{r(\rho')}q, & 0 < R_1 < \rho',
\end{array}  \right. , \ \Pi_2 = +\frac{2\pi a }{r(\rho')}q, \rho  = - R_2. 
\end{equation}
Therefore, the total flux is zero if $R_1 < \rho'$ because the charge is out of spheres and total flux is $4\pi q$ if $R_1 > \rho'$ because the charge is inside spheres and the Gauss theorem (\ref{eq:gauss}) is satisfied.  

Let us consider the case when the observation point, $\rho$, is far from both the wormhole throat and from the position of a particle, $\rho'$, that is $\rho \gg  a, \rho'$. In this limit we have
\begin{eqnarray}
\rho>0 &:& V = \frac q{\rho} - \frac {aq}{2\rho(a+\rho')} + O(\rho^{-2}),\nonumber \\
\rho<0 &:& V = -\frac {qa}{2\rho(a+\rho')} + O(\rho^{-2}).\label{eq:afar}
\end{eqnarray}
From the first of these expressions, we observe that the potential of the particle contains an additional term with the same $\rho$ dependence in addition to the standard Coulomb part. 

The same problem arose for a particle in the Schwarzschild space-time \cite{Linet:1976:eamitsm,Vilenkin:1979:siocpitgf}. The electric field of a test particle at rest in the Schwarzschild black hole \eqref{Schwarzschild} was firstly found by Copson in Ref. \cite{Copson:1928:oeiagf} (see Eq. \eqref{Copson}). The potential $V = A^t$ of probe charge far from the black hole, $r\to \infty$, has additional term 
\begin{equation}
V \approx \frac{q}{r} - \frac{M}{rr'}. \label{eq:bhatinfty}
\end{equation} 
Therefore, the flux through a sphere with radius $r = const$ reads
\begin{equation}
\Pi = \int r^2 \sin\theta \partial_r A_t d\theta \wedge d\varphi = \left\{
\begin{array}{ll}
4\pi q - 4\pi q \frac{ M}{r'}, & r > r'\\ 
 - 4\pi q \frac{ M}{r'}, & r < r'
\end{array} \right. . 
\end{equation}
The origin of the additional term in flux is bad behavior (\ref{eq:bhatinfty}). To solve this problem, it was suggested to add a solution of the corresponding homogeneous equation to the solution with a "bad"\ asymptotic behavior \cite{Linet:1976:eamitsm} (see Eq. \eqref{CopsonLinet}). Then the flux of this field reads
\begin{equation}
\Pi =  \left\{
\begin{array}{ll}
4\pi q , & R > r'\\ 
0, & R < r'
\end{array} \right.  
\end{equation}
and the Gauss theorem is satisfied. At the horizon the vector potential has constant value $A_t(r=2M) = -\frac{q}{r'}$.

However, in our case of wormhole spacetime, there is no need for such modification: the potential (\ref{eq:afar}) correctly describes the real potential of the particle at rest, and the Gauss theorem is satisfied. In fact, the observation of a charge $q$ from a large distance would yield the result $q(1- \frac a{2(\rho'+a)})$. The explanation is very easy -- some electric field lines have gone through the throat to another, invisible domain of space-time and we must take these field lines into account to formulate the Gauss theorem correctly. 

Now let us restrict ourselves to the $"+"$ part of space-time and consider the situation when $\Omega=\Omega'$. After renormalization, we obtain electric potential
\begin{equation*}
V^{\textrm{ren}} = \frac q{2 a} \ln \left|1-\frac{a^2}{r(\rho)r(\rho')}\right|.
\end{equation*}
The self-energy is
\begin{equation}
U_e = \frac{q^2}{4 a} \ln \left|1-\frac{a^2}{(a+\rho)^2}\right|.\label{eq:selfD}
\end{equation}
Some limiting cases are:
\begin{equation}
U_e |_{\rho\to 0} = \frac{q^2}{4a} \ln \frac{2\rho} a,\ U_e |_{\rho\to\infty} = - \frac{q^2a}{4\rho^2} + \frac{a^2q^2}{2\rho^3}, \ U_e |_{a\to 0} = -\frac{a q^2}{4(a+\rho)^2}. \label{eq:limits}
\end{equation}
The self-force
\begin{equation*}
\mathbf{\mathcal{F}} = -\nabla U_e, 
\end{equation*}
has the radial component only
\begin{equation}\label{eq:infthinforce}
\mathcal{F}^\rho = -\partial_\rho U_e = -\frac{a q^2}{2r(\rho)^3}\frac{\mathrm{sgn}(\rho)}{1-\frac{a^2}{r(\rho)^2}}.
\end{equation}
The self-force is always attractive, it turns into infinity at the throat and goes down monotonically to zero as $\rho \to \infty$. We may compare this expression with its analog for Schwarzschild space-time with Schwarzschild radius $r_s = a$:
\begin{equation*}
\mathcal{F}^r = +\frac{a q^2}{2r^3}\sqrt{1-\frac{a^2}{r^2}}.
\end{equation*}
Important observations are

1) The self-force in the wormhole space-time has an opposite sign -- it is attractive.

2) Far from the wormhole throat, $\rho \to \infty$, and from the black hole we have the same results but with opposite signs
\begin{equation*}
\left.\mathcal{F}^\rho_{wh}\right|_{\rho\to\infty} = -\frac{a q^2}{2\rho^3},\ \left.\mathcal{F}^\rho_{bh}\right|_{\rho\to\infty} = +\frac{a q^2}{2\rho^3}.
\end{equation*}

3) At the Schwarzschild radius $r_s = a$ the self-force equals zero, whereas at the wormhole throat it tends to infinity. The latter discrepancy originates in the selected throat profile function that leads to the curvature singularity at the throat. 

One comment is in order. If we take into account the suggestion in Ref. \cite{Krasnikov:2008:eioapcwaw} to consider flux through the wormhole's throat as the fixed parameter -- "charge" of the wormhole then we have to demand the flux through the throat due to external charge has to be zero. With this definition, the self-energy will differ from the obtained above (\ref{eq:selfD}) 
\begin{equation}\label{eq:KrasnikovCoerrection}
U^*_e=\frac{q^2}{4 a} \ln \left|1-\frac{a^2}{r^2(\rho)}\right| + \frac{q^2a}{4r^2(\rho)}.
\end{equation}
In this case, the self-energy falls as $\rho^{-4}$ far from the throat.   \bigskip

\textbf{II. Profile} $r = \sqrt{a^2+\rho^2}$
\bigskip

In this case, the electric potential reads \cite{Taylor:2013:sfoaacsspiawst,Khusnutdinov:2008:saoapcitstoawwatozl,Khusnutdinov:2007:Scpwst}
\begin{equation}
V = \frac{q}{\pi \sqrt{\rho^2 + \rho'^2 - 2\rho\rho' \cos\gamma + a^2 \sin^2\gamma}} \arccos \left(-\frac{\rho\rho' + a^2 \cos\gamma}{\sqrt{\rho^2 + a^2} \sqrt{\rho'^2 +a^2}}\right). \label{eq:taylor}
\end{equation}
In particular case of coincidence angle variables, $\gamma =0$, the potential was obtained in Ref. \cite{Khusnutdinov:2008:saoapcitstoawwatozl} ($\rho' >0$):
\begin{eqnarray}
V &=& \frac 1{|\rho-\rho'|} -\frac 1{\pi} \frac{\arctan \frac{\rho}a - \arctan \frac{\rho'}a}{\rho - \rho'}, \rho > \rho', \nonumber \\
V &=& -\frac 1{\pi}\frac{-\arctan \frac{\rho}a + \arctan \frac{\rho'}a - \pi}{-\rho + \rho'}, \rho <0.\label{eq:drainholegreen}
\end{eqnarray}

The potential far from the charge $|\rho| \gg \rho'$ has the following form
\begin{eqnarray*}
V &=& \frac q{\rho} + \frac q{\rho}\left(-\frac 12 + \frac 1{\pi} \arctan \frac{\rho'}a \right), \rho >0\\
V &=& \frac q{\rho}\left(-\frac {1}2 + \frac 1\pi \arctan \frac{\rho'}a\right), \rho <0,
\end{eqnarray*}
and again it obeys the Gauss theorem (\ref{eq:gauss}). 

\begin{figure}[ht]
\begin{center}
\includegraphics[width=6truecm]{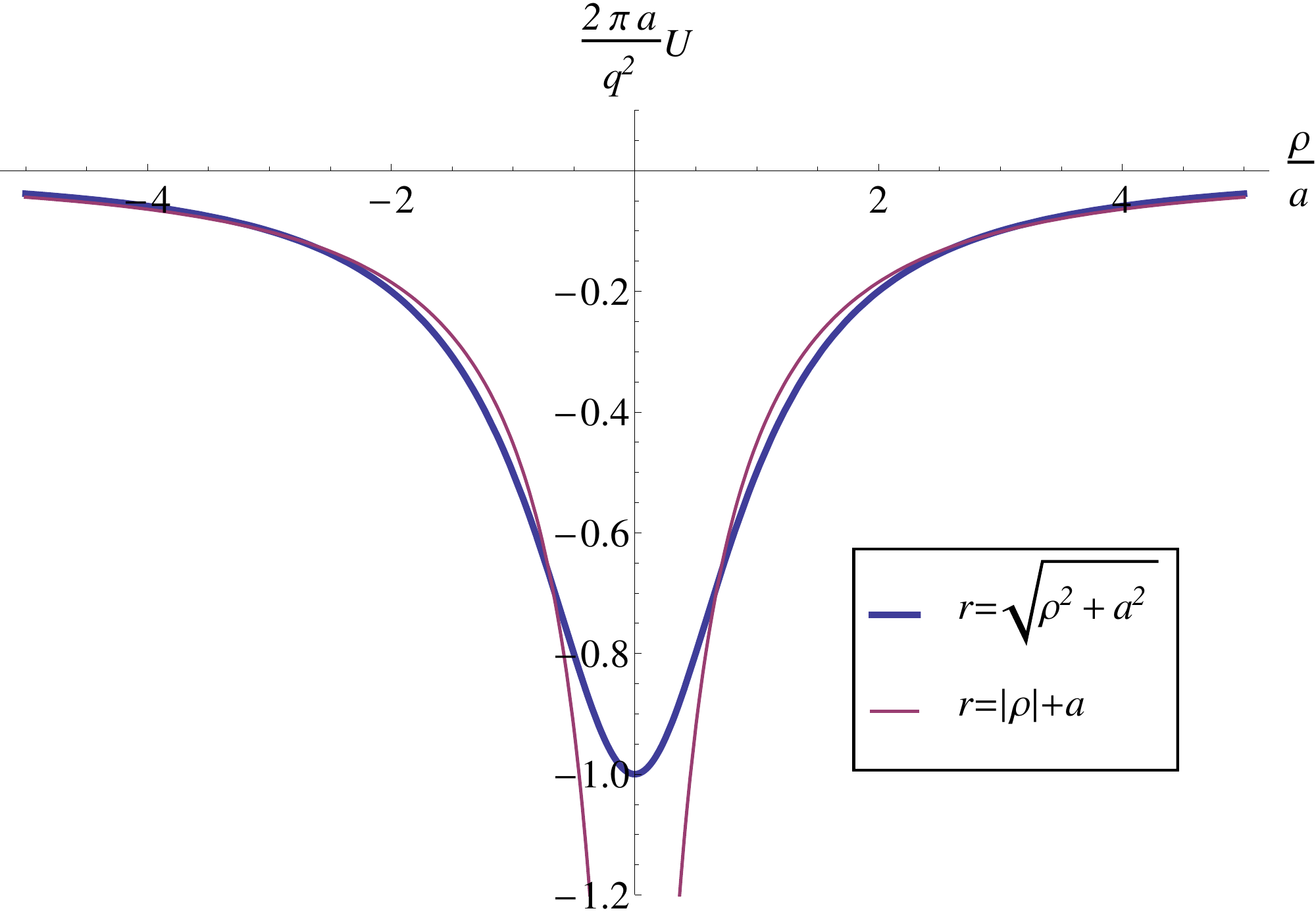}\includegraphics[width=6truecm]{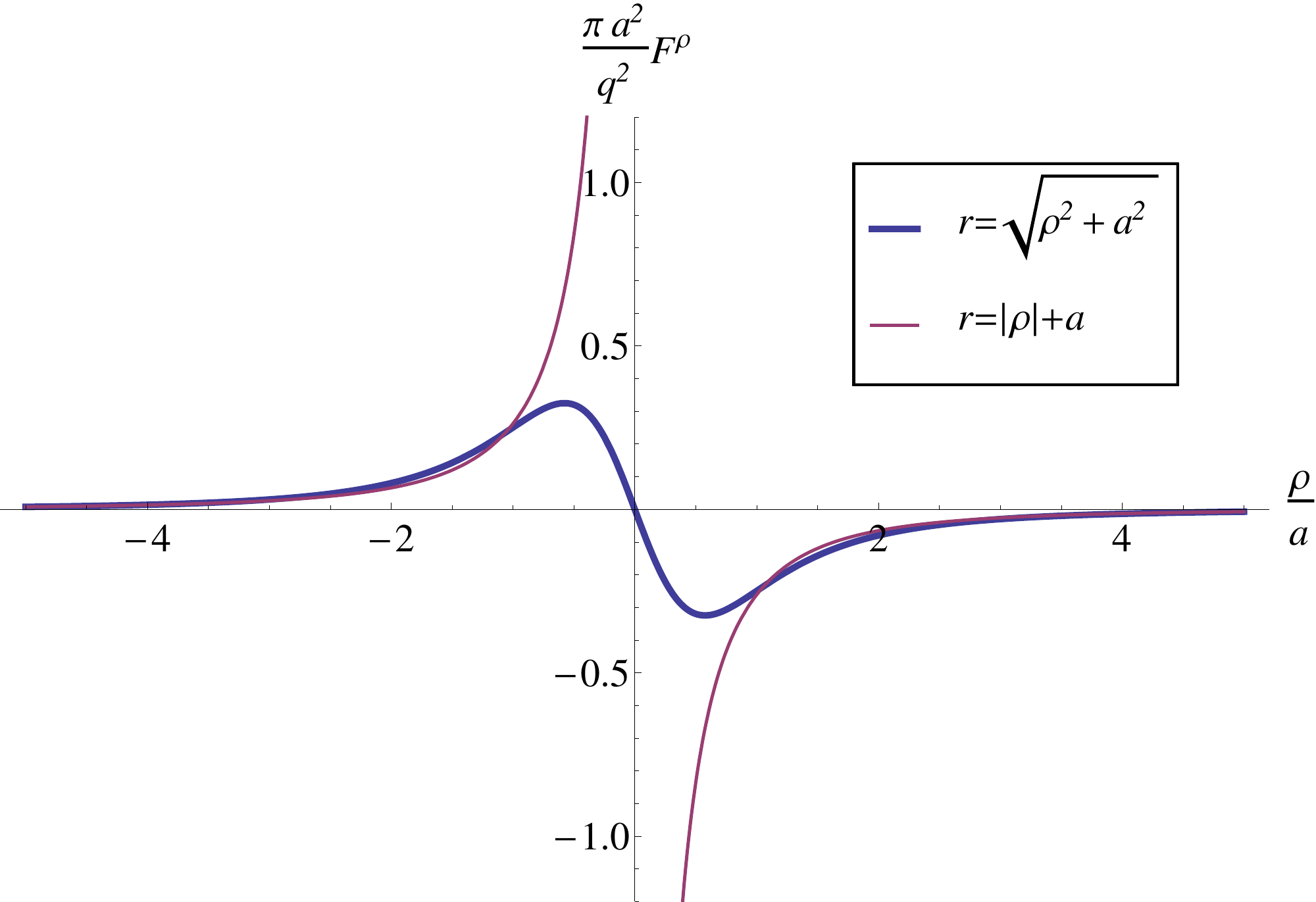}
\end{center}
\caption{The self-energy  (left panel) and self-force (right panel) for two models of throat profile.} \label{fig:pf}
\end{figure}

Therefore, we obtain the self-potential
\begin{equation}
U_e =-\frac{q^2}{2\pi} \frac a{\rho^2 +a^2},\label{eq:selfenergy}
\end{equation}
and the self-force
\begin{equation}
\mathcal{F}^\rho = -\partial_\rho U_e = -\frac{q^2}{\pi}\frac{a\rho}{(\rho^2 +a^2)^2}.\label{eq:force}
\end{equation}
As expected the self-force is everywhere finite and equals zero at the throat.  Far from the wormhole we have 
\begin{equation*}
\mathcal{F}^\rho|_{\rho\to\infty} \approx -\frac{q^2}{\pi} \frac{a}{\rho^3}.
\end{equation*}
Thus, the self-force is always attractive. It has the maximum value at the distance $\rho^*=a/\sqrt{3}$ with magnitude $|F^\rho_{\textrm{max}}| = 3\sqrt{3}q^2/16\pi a^2$. Plots of the potential and the self-force are shown in Fig. \ref{fig:pf} for infinitely short and drainhole models of the throat.  

The same expression for self-energy (\ref{eq:selfenergy}) was obtained also in Refs. \cite{Linet:2007:Ewg,Boisseau:2013:eiaswr} by a different method developed in Ref. \cite{Linet:2005:bhiwteoseisicf}. \bigskip

\textbf{III. General profile of the throat}
\bigskip

The Green function may be expressed in terms of the full set of solutions to the corresponding equation. Due to spherical symmetry, we deal only with the radial equation:
\begin{equation}\label{eq:radial_gen}
z_l'' + \frac{2 r'}r z_l' - \frac{l(l+1)}{r^2} z_l = 0.
\end{equation}
Let us perform WKB analysis  of this equation \cite{Khusnutdinov:2007:Scpwst} using $(l+1/2)^{-1}$ as WKB parameter. 

The contribution with $l=0$ has to be considered separately. For $l=0$ the equation (\ref{eq:radial_gen}) simplifies considerably ($\varphi$ stands here for the zero mode only) and has the following solutions 
\begin{equation}
\varphi_1 = \int_\rho^\infty \frac{d\rho}{r^2},\ \varphi_2 = 1, \label{eq:phizero}
\end{equation}
with Wronskian $W(\varphi_1,\varphi_2) = 1/r^2$. 

To perform WKB analysis for modes $l\geq 1$ we represent the solution of this equation in the exponential form $z_l = e^{S}$ and expand $S$ in the following power series in $\nu = l+1/2$:
\begin{equation}
S = \sum_{n=-1}^\infty \nu^{-n} S_n. \label{eq:sexp}
\end{equation}

The calculations have performed in Ref. \cite{Khusnutdinov:2007:Scpwst}. The self-energy reads
\begin{equation}\label{U_renorm'd}
U_e = \frac{q^2}{2} \left[- \frac 1{r(\rho)} + \frac 1{r(\rho)} \sum_{k=1}^\infty \zeta_H\left(2k,\frac 32\right) j_{2k} (\rho)+ \frac 1a \varphi_1(\rho) - \frac 1{2a} \frac{\varphi_1^2(\rho)}{\varphi_1(0)}\right],
\end{equation}
where $\zeta_H$ is the Hurwitz zeta-function \cite{Bateman:1953:Htf-1}. The first two functions $j_k$ are listed below 
\begin{eqnarray*}
\fl j_2  &=& - \frac{-1+r'^2 + 2r r''}{8}= \frac{3}{8} a_1r^2,\\
\fl	j_4  &=& \frac 1{128} \left(3+3 r'^4 - 12 r r'' - 4r^2 r''^2 -2 r'^2 (3+2r r'') - 32 r^2 r' r^{(3)} - 8 r^3 r^{(4)}\right),
\end{eqnarray*}
where $a_1$ stands for the first heat kernel coefficient (see, for example, review \cite{Vassilevich:2003:hkeum}). This expression is exact and may be used for an arbitrary throat profile. It was shown \cite{Khusnutdinov:2007:Scpwst} that this series is fast convergent we may write out an approximate expression for the self-potential in which there is the contribution from first heat kernel coefficient
\begin{equation}
U_e \approx \frac{q^2}{2} \left[- \frac 1r +\left( \frac{\pi^2}2 -4 \right) \frac 38 a_1 r^2 + \frac 1a \varphi_1(\rho) - \frac 1{2a} \frac{\varphi_1^2(\rho)}{\varphi_1(0)}\right].
\end{equation}

Far from the wormhole throat, the following expression for the self-potential has obtained
\begin{equation}
U_e|_{\rho\to\infty} = -\frac {q^2}{4\rho^2}\frac a{\varphi_1(0)} = -\frac {q^2}{4\rho^2} \left[\int_0^\infty \frac{d\rho}{r^2(\rho)}\right]^{-1}.
\end{equation}
Note that it is always negative, hence the self-force is an attractive force for any profile of the throat. All information about the specific throat profile is encoded in the factor
\begin{equation*}
\int_0^\infty \frac{d\rho}{r^2(\rho)}.
\end{equation*}
\subsubsection{Electromagnetic self-force in the space-time of massive wormhole}

The potential of a particle at rest in the background of the massive wormhole with metric Eq. \eqref{eq:metricBE} has obtained in Ref. \cite{Khusnutdinov:2010:Spcsmw}. The potential reads
\begin{equation}
A_t = - \frac{q e^{\mu (\arctan x + \arctan x' - \pi)}}{ a_m\sqrt{x^2 + x'^2 - 2x x' \cos\gamma + \sin^2\gamma}} \frac{\sinh \left[\mu \arccos \left(-\frac{x x' + \cos\gamma}{\sqrt{x^2 + 1} \sqrt{x'^2 + 1}}\right)\right]}{\sinh \pi\mu}, \label{eq:taylor1}
\end{equation}
where
\begin{equation*}
x = \frac{\rho}{a_m},\ \mu = \frac{m}{a_m}, \ a_m= \sqrt{n^2 -m^2},
\end{equation*}
and $A_\rho = A_\theta = A_\varphi = 0$. 

Taking into account renormalization procedure discussed in Sec. \ref{sec:renormalization} the following singular part (\ref{eq:singBE}) of tetrad component
\begin{equation}
A_{(t)}^{\textrm{DS}} = -\frac{q}{\rho -\rho'}e^{- \frac{1}{2}\alpha(\rho')},
\end{equation}
has to be extracted. Therefore, we obtain the expression for the self-energy
\begin{equation}
U_e = - \frac q2 A_{t}^{ren} = -\frac{q^2}{\rho^2 + n^2 - m^2} \frac{m e^{-\alpha}}{2\tanh\pi \mu},
\end{equation}
and tetrad component of the self-force
\begin{equation}\label{FselfWH}
\mathcal{F}^{(\rho)} = -\partial_\rho U_e = - \frac{q^2}{(\rho^2 + n^2 - m^2)^2}   \frac{m (\rho - m)e^{-\alpha}}{\tanh\pi \mu}.
\end{equation}
For massless wormhole, $m\to 0$, we recover results obtained in Ref. \cite{Khusnutdinov:2007:Scpwst} for drainhole wormhole. Far from the wormhole throat, $\rho\to \pm \infty$, the force falls down as third power but with different coefficients 
\begin{equation}
\mathcal{F}^{(\rho)}_{\rho\to +\infty} = - \frac{1}{\rho^3}   \frac{m q^2}{\tanh\pi \mu},\ \mathcal{F}^{(\rho)}_{\rho\to -\infty} = - \frac{1}{\rho^3}   \frac{m q^2 e^{-2\pi \mu}}{\tanh\pi \mu}.
\end{equation}
The maximum of self-energy appears at throat for $\rho_u =m$ with value 
\begin{equation}
U_{e,\textrm{max}} = -\frac{q^2}{n^2} \frac{m e^{-\alpha (m)}}{2\tanh\pi \mu},
\end{equation}
and the self-force has extrema at $\rho_f = m \pm \frac{n}{\sqrt{3}}$ with values 
\begin{equation}
\mathcal{F}^{(\rho)}_{\textrm{max}} = \mp \frac{q^{2}m 3\sqrt{3} e^{-\alpha (m \pm \frac{n}{\sqrt{3}})}}{4n (\sqrt{3} m -2n)^2 \tanh\pi \mu}. 
\end{equation}
For extremal wormhole (Case II in Ref. \cite{Ellis:1973:eftadapmigr}) when $m\to n$ the maximum of self-energy and self-force admit constant values
\begin{equation}
U_{\textrm{max}} = - \frac{q^2}{2e^2 n} = - \frac{q^2}{n} 0.067\ldots,\ \mathcal{F}^{(\rho)}_{\textrm{max}} = \frac{q^2}{n^2} \left(9 \mp \frac{21 \sqrt{3}}{4}\right) e^{-3 \pm \sqrt{3}} = \frac{q^2}{n^2}  \genfrac\{\}{0pt}{0}{- 0.026\ldots}{+0.159\ldots}.
\end{equation}

\begin{figure}[ht]
	\begin{center}
		\includegraphics[width=6truecm]{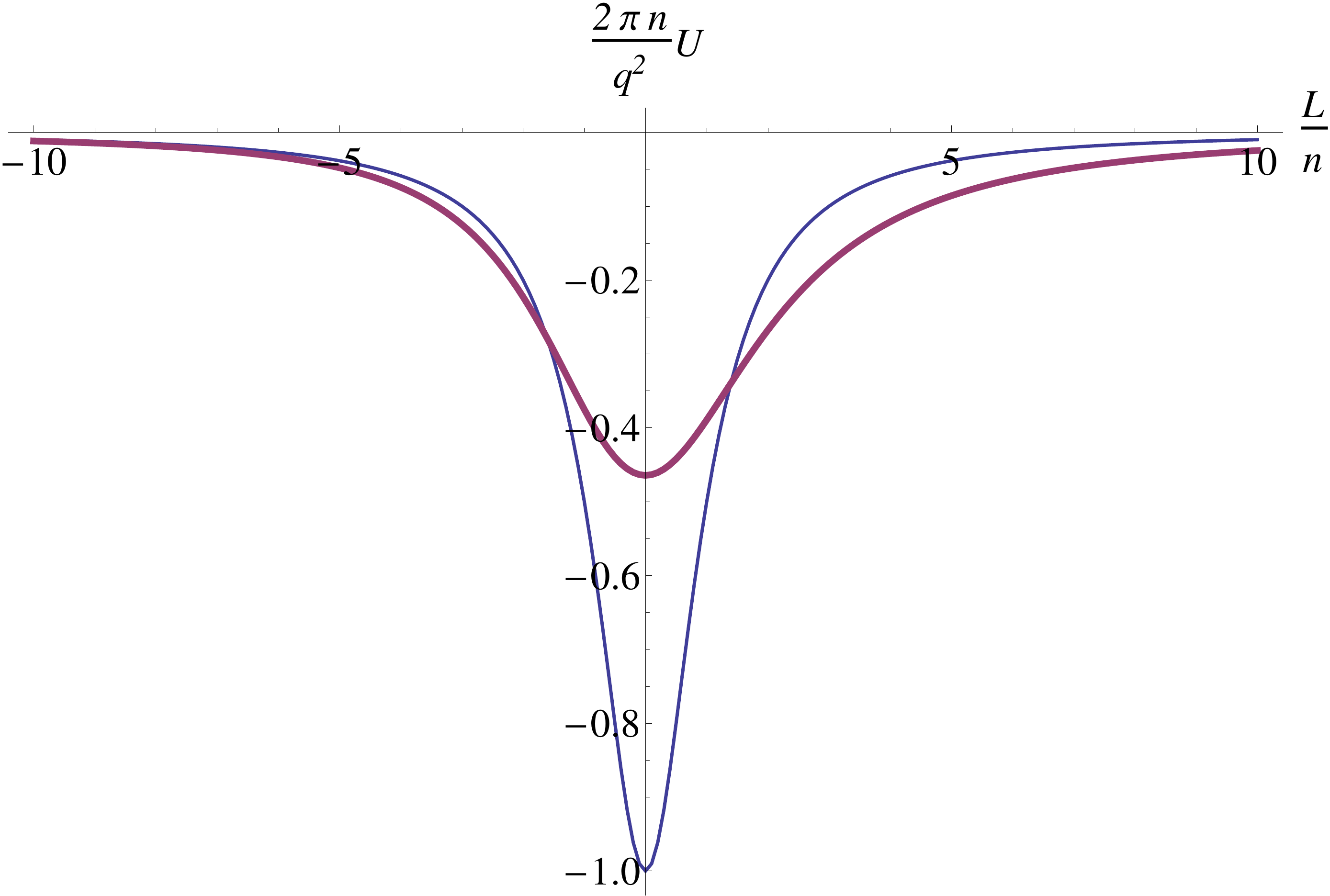}\includegraphics[width=6truecm]{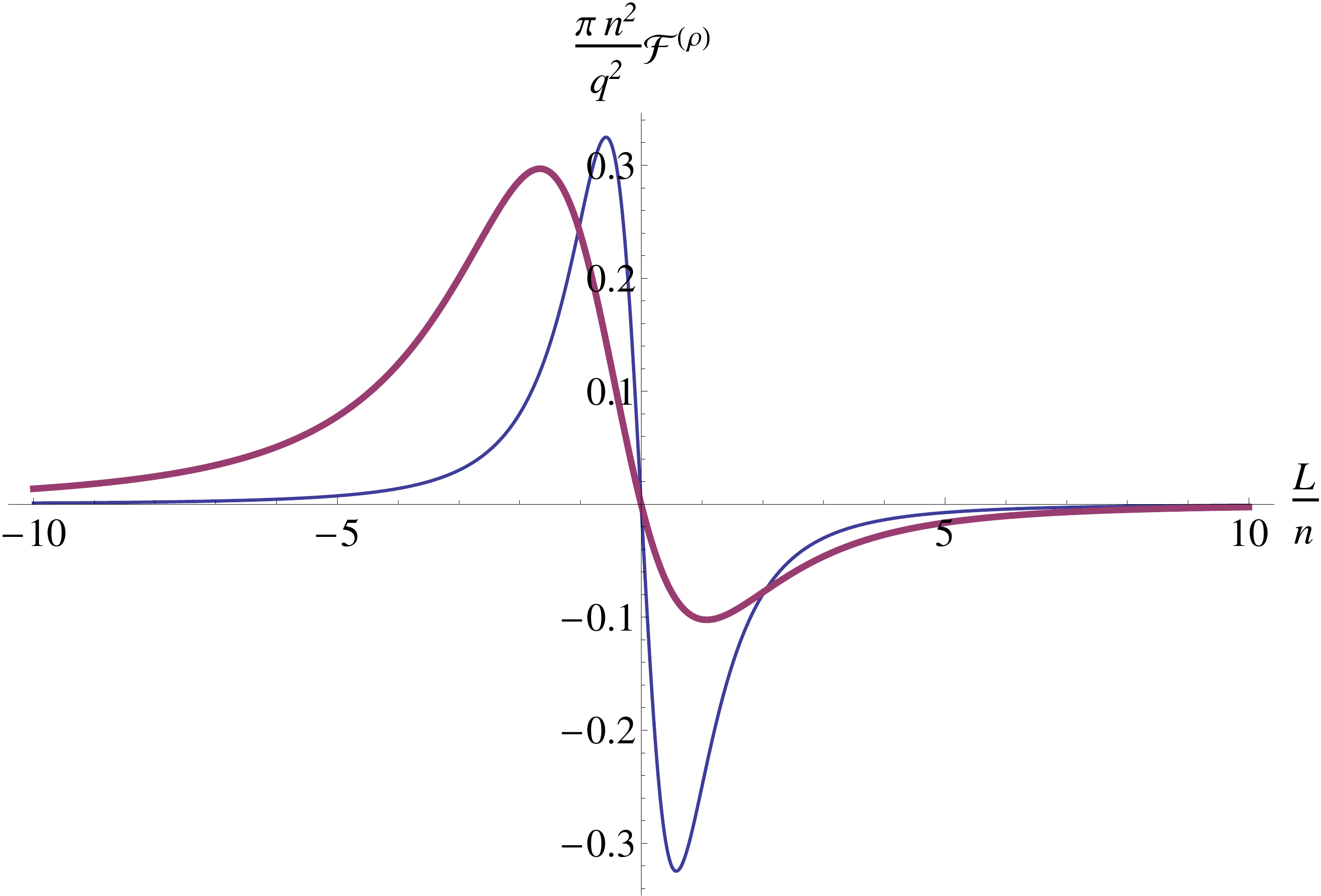}
	\end{center}
	\caption{Left panel: the self-energy, right panel: the self-force for massless (thin line) and massive ($\frac{m}{n} = 0.7$) wormholes as functions of dimensionless proper distance $L$. The function for the massive case are asymmetric.} \label{fig:sf45}
\end{figure}

The self-energy and self-force are plotted in Fig. \ref{fig:sf45} 
as a function of proper distance $L$  defined by Eq. (\ref{eq:l}). 

In all considered above cases the self-force is an attractive force. In the case of thin-shell cylindrical \cite{Celis:2012:pgaoagbtsfoacctsw} or spherical \cite{RubindeCelis:2013:pgaoagbtsfoacstsw} wormholes, the self-force may be attractive or repulsive, depending on parameters of the model and distance to wormholes throat.  
\subsubsection{The scalar self-force in the space-time of massless wormholes} \label{Sec:WHS}


In the background of metric given by line element (\ref{eq:metricD}) the field equation for the static field reads 
\begin{equation}
\Phi'' + \frac{2r'}{r}  \Phi' + \hat{L}^2 \Phi - \left(m_s^2 + \xi \mathcal{R} \right) \Phi =- 4 \pi q_s \frac{\delta^{(3)}(\vx - \vx')}{r^2\sin\theta}.
\end{equation}
Due to spherical symmetry, we extract angular and radial parts
\begin{equation}
 \Phi(\mathbf{x};\mathbf{x'}) = q_s\sum_{l=0}^\infty (2l+1) P_l(\cos \gamma )z_l(\rho,\rho').
\end{equation}
The radial part, $z_l(\rho,\rho')$, may be expressed in terms of solutions to the homogeneous equation
\begin{equation}\label{eq:scalar_radial}
	z_l'' + \frac{2 r'}r z_l' - \left(m_s^2 + \frac{l(l+1)}{r^2} + \xi \mathcal{R}\right)z_l = 0, 
\end{equation}
where the scalar curvature is given by Eq. \eqref{eq:R}. As noted by Linet \cite{Linet:1986:otweitstoacs} for the cosmic string background the self-force for the massive field is depressed by an exponential factor $e^{-Mr}$ and therefore of particular interest is the massless case. 
\bigskip

\textbf{I. Profile} $r = |\rho| + a$
\bigskip

Because the curvature has the delta-like form \eqref{eq:infshort_curvature} the boundary conditions at the throat read 
\begin{eqnarray}
z_l(+0) - z_l(-0) &=& 0,\nonumber\\
z'_l(+0) - z_l'(-0) &=& -\frac{8\xi}{a} z_l(0). \label{eq:bc_delta}
\end{eqnarray}

Let us first of all consider massless field $m_s=0$. In the limit of angle coincidence $\Omega' = \Omega$ the potential and self-potential read \cite{Bezerra:2009:Sspwst} ($\rho > \rho' >0$)
\begin{eqnarray}
 \Phi &=& \frac{q_s}{\rho -\rho'} - \frac{q_s a(1-8\xi)}{2 r r'} F\left(\frac{a^2}{rr'},1,1-4\xi\right),\nonumber \\
U_s &=& - \frac{aq_s^2(1-8\xi)}{4 r^2} F\left(\frac{a^2}{r^2},1,1-4\xi\right), \label{eq:scalar_short}
\end{eqnarray}
where the Lerch  function 
\begin{equation}
F(z,s,v) = \sum_{n=0}^\infty \frac{z^n}{(n+v)^s}
\end{equation}
may be found in Ref. \cite{Bateman:1953:Htf-1}. In contrast to the black hole case the self-energy for minimal coupling $\xi =0$ is not zero and it coincides with the electromagnetic case (\ref{eq:selfD}). We observe also that the self-energy has simple pole at specific values of $\xi_l = \frac{l+1}{4}\ (l=0,1,\ldots)$, and it is zero for $\xi = \frac{1}{8}$. The last is explained by the simple observation that the $3D$ section $t=const$ is conformally flat (see Eq. \ref{eq:conf_flat}). For a conformally flat case the self-force is zero \cite{Hobbs:1968:rdicfu}.

For the case of massive field, $m_s\not = 0$ one has ($\nu = l+ 1/2$ and $\rho > \rho' >0$) 
\begin{eqnarray}
\fl \Phi &=& \frac{q_s e^{-m_s (\rho - \rho')}}{\rho - \rho'}  -  2q_s\sum_{l=1}^\infty \nu \left.\frac{m_s a(I_\nu K_\nu' + I_\nu'K_\nu) + (8\xi-1) I_\nu K_\nu}{2m_s a K_\nu K_\nu' + (8\xi-1) K_\nu^2}\right|_{m_s a} \nonumber\\ 
\fl &\times&\frac{K_\nu(m_s r)K_\nu(m_sr')}{\sqrt{rr'}}, \nonumber\\
\fl U_s &=&   -q_s^2 \sum_{l=0}^\infty \nu \left.\frac{m_s a(I_\nu K_\nu' + I_\nu'K_\nu) + (8\xi-1) I_\nu K_\nu}{2m_s a K_\nu K_\nu' + (8\xi-1) K_\nu^2}\right|_{m_s a} \frac{K_\nu^2(m_s r)}{r}, \label{eq:us_infshort}
\end{eqnarray} 
where $I_\nu,K_\nu$ are modified Bessel functions. 

The self-energy has an infinite number of simple poles at points
\begin{equation}
\widetilde{\xi}_l = \xi_l - \frac{m_sa}{4}\frac{K_{\nu-1}(m_sa)}{K_\nu (m_sa)},\ \xi_l = \frac{l+1}{4}. \label{eq:xi_1}
\end{equation}
The origin of these poles connects with bound states \cite{Taylor:2014:ptpsfcwst,Taylor:2017:Eptpsfcwst}. Indeed, let consider bound states of homogeneous radial equation (\ref{eq:scalar_radial}) where for this kind of profile $\mathcal{R} = - 8\xi \delta (\rho)/a$. We take the solution localized at origin in the form $z_l = c_1 K_\nu (a+\rho)$ for $\rho >0$ and $z_l = c_2 K_\nu (a - \rho)$ for $\rho < 0$. The boundary conditions (\ref{eq:bc_delta}) have a solution if the main determinant equals zero. This condition exactly gives above special values of $\xi$ given by Eq. (\ref{eq:xi_1}). 

The self-potential is logarithmically divergent at the throat, $\rho =0$, for any $\xi$ except $3D$ conformally flat value $\xi = \frac{1}{8}$. Indeed, the uniform expansion of self-energy (see \cite{Abramowitz:1970:eHMFFGMT} for Debye uniform expansion ob Bessel functions)
\begin{equation}
U_s = -q_s^2 \sum_{l=0}^\infty \frac{a^\nu}{r^{\nu+1}} \left( \frac{8\xi -1}{4\nu} + O(\nu^{-2}) \right)
\end{equation}
is conversion on the throat, $r=a$, only for $3D$ conformal coupling value $\xi = \frac{1}{8}$.

\begin{figure}[ht]
	\begin{center}
		\includegraphics[width=6truecm]{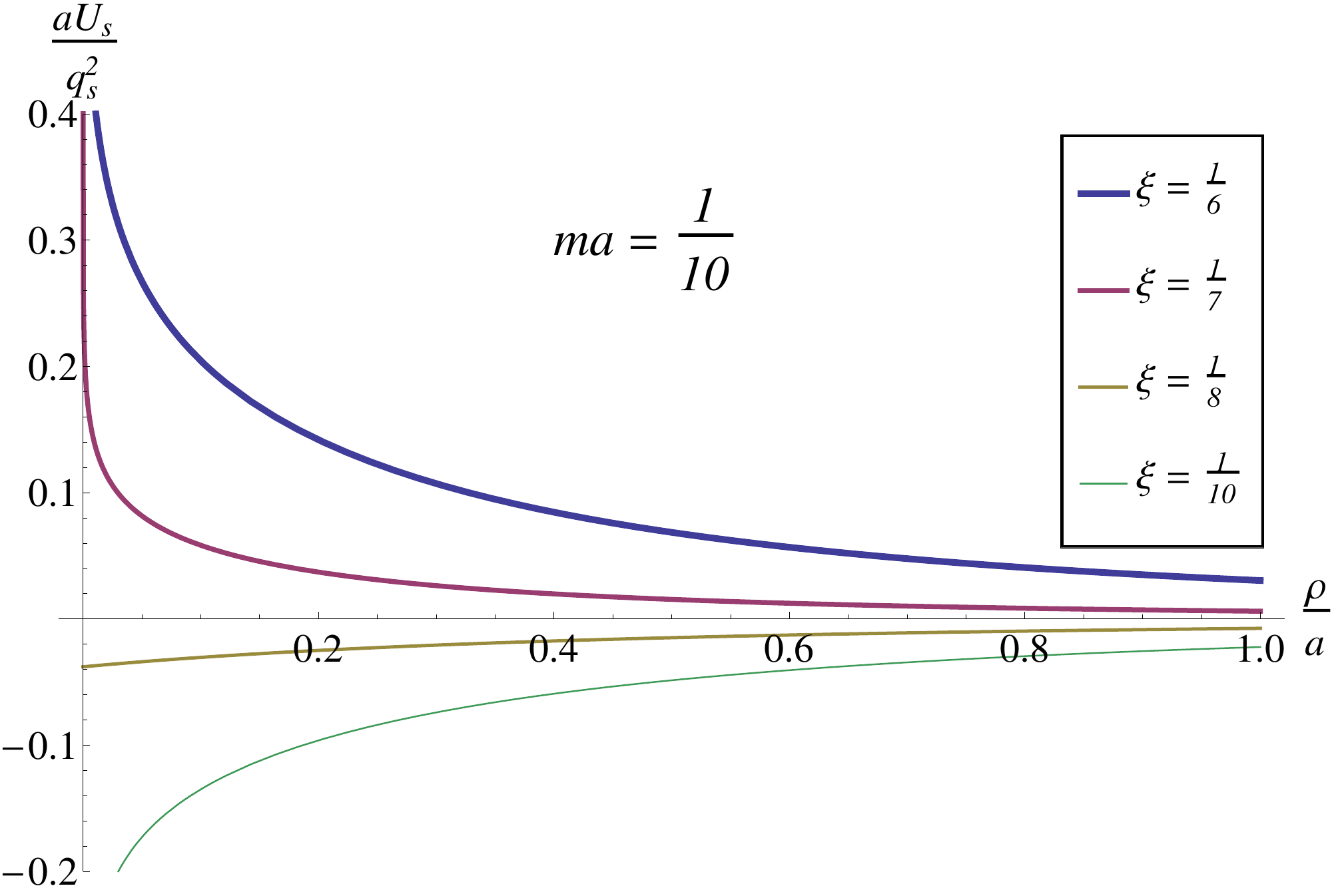}\includegraphics[width=6truecm]{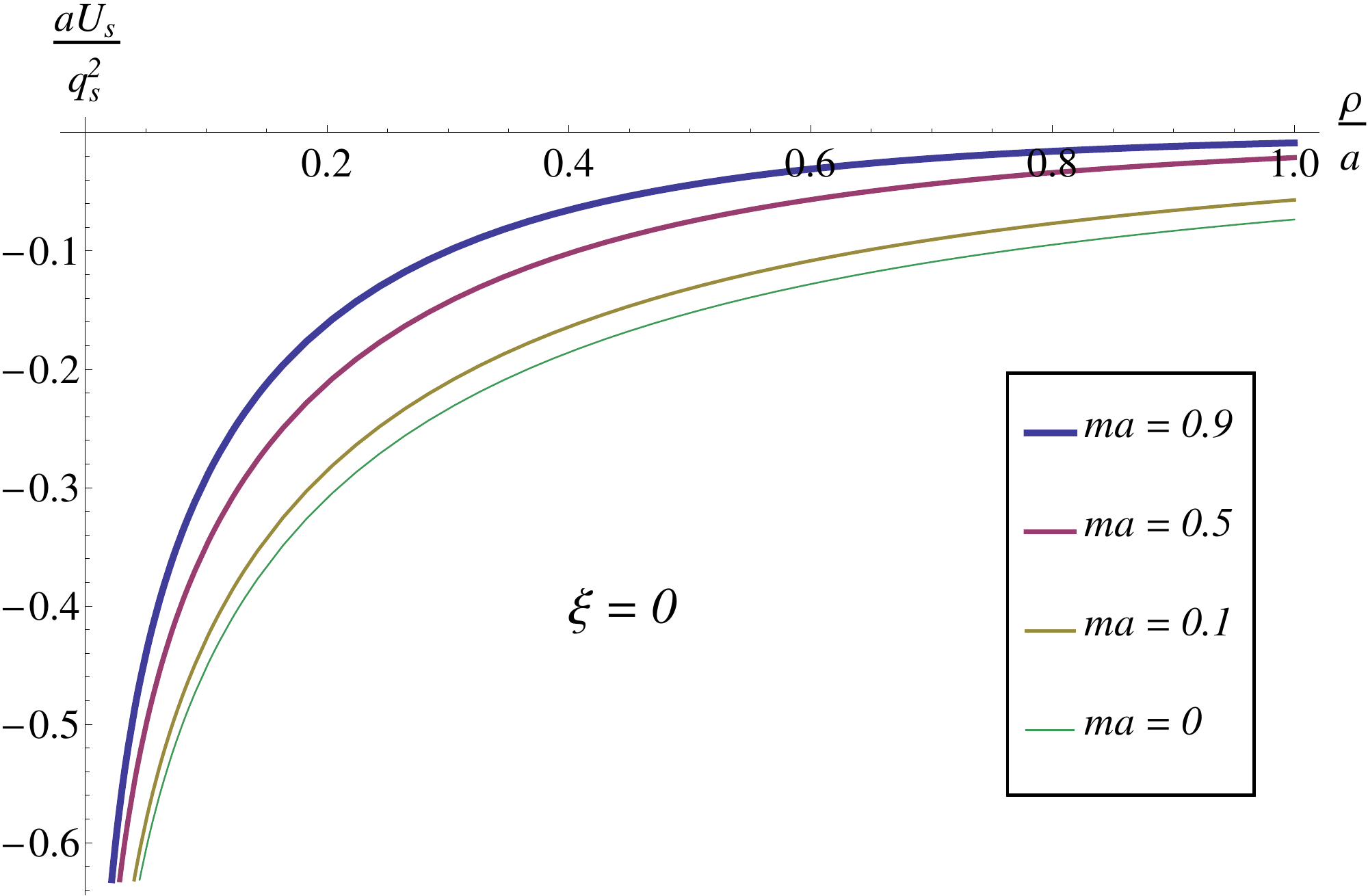}
	\end{center}
	\caption{Left panel: the self-energy for a fixed mass of fields and different values of $\xi$. Right panel: the self-force fixed non-minimal coupling  $\xi$ and different masses of fields. The greater mass of fields the closer to wormhole's throat where the self-energy is located in agreement with Linet \cite{Linet:1986:otweitstoacs}. } \label{fig:sf67}
\end{figure}

In Fig. \ref{fig:sf67} we show the numerical evaluation of Eq. (\ref{eq:us_infshort}) for different values of $\xi$ and fixed mass $m_sa = 0.1$ (left panel) and different values of $ma$ and fixed $\xi =0$ (right panel).
We observe that the self-energy for a fixed mass of field changes sign for a specific value of $\xi$. The main contribution comes from zero terms and it changes sign for 
\begin{equation*}
\xi - \frac{1}{4} - \frac{m_sa}{8} (1 - \coth m_sa) = 0.
\end{equation*}
For $m_sa = 0.1$ (left panel in Fig. \ref{fig:sf67}) we obtain $\xi = 0.137\ldots$. For fixed $\xi$ the self-force is concentrated at the particle because it is suppressed by factor $e^{-m_sr}$ (see Ref. \cite{Linet:1986:otweitstoacs}). 
\bigskip

\textbf{II. Profile} $r = \sqrt{\rho^2 + a^2}$
\bigskip

For massless case, $m_s=0$, the potential and self-energy read \cite{Taylor:2013:sfoaacsspiawst}
\begin{eqnarray}
\fl \Phi &=& \frac{q_s }{ \sqrt{\rho^2 + \rho'^2 - 2\rho\rho' \cos\gamma + a^2 \sin^2 \gamma}} \frac{\sin \left[\sqrt{2\xi} \arccos \left(-\frac{\rho\rho' + a^2 \cos \gamma}{\sqrt{\rho^2 + a^2} \sqrt{\rho'^2 + a^2}}\right)\right]}{\sin (\pi\sqrt{2\xi})},\nonumber\\
\fl U_s &=& -\frac{q_s^2 a}{2 r^2} \sqrt{2\xi} \cot (\pi \sqrt{2\xi}). \label{eq:taylor2}
\end{eqnarray}
The series expansion of the potential was obtained in Ref. \cite{Bezerra:2009:Sspwst}. 

The potential has simple poles at points $ \xi = \frac{n^2}{2},\ n = 1, 2, \ldots$ and equals to zero for $\xi = \frac{(2n-1)^2}{8},\ n = 1, 2, \ldots$. The origin of these poles is the same as for profile $r = |\rho| + a$ -- for these values of $\xi$ the bound states appear. 
\bigskip

\textbf{III. General profile of the throat}
\bigskip

The self-energy reads \cite{Bezerra:2009:Sspwst}
\begin{equation}
U_s = \frac{q_s^2}{2} \left\{- \frac 1r + \frac 1r \sum_{k=1}^\infty \zeta_H(2k,\frac 32) j_{k} + g_0\right\},
\end{equation}
where 
\begin{equation}
g_0 = \frac{\varphi^2}{(\varphi^1 \varphi^2)' r^2}\left\{ - \varphi^1 + \left. \frac{(\varphi^1 \varphi^2)'}{2\varphi^2 \varphi^2{}'}\right|_{\rho = 0} \varphi^2 \right\},
\end{equation}
and the functions $\varphi^{1,2}$ are the two independent solutions of the radial equation for zero modes
\begin{equation}
\varphi'' + \frac{2 r'}r \varphi'  - \xi \mathcal{R} \varphi = 0.
\end{equation}
The functions $j_{k}$ may be found in Ref. \cite{Bezerra:2009:Sspwst}. 

In contrast to the electromagnetic field case, the solution of the above equation for zero modes can not be obtained in closed form for any $\xi$ and, for this reason, we can not make a conclusion about the sign of the self-force far form the wormhole throat. Far from the throat, the self-energy falls as second power
\begin{equation}
U_s \approx \frac{q_s^2}{\rho^2} A, 
\end{equation}
where the sign of the constant $A$ depends on the profile of the throat and non-minimal coupling $\xi$.  

\subsection{The self-force in the space-time of thin-shell wormholes} \label{Sec:WHTS}

The different kind of wormhole space-times may be constructed by the cutting-gluing procedure -- we cut out two $3D$ domains and glue them together along the boundaries. Usually, this shell appears in the place of gluing together \cite{Visser:1995:LWfEtH}. In this way the cylindrical wormhole space-time \cite{Eiroa:2004:ctsw,Bronnikov:2009:cw,Bronnikov:2019:cwasfvpfmig} was obtained. The special case of thin-shell cylindrical wormhole \cite{Eiroa:2004:ctsw} was used for the calculation of self-force \cite{Celis:2012:pgaoagbtsfoacctsw,Celis:2016:eoactsomotesfoac}. This wormhole is, in fact, two copies of infinitely thin cosmic string space-time with metric \eqref{ds^2infhin} glued together along with cylinders of non-zero radius $r_c$. The spherical-symmetric wormhole with Schwarzschild exterior may be constructed in the same way \cite{Frolov:1990:peiwatm,Visser:1995:LWfEtH} -- two copies of Schwarzschild black holes metric \eqref{Schwarzschild} which cut for $r_c > 2M$ and glued together along with these spheres or by intermediate space-time. Locally these kinds of wormholes are cosmic string or Schwarzschild space-times but globally not -- there exist cylinder or sphere of minimal radii. Globally they are wormholes. 

In the Ref. \cite{RubindeCelis:2013:pgaoagbtsfoacstsw} the Schwarzschild thin-shell wormhole was considered. It consists of two copies of Schwarzschild space-time \eqref{Schwarzschild} with the same mass and $r=r_c > 2m$ which glued along with spheres $r=r_c$. The tetrad component of self-force is radial and reads
\begin{equation}
\mathcal{F}^{(r)} = \mathcal{F}^{(r)}_{\mathrm{bh}} + q^2 \sum_l f'_l(r) \frac{f_l(r)}{f_l(r_c)} \left[\frac{2l+1}{2r_c^2 f'_l(r_c)} + \gs_l(r_c)\right],
\end{equation} 
where $f_l$ and $\gs_l$ are two independent solutions of the radial equation in this background which firstly were found by Israel in Ref. \cite{Israel:1968:ehisest}. Far from the wormhole $\gs_l \to r^l$ and $f_l \to r^{-l-1}$. The first term $\mathcal{F}^{(r)}_{\mathrm{bh}}$ in the above equation is the self-force in the Schwarzschild space-time given by Eq. \eqref{eq:bhself} and the second one is the contribution from non-trivial topology. In the limit $M\to 0$ the infinitely short throat wormhole space-time \eqref{eq:infshort} appears. The self-force in this limit coincides exactly \cite{RubindeCelis:2013:pgaoagbtsfoacstsw} with that obtained in Ref. \cite{Khusnutdinov:2007:Scpwst} and given by Eq. \eqref{eq:infthinforce}. Far from the wormhole's throat is always attractive
\begin{equation}
\mathcal{F}^{(r)} = - \frac{q^2}{2r^3} (r_c - 2M) + O(r^{-5}).
\end{equation}
If the radius of throat $r_c$ is below $3M$, the domain close to the throat appears where self-force becomes repulsive \cite{RubindeCelis:2013:pgaoagbtsfoacstsw}. 

The thin-shell cylindrical wormhole \cite{Eiroa:2004:ctsw} was considered in Ref. \cite{Celis:2012:pgaoagbtsfoacctsw}. The space-time is two copies of infinitely thin cosmic string space-time with metric \eqref{ds^2infhin} with $r>r_c$ glued along with cylinders $r=r_c$. The self-force in this background has found in Ref. \cite{Celis:2012:pgaoagbtsfoacctsw} and it reads
\begin{equation}
\mathcal{F}^{r} = \mathcal{F}^{r}_{\mathrm{cs}} + \frac{4q^2 \nu}{\pi}\int_0^\infty k dk \sum_{n=0}^\infty{}' \frac{K_{n\nu }(kr)}{K_{n\nu }(kr_c)} K'_{n\nu }(kr) \left[I_{n\nu}(kr_c) + \frac{1}{2kr_c K'_{n\nu }(kr_c)}\right],
\end{equation} 
where $K_{n\nu}(x)$ is the modified Bessel function and prime of sum means one-half of the zero terms. The first term is self-force in the space-time of infinitely thin cosmic string given by Eq. \eqref{InThStringSF} and the second term is the contribution due to non-trivial topology. For $\nu =1$ the $\mathcal{F}^{r}_{\mathrm{cs}} = 0$, but the second part is not zero. The resulting self-force comes from a non-trivial topology contribution, and it is always attractive. For $\nu >1$ the behavior of the self-force is changed drastically. The first term, due to angle deficit appears and plays an important role. The force becomes always repulsive far from the throat and attractive close to the throat. The greater $\nu$, the closer to the throat the attractive domain. 

It is possible to construct the cylindrical wormhole \cite{Celis:2016:eoactsomotesfoac} by using two cosmic string space-times \eqref{ds^2infhin} with different parameters, $\nu_1, \nu_2$, which are cut on the radial coordinates $r_{1c}, r_{2c}$. Due to continuity the induced metric $r_{1c}/\nu_1 = r_{2c}/\nu_2$. The self-energy may be divided on the bulk term given by Eq. \eqref{InThStringSF} with corresponding parameter $\nu$ and additional term. The latter may be interpreted as energy interaction of the charge $q$ and specific charge
\begin{equation}
Q_{1,2} = - q\frac{\nu_{1,2}}{\nu_1 + \nu_2},
\end{equation}   
where the index corresponds to a position of the charge in the first or second cosmic string space-time.  

\subsection{The self-force in the space-time of wormholes with long throat} \label{Sec:WHLT} 

The metric of this space-time has the form of massless "drainhole" \eqref{eq:metricD}. The profile function $r(\rho)$ is characterized by two parameters -- the radius of throat $a$ and the length of throat $\tau$  with the relation $a \ll \tau$ holds. In fact, we have a long tube that fast passes into flat space-time.  In Refs. \cite{Popov:2010:sfoaspcitlt,Popov:2013:sfoascitltoaw,Popov:2015:ssfoscialt}, the self-force was considered for particles inside of this tube where $\rho < \tau$. The massless \cite{Popov:2010:sfoaspcitlt} and massive \cite{Popov:2015:ssfoscialt} scalar and Maxwell \cite{Popov:2013:sfoascitltoaw} fields were considered in WKB approximation. The all above conditions of the long throat are encoded in single WKB parameter $\varepsilon_{WKB} = L_*(\rho)/L(\rho) \ll 1$, where $L_* = r(\rho)$ for Maxwell field, and 
\begin{equation}
L_*(\rho) = \frac{r(\rho)}{2\xi + m^2_{\mathrm{sc}} r^2(\rho)},
\end{equation} 
for the scalar massive field. For scalar field 
\begin{equation}
\frac{1}{L(\rho)} = \max \left\{\left|\frac{r'}{r}\right|, \left|\frac{r'}{r}\sqrt{|\xi|}\right|, \left|\frac{r''}{r}\right|, \ldots\right\}, 
\end{equation}
and one should set $\xi=0$ for Maxwell field.  

The self-force for the Maxwell field has two contributions \cite{Popov:2013:sfoascitltoaw}. The first one comes from zero momentum term, $l=0$ and other terms, $l>0$, in the WKB approximation
\begin{equation}
\mathcal{F}_\rho^{\mathrm{em}} = \mathcal{F}_\rho^{l=0} + \mathcal{F}_\rho^{\mathrm{WKB}}.
\end{equation}
Here, 
\begin{eqnarray}
\mathcal{F}_\rho^{l=0} &=&\left. \frac{q^2}{2r^2} \int_0^\rho \frac{d\rho}{r^2}\right/\int_0^\infty \frac{d\rho}{r^2}\nonumber\\
\mathcal{F}_\rho^{\mathrm{WKB}} &=& - A \frac{q^2 r'}{2r^2}(1+ O(\varepsilon_{\mathrm{WKB}}^2)),
\end{eqnarray}
where $A = 0.877\ldots$. 

For scalar massive filed \cite{Popov:2010:sfoaspcitlt,Popov:2015:ssfoscialt} the self-force reads
\begin{eqnarray}
\fl \mathcal{F}_\rho^{\mathrm{sc}} &=& - \frac{q^2r'}{2r^2}\left\{\mu(\rho) - 2 \int_0^{\mu(\rho)} \frac{xdx}{(1+e^{2\pi x}) \sqrt{\mu^2(\rho) - x^2}}\right.\nonumber \\
\fl &-& \left. 4\pi m^2 r^2\int_0^{\mu(\rho)} \frac{e^{2\pi x} dx}{(1+e^{2\pi x})^2 \sqrt{\mu^2(\rho) - x^2}}\right\}(1+ O(\varepsilon_{\mathrm{WKB}}^2)),
\end{eqnarray}
where $\mu^2(\rho) = 2(\xi - \frac{1}{8}) + m^2_{\mathrm{sc}} r^2$.

\section{Concluding remarks}

We have considered gravitationally induced self-interaction force for particles in the different curved backgrounds. For a particle at rest, the scalar field or time component of the vector potential are proportional to the corresponding Green function and the self-energy may be recognized as a renormalized Green function. The DeWitt--Schwinger renormalization procedure is an appropriate scheme. The Green function has all information about the local and global structure of the space-time. Therefore, the presence of horizons, singularities, or non-trivial topological structures of space-time have a significant impact on self-force, its value, and sign and may be considered for probing the global structure of space-time. The space-time of Schwarzschild black hole and the star with corresponding mass are equal out of the star's surface. In the latter case there is no horizon and self-force "feels" this difference close to the star. Far from the star, the main term of self-force is the same, but the next term depends on the presence of the horizon. 

The space-times of topological defects namely cosmic strings and global monopoles are other examples where self-force "feels" the global structure of the space-time. The space-time of infinitely thin cosmic string and Minkowski space-time are locally identical but globally not. Any circle out of string has the same length as in Minkowski space-time, But a circle around the string has another length because the space-time of cosmic string is the multiplication of cone space and two-dimensional Minkowski space-time $\mathbb{M}_2 \otimes \mathbb{C_\nu}$. In Minkowski space-time, the self-force is zero for a particle at rest, but in the cosmic string, it is not zero. It is repulsive and tends to infinity on the string. In the space-time of cosmic string with the non-trivial internal structure, the self-force admits finite value at the center of the string. 

The space-times of wormholes have a non-trivial topological structure and have no horizons. Far from the wormhole's throat, the space-time is identical to the Schwarzschild space-time with specific parameters, but globally not. We have a bridge to another universe instead of a horizon. As a consequence, far from the black hole and the wormhole with the same radius of the throat as horizon radius, the self-force has an opposite sign but the same modulo. A particle has to be attracted to the wormhole and repelled from black-hole. From the astrophysical point of view, a wormhole's throat in the universe has to be surrounded by matter.           

\ack \label{Sec:Ack}

I would like to thank A. Popov, D. Vassilevich, M. Volkov, and A. Zelnikov for helpful comments.  The work was supported in parts by the grants 2016/03319-6 and 2019/10719-9 of the S\~ao Paulo Research Foundation (FAPESP), by the RFBR projects 19-02-00496-a.

\section*{References}\label{Sec:Ref}
\addcontentsline{toc}{section}{\nameref{Sec:Ref}}
\providecommand{\newblock}{}

\bibliographystyle{iopart-num}
\end{document}